\documentclass{aa}

\usepackage{graphicx}
\usepackage{txfonts}
\usepackage{epsfig}
\usepackage{natbib}
\newcommand{\et}    {et al.}
\newcommand{\eg}    {e.g.,}
\newcommand{\ie}    {i.e.,}
\newcommand{\jpb}   {Jy~beam$^{-1}$}
\newcommand{\kms}   {km~s$^{-1}$}
\newcommand{\supa}  {$^\mathrm{a}$}
\newcommand{\supb}  {$^\mathrm{b}$}
\newcommand{\supc}  {$^\mathrm{c}$}

\newcommand{\lo}    {$L_{\sun}$}
\newcommand{\mo}    {$M_{\sun}$}
\newcommand{\phn}   {\phantom{0}}

\newcommand{\phs}   {\phantom{$^\mathrm{a}$}}
\newcommand{\hii}   {\ion{H}{ii}}
\begin{document} 

   \title{A necklace of dense cores in the high-mass star forming region G35.20$-$0.74\,N: ALMA observations}
%   \titlerunning{A necklace of high-mass star-forming cores: ALMA observations of G35.20$-$0.74\,N}

   \author{\'A. S\'anchez-Monge
          \inst{1,2}\fnmsep\thanks{E-mail: sanchez@ph1.uni-koeln.de (\'A.\ S.-M.)}
          \and
          M.~T. Beltr\'an\inst{2}
          \and
          R. Cesaroni\inst{2}
          \and
          S. Etoka\inst{3,4}
          \and
          D. Galli\inst{2}
          \and
          M.~S.~N. Kumar\inst{5}
          \and
          L. Moscadelli\inst{2}
          \and
          T. Stanke\inst{6}
          \and
          F.~F.~S. van der Tak\inst{7,8}
          \and
          S. Vig\inst{9}
          \and
          C.~M. Walmsley\inst{2,10}
          \and
          K.-S. Wang\inst{11}
          \and
          H. Zinnecker\inst{12}
          \and
          D. Elia\inst{13}
          \and
          S. Molinari\inst{13}
          \and
          E. Schisano\inst{14}
          }

   \institute{I.\ Physikalisches Institut, Universit\"at zu K\"oln, Z\"ulpicher Str.\ 77, D-50937, K\"oln, Germany
              \and
              Osservatorio Astrofisico di Arcetri, INAF, Largo Enrico Fermi 5, I-50125, Firenze, Italy
              \and
              Hamburger Sternwarte, Gojenbergsweg 112, D-21029, Hamburg, Germany
              \and
              Jodrell Bank Centre for Astrophysics, School of Physics and Astronomy, University of Manchester, Manchester M13 9PL, UK
              \and
              Centro de Astrof\'isica da Universidade do Porto, Rua das Estrelas, 4150-762 Porto, Portugal
              \and
              ESO, Karl-Schwarzschilk-Strasse 2, D-85748 Garching bei M\"unchen, Germany
              \and
              SRON Netherlands Institute for Space Research, PO Box 800, 9700, AV Groningen, The Netherlands
              \and
              Kapteyn Astronomical Institute, University of Groningen, 9700 AV Groningen, The Netherlands
              \and
              Dpt.\ of Earth and Space Science, Indian Institute of Space Science and Technology, Thiruvananthapuram, 695 547 Kerala, India
              \and
              Dublin Institute for Advanced Studies (DIAS), 31 Fitzwilliam Place, Dublin 2, Ireland
              \and
              Leiden Observatory, Leiden University, PO Box 9513, 2300 RA Leiden, The Netherlands
              \and
              SOFIA Science Center, NASA Ames Research Center, Mailstop 232-12, Moffett Field, CA 94035, USA
              \and
              Istituto di Astrofisica e Planetologia Spaziali (IAPS-INAF), via Fosso del Cavaliere 100, I-00133 Roma, Italy
              \and
              Infrared Processing and Analysis Center, Institute of Technology, Pasadena, CA 91125, USA
             }

   \date{Received; accepted}

% \abstract{}{}{}{}{} 
% 5 {} token are mandatory
 
  \abstract
   {The formation process of high-mass stars (with masses $>$8~$M_\sun$) is still poorly understood, and represents a challenge from both the theoretical and observational points of view. The advent of the Atacama Large Millimeter Array (ALMA) is expected to provide observational evidence to better constrain the theoretical scenarios.}
   {The present study aims at characterizing the high-mass star forming region G35.20$-$0.74\,N, which is found associated with at least one massive outflow and contains multiple dense cores, one of them recently found associated with a Keplerian rotating disk.}
   {We used the radio-interferometer ALMA to observe the G35.20$-$0.74\,N region in the submillimeter continuum and line emission at 350~GHz. The observed frequency range covers tracers of dense gas (\eg\ H$^{13}$CO$^+$, C$^{17}$O), molecular outflows (\eg\ SiO), and hot cores (\eg\ CH$_3$CN, CH$_3$OH). These observations were complemented with infrared and centimeter data.}
   {The ALMA 870~$\mu$m continuum emission map reveals an elongated dust structure ($\sim$0.15~pc long and $\sim$0.013~pc wide; full width at half maximum) perpendicular to the large-scale molecular outflow detected in the region, and fragmented into a number of cores with masses $\sim$1--10~$M_\sun$ and sizes $\sim$1600~AU (spatial resolution $\sim$960~AU). The cores appear regularly spaced with a separation of $\sim$0.023~pc. The emission of dense gas tracers such as H$^{13}$CO$^+$ or C$^{17}$O is extended and coincident with the dust elongated structure. The three strongest dust cores show emission of complex organic molecules characteristic of hot cores, with temperatures around 200~K, and relative abundances 0.2--2$\times10^{-8}$ for CH$_3$CN and 0.6--5$\times10^{-6}$ for CH$_3$OH. The two cores with highest mass (cores~A and B) show coherent velocity fields, with gradients almost aligned with the dust elongated structure. Those velocity gradients are consistent with Keplerian disks rotating about central masses of 4--18~$M_\sun$. Perpendicular to the velocity gradients we have identified a large-scale precessing jet/outflow associated with core~B, and hints of an east-west jet/outflow associated with core~A.}
   {The elongated dust structure in G35.20$-$0.74\,N is fragmented into a number of dense cores that may form high-mass stars. Based on the velocity field of the dense gas, the orientation of the magnetic field, and the regularly spaced fragmentation, we interpret this elongated structure as the densest part of a 1D filament fragmenting and forming high-mass stars.}

   \keywords{stars: formation --
             stars: massive --
             ISM: molecules --
             ISM: jets and outflows --
             ISM: individual objects: G35.20$-$0.74\,N}

   \maketitle
%
%________________________________________________________________

%----------------------------------------------------------------------
\section{Introduction\label{s:intro}}

High-mass stars (O- and B-type stars with masses $\ga$8~\mo) are crucial for the understanding of many physical phenomena in the Galaxy; however, their first stages of formation are still poorly understood and represent a challenge from both a theoretical and observational point of view. A number of different theoretical scenarios have been proposed to explain the formation of OB-type stars (see \citealt{zinneckeryorke2007} and \citealt{tan2014} for a review). Even with their differences, the most accepted theories (monolithic collapse in a turbulence-dominated core --- \citealt{krumholz2009}; competitive accretion driven by a stellar cluster --- \citealt{bonnellbate2006}; Bondi-Hoyle accretion --- \citealt{keto2007}) agree in the prediction of the formation of circumstellar disks around stars of all masses. However, while in the low-mass regime, circumstellar disks have been extensively identified and studied \citep[\eg][]{simon2000}, there are only a few convincing disks (with sizes $\le$1000~AU) found around B-type protostars \citep[\eg][]{schreyer2002, cesaroni2005, cesaroni2007, carrascogonzalez2012, wang2012}. Moreover, convincing evidence of centrifugally supported disks around O-type protostars still remains elusive. Infrared interferometric observations of the CO overtone bands suggest that disks might be associated with O-type stars \citep[\eg][]{kraus2010, dewit2011, boley2013}, but (sub)millimeter images have only revealed the existence of huge ($\sim$0.1~pc) and massive ($\sim$100~\mo) cores undergoing solid-body rotation, the so-called `toroids' \citep[\eg][]{beltran2011}. The detection of circumstellar disks, with expected sizes $\le$1000~AU \citep[\eg][]{cesaroni2007}, requires angular resolutions $<0\farcs1$ (assuming a typical distance of 5~kpc). Thus, the lack of observed disks is expected to dramatically change with the advent of the Atacama Large Millimeter Array (ALMA), with which resolutions $0\farcs1$ will be easily obtained.

In a progressive effort to confront the theoretical scenario of the formation of massive stars through disk-mediated accretion, and to set the stage for a follow-up quest for the more challenging disks around O-type stars, we carried out ALMA Cycle~0 observations towards two IR disk candidates around B-type protostars in G35.20$-$0.74\,N and G35.03$+$0.35. Results on G35.03$+$0.35 are presented in a parallel paper by Beltr\'an \et\ (A\&A, submitted), while first results on G35.20$-$0.74\,N have been already published in \citet{sanchezmonge2013b}. In this work, we present the ALMA results on G35.20$-$0.74\,N in a more extensive and detailed way.

\begin{figure*}[t!]
\begin{center}
\begin{tabular}[b]{c}
 \epsfig{file=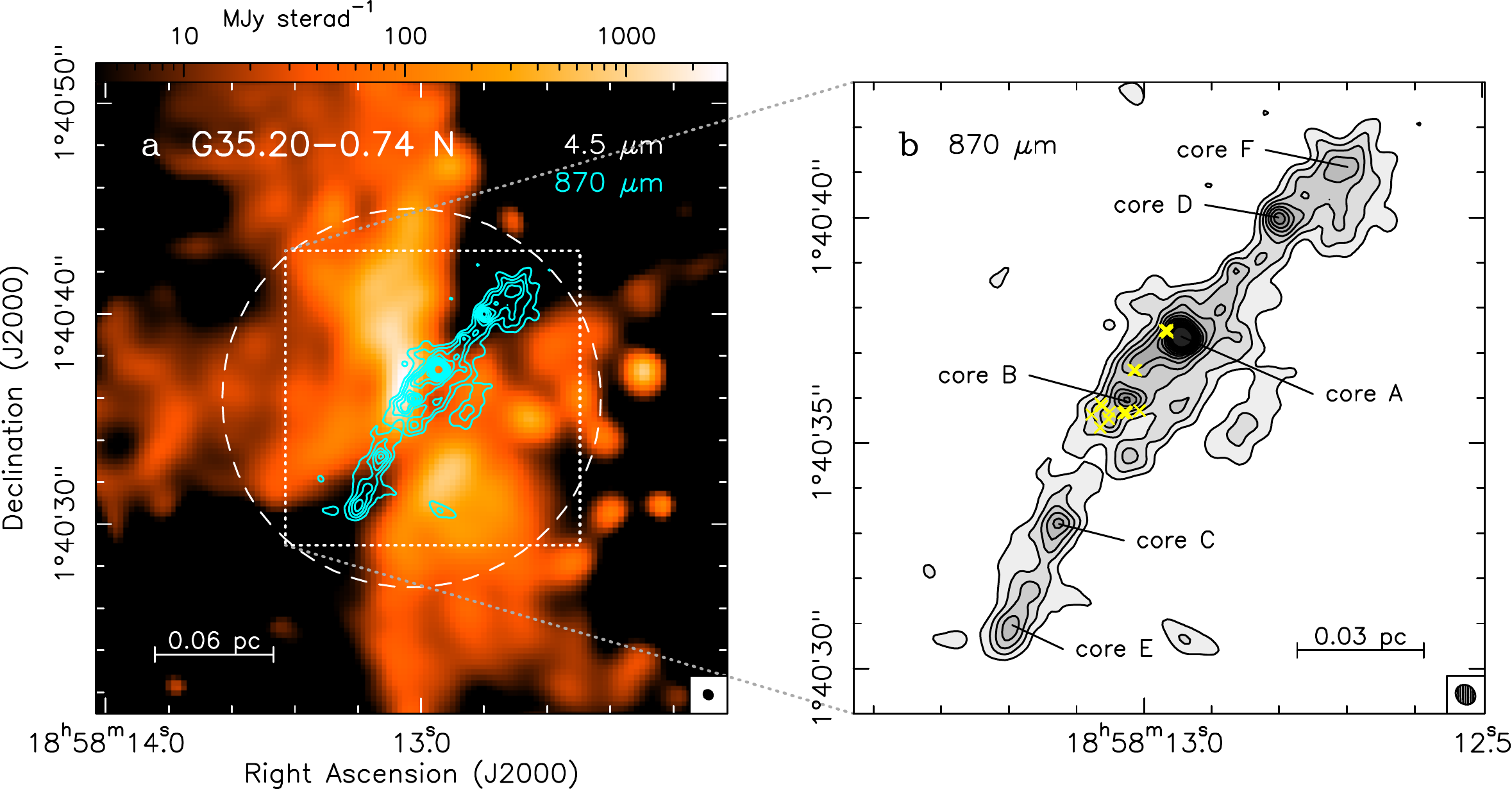, width=0.85\textwidth, angle=0} \\
\end{tabular}
\caption{\textbf{a)} \emph{Spitzer} 4.5~$\mu$m image (in logarithmic scale) of the star forming region G35.20$-$0.74\,N, overlaid with a contour map of the 870~$\mu$m (350~GHz) continuum emission obtained with ALMA. The IR image has been obtained by applying HiRes deconvolution \citep{velusamy2008} to the \emph{Spitzer}/IRAC data. The submillimeter map has been corrected for primary beam attenuation. Contours start at 5$\sigma$, increasing in steps of 6$\sigma$ (with $\sigma$=1.8~m\jpb), up to a maximum level of 201.7~m\jpb. The dashed white circle denotes the primary beam of the ALMA 12-m antennas ($\sim$9\arcsec), while the ALMA synthesized beam ($0\farcs474\times0\farcs411$, PA=46\degr) is shown in the bottom-right corner of the panel. \textbf{b)} Close-up of the central region that shows the 870~$\mu$m continuum emission map obtained with ALMA. Six cores (from A to F) have been identified and marked in the image (see Table~\ref{t:continuum}). Yellow crosses mark the position of OH masers \citep{hutawarakorncohen1999}. The ALMA synthesized beam is shown in the bottom-right corner of the panel. The spatial scale is also indicated in both panels.}
\label{f:continuum}
\end{center}
\end{figure*}

G35.20$-$0.74 is a star forming complex located at a distance of $2.19^{+0.24}_{-0.20}$~kpc \citep{zhang2009}. The main site of high-mass star formation activity, known as G35.20$-$0.74\,N (hereafter G35.20N), is associated with the IRAS source 18556$+$0136 and has a bolometric luminosity $\sim$1--$10\times10^4$~\lo\ \citep[\eg][]{gibb2003, sanchezmonge2013b, zhang2013}. In the infrared, the region is dominated by the presence of a butterfly-shaped nebula oriented NE--SW (see Fig.~\ref{f:continuum}a). This direction is coincident with the orientation (P.A.$\approx$58\degr) of a bipolar outflow observed in $^{12}$CO \citep[\eg][]{dent1985a, gibb2003, birks2006, lopezsepulcre2009, qiu2013}. The $^{12}$CO\,(1--0) line emission appears to trace also a N--S collimated flow \citep[see][]{birks2006}, coinciding with a thermal radio jet \citep[\eg][]{heatonlittle1988, gibb2003} seen also at IR wavelengths \citep[\eg][]{dent1985b, walther1990, fuller2001, debuizer2006, zhang2013}. The different orientations between the poorly collimated NE--SW outflow and the N--S jet have been interpreted as manifestations of the same flow undergoing precession \citep[\eg][]{little1998, gibb2003}, or as emission coming from a number of flows driven by the different sources forming in the G35.20N complex \citep[\eg][]{gibb2003, birks2006}.

Emission from dense gas tracers (\eg\ NH$_3$, C$^{18}$O, CS, HCN, H$^{13}$CO$^+$), as well as continuum dust emission at millimeter wavelengths \citep[\eg][]{little1985, brebner1987, gibb2003, lopezsepulcre2009}, have revealed the presence of a dense clump elongated perpendicular to the NE--SW outflow, with a velocity gradient from NW ($\sim$32~\kms) to SE ($\sim$36~\kms). This velocity gradient was first thought to originate from a large ($\sim$0.6~pc) flattened structure rotating about the axis of the NE--SW outflow. Subsequently, on the basis of their H$^{13}$CO$^+$ and H$^{13}$CN observations, \citet{gibb2003} propose that this is actually a fragmented rotating envelope containing multiple young stellar objects. The high angular resolution map at 870~$\mu$m obtained with ALMA \citep{sanchezmonge2013b} reveals an elongated structure containing a chain of compact cores (see Fig.~\ref{f:continuum}b), and confirms a significant level of clumpiness. The two strongest cores (named core~A and core~B in \citealt{sanchezmonge2013b}) are found close to the center of the elongated structure (core~A to the NW and core~B to the SE, separated by $\sim$2\arcsec). These cores, detected in several complex organic molecules, show velocity gradients roughly aligned with the elongated structure, \ie\ perpendicular to the large-scale NE--SW outflow, but opposite to each other. The kinematics of core~B, which is located at the center of symmetry of the outflow/jet, have been modeled as a Keplerian disk rotating about a central mass of $\sim$18~\mo\ \citep{sanchezmonge2013b}.

%----------------------------------------------------------------------
\begin{table}[t!]
\caption{\label{t:beams}Molecular transitions studied in this paper}
\begin{tabular}{l c c c c c}
\hline\hline
\cline{4-5}
&Frequency\supa
&$E_\mathrm{upper}/k$\supa
\\
Observation
&(MHz)
&(K)
\\
\hline
CH$_3$OCHO\,(27$_{7,21}$--26$_{7,20}$)~A		&335\,015.90\phs		&443.5		\\
CH$_3$OH\,(2$_{2,1}$--3$_{1,2}$)				&335\,133.57\phs		&\phn44.7	\\
$^{13}$CH$_3$OH\,(12$_{1,11}$--12$_{0,12}$)	&335\,560.21\phs		&192.7		\\
CH$_3$OH\,(7$_{1,7}$--6$_{1,6}$)				&335\,582.02\phs		&\phn79.0	\\
CH$_3$OCHO\,(27$_{9,18}$--26$_{9,17}$)~A		&336\,111.32\phs		&277.9		\\
CH$_3$OH\,(14$_{7,8}$--15$_{6,9}$)			&336\,438.21\phs		&488.2		\\
HC$_3$N\,(37--36)							&336\,520.08\phs		&306.9		\\
CH$_3$OH\,(12$_{1,11}$--12$_{0,12}$)			&336\,865.15\phs		&197.1		\\
CH$_3$OCHO\,(26$_{6,20}$--25$_{6,19}$)~E		&336\,889.20\phs		&235.5		\\
CH$_3$OCHO\,(26$_{6,20}$--25$_{6,19}$)~A		&336\,918.18\phs		&235.5		\\
C$^{17}$O\,(3--2)							&337\,061.05\phs		&\phn32.4	\\
C$^{34}$S\,(7--6)							&337\,396.69\phs		&\phn50.2	\\
CH$_3$OH\,(7$_{0,7}$--6$_{0,6}$)~$v_t=1$		&337\,748.83\phs		&488.5		\\
CH$_3$OH\,(7$_{1,6}$--6$_{1,5}$)~$v_t=1$		&337\,969.44\phs		&390.1		\\
H$_2$CS\,(10$_{1,10}$--9$_{1,9}$)			&338\,083.20\phs		&102.4		\\
SO$_2$\,(18$_{4,14}$--18$_{3,15}$)			&338\,305.99\phs		&196.8		\\
CH$_3$OH\,(7$_{3,4}$--6$_{3,3}$)				&338\,583.22\phs		&112.7		\\
CH$_3$OH\,(7$_{2,5}$--6$_{2,4}$)				&338\,639.80\phs		&102.7		\\
H$^{13}$CO$^+$\,(4--3)						&346\,998.34\phs		&\phn41.6	\\
$^{13}$CH$_3$OH\,(14$_{1,13}$--14$_{0,14}$)	&347\,188.28\phs		&254.3		\\
SiO\,(8--7)									&347\,330.63\phs		&\phn75.0	\\
CH$_3$OCHO\,(28$_{10,19}$--27$_{10,18}$)~E	&347\,628.34\phs		&306.8		\\
CH$_3$OCHO\,(28$_{6,23}$--27$_{6,2}$)~E		&348\,049.89\phs		&266.1		\\
CH$_3$OCHO\,(28$_{6,23}$--27$_{6,2}$)~A		&348\,065.97\phs		&266.1		\\
H$_2$CS\,(10$_{1,9}$--9$_{1,8}$)				&348\,534.36\phs		&105.2		\\
CH$_3$OH\,(14$_{1,13}$--14$_{0,14}$)			&349\,107.00\phs		&260.2		\\
CH$_3$$^{13}$CN\,(19$_K$--18$_K$)			&349\,280.77\supb	&\ldots		\\
CH$_3$CN\,(19$_K$--18$_K$)					&349\,453.70\supb	&\ldots		\\
HNCO\,(16$_{1,16}$--15$_{1,15}$)				&350\,333.35\phs		&186.2		\\
CH$_3$CN\,($K,l$)~$v_8=1$					&350\,552.44\supc	&\ldots		\\
\hline
\end{tabular}

\supa The frequencies are obtained from the CDMS (Cologne Database for Molecular Spectroscopy; \citealt{muller2001}) and JPL (Jet Propulsion Laboratory; \citealt{pickett1998}) catalogs. $E_\mathrm{upper}$ is the upper level energy, and $k$ the Boltzmann constant.\\
\supb The frequency is that of the $K$=0 component. \\
\supc The frequency is that of the $K,l$=2,$+1$ component. 
\end{table}
%----------------------------------------------------------------------
%----------------------------------------------------------------------
%----------------------------------------------------------------------
%----------------------------------------------------------------------
\begin{table*}[t!]
\caption{\label{t:continuum}Parameters of the cores detected in the 870~$\mu$m ALMA continuum map (see Fig.~\ref{f:continuum}).}
\begin{tabular}{c c c c c c c c}
\hline\hline
&\multicolumn{2}{c}{Peak position}
&
&
&\multicolumn{2}{c}{Source Diameter\supb}
\\
\cline{2-3}\cline{6-7}
&$\alpha\mathrm(J2000)$
&$\delta\mathrm(J2000)$
&$I_\mathrm{peak}$\supa
&$S_\nu$\supa
&$\theta_\mathrm{S}$
&$D_\mathrm{S}$
\\
Core
&h m s
&$\degr$ $\arcmin$ $\arcsec$
&(\jpb)
&(Jy)
&(arcsec)
&(AU)
\\
\hline  %Fluxes measured at 5sigma level, and sizes from the 50% contour level polygon.
A	&18 58 12.945	&01 40 37.36		&0.208	&$0.692\pm0.023$		&0.414	&\phn900	\\ 
B	&18 58 13.025	&01 40 35.92		&0.076	&$0.333\pm0.008$		&0.838	&1800	\\ 
C	&18 58 13.126	&01 40 33.20		&0.056	&$0.190\pm0.006$		&0.717	&1600	\\ 
D	&18 58 12.801	&01 40 40.00		&0.083	&$0.229\pm0.007$		&0.571	&1300	\\ 
E	&18 58 13.195	&01 40 30.96		&0.048	&$0.162\pm0.005$		&0.842	&1900	\\ 
F	&18 58 12.699	&01 40 41.12		&0.049	&$0.276\pm0.006$		&1.181	&2600	\\ 
\hline
\end{tabular}

\supa\ Primary beam corrected peak intensity ($I_\mathrm{peak}$) and flux density ($S_\nu$) integrated within the 5$\sigma$ contour level.\\
\supb\ Deconvolved average diameter of the 50\% contour level \citep[following][]{sanchezmonge2013a}.\\
\end{table*}
%----------------------------------------------------------------------

Finally, emission from different maser species (H$_2$O, OH, CH$_3$OH) has been detected toward the G35.20N region \citep[\eg][]{brebner1987, forstercaswell1989, forstercaswell1999, hutawarakorncohen1999, vlemmings2008, sugiyama2008, surcis2012}. The highest angular-resolution observations show two main regions for the maser emission, separated by about $2\farcs3$ ($\sim$5000~AU), and oriented SE-NW. \citet{hutawarakorncohen1999} measured magnetic field strengths (through 1665~MHz OH Zeeman-splitting observations) between $-$2.5~mG to the NW, and $+$5.2~mG to the SE. Polarimetric observations of 6.7~GHz CH$_3$OH masers \citep{surcis2012}, indicate strong (5--12\%) linear polarization, with the magnetic field oriented almost E--W, P.A.$\approx$80--110\degr. Recent polarimetric dust observations carried out with the SMA \citep{qiu2013}, revealed that the magnetic field is aligned with the long axis of the elongated dust structure to the NW roughly following the large-scale field (PA$\approx$56\degr; \citealt{valleebastien2000}), but changes its direction drastically by approximately 90\degr\ when moving to the south of the filament (at a position between cores~B and C; see Fig.~2 in \citealt{qiu2013}). The field strength is estimated to be $\sim$1~mG, \ie\ consistent with the strength measured via Zeeman-splitting observations \citep{hutawarakorncohen1999}.

Given the complexity of this region and the (sometimes contradictory) findings of previous observations, the main goal of our ALMA observations was to address some fundamental questions such as the fragmentation level of the elongated structure, the kinematics of the dense gas at different scales and the number of outflows and powering sources. This paper is organized as follows. In Sect.~\ref{s:obs}, we describe our ALMA observations. In Sect.~\ref{s:res}, we present the main results and reconstruct the spectral energy distribution of G35.20N from the millimeter to the near IR. In Sect.~\ref{s:analysis}, we analyze and discuss our findings focusing on the kinematical properties of the structures detected. Finally, in Sect.~\ref{s:conclusions} we summarize the main results and draw the conclusions.

%----------------------------------------------------------------------
\begin{figure*}[htp!]
\begin{center}
\begin{tabular}[b]{c c c}
 \epsfig{file=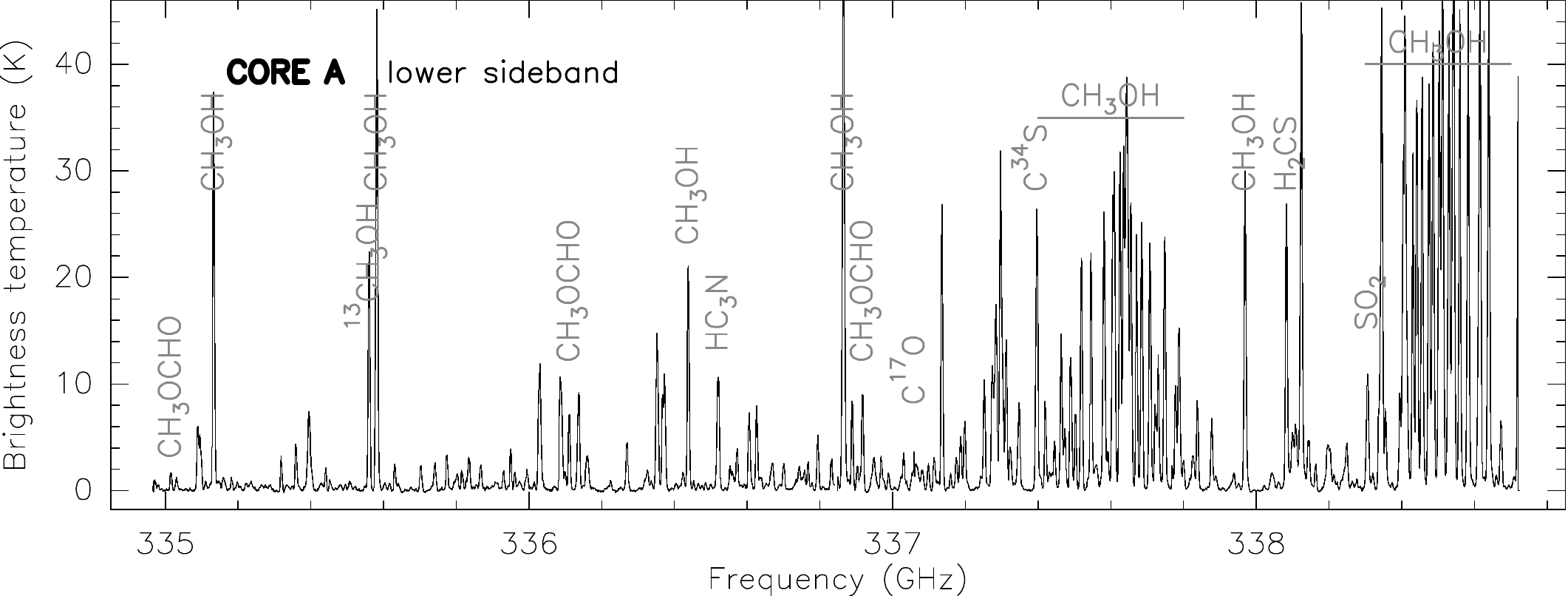, width=0.85\columnwidth, angle=0} &&
 \epsfig{file=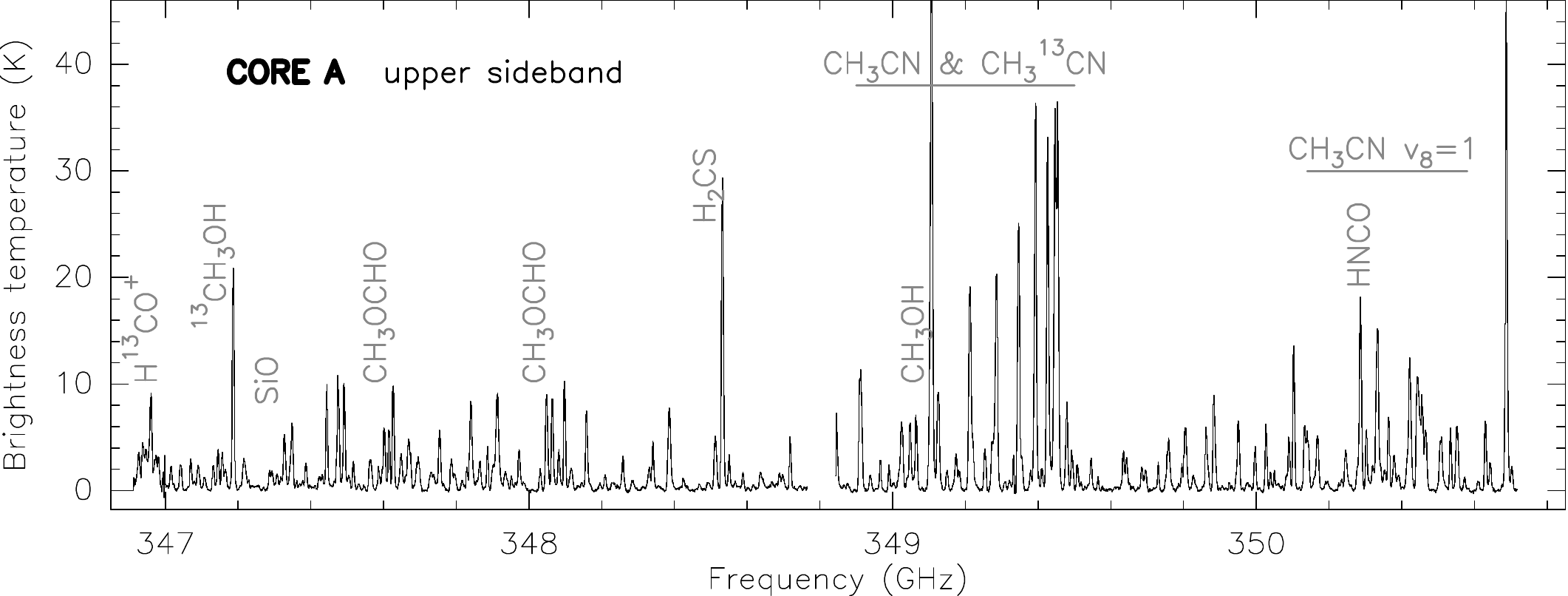, width=0.85\columnwidth, angle=0} \\
 \epsfig{file=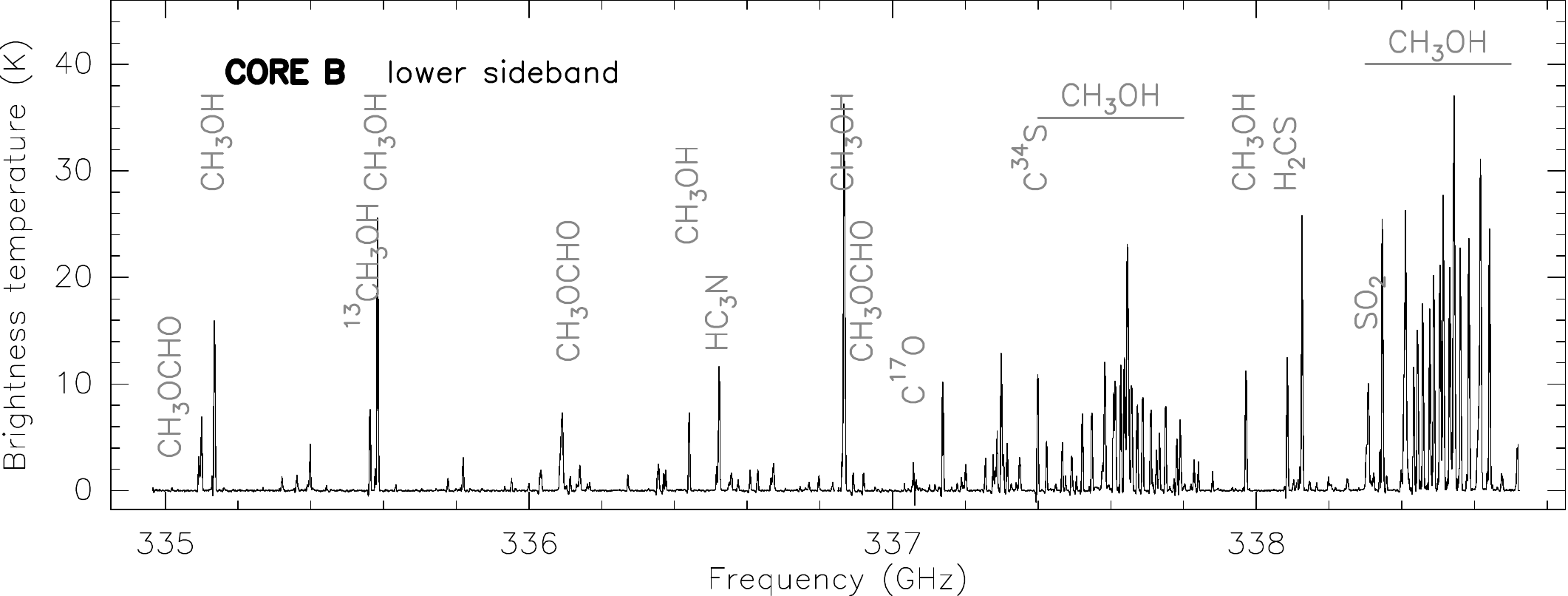, width=0.85\columnwidth, angle=0} &&
 \epsfig{file=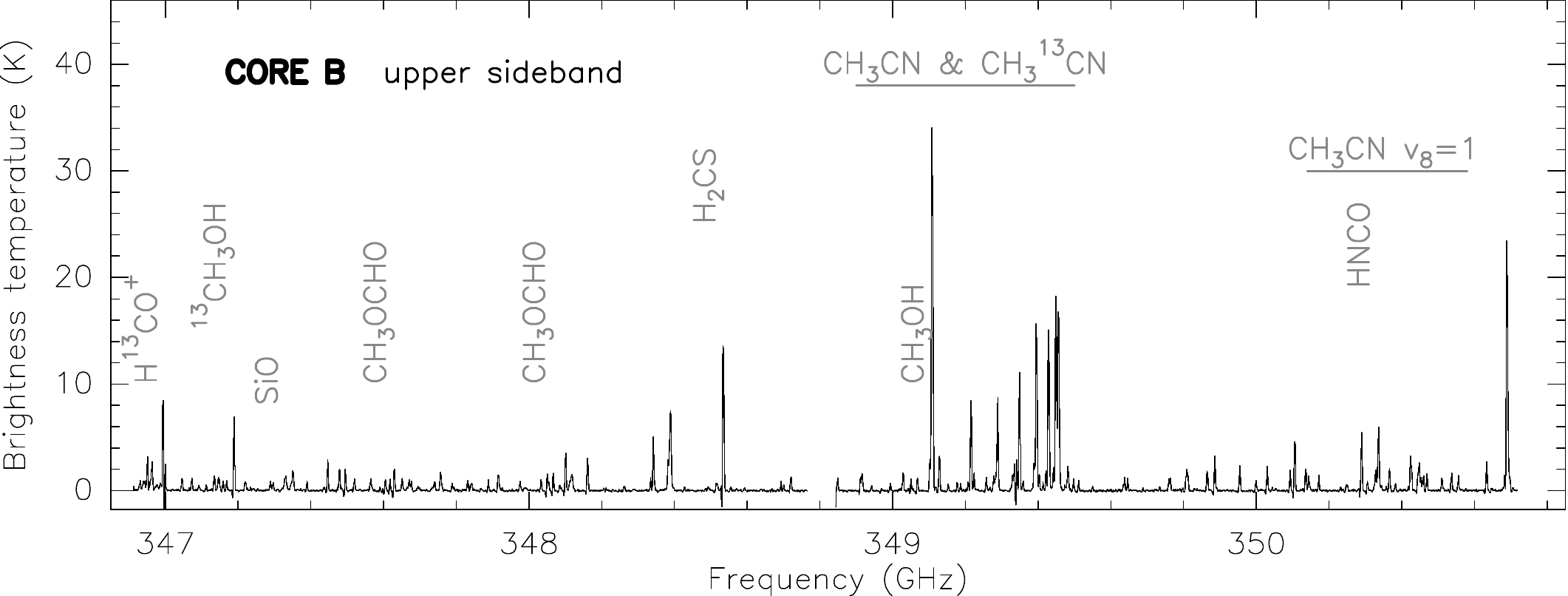, width=0.85\columnwidth, angle=0} \\
 \epsfig{file=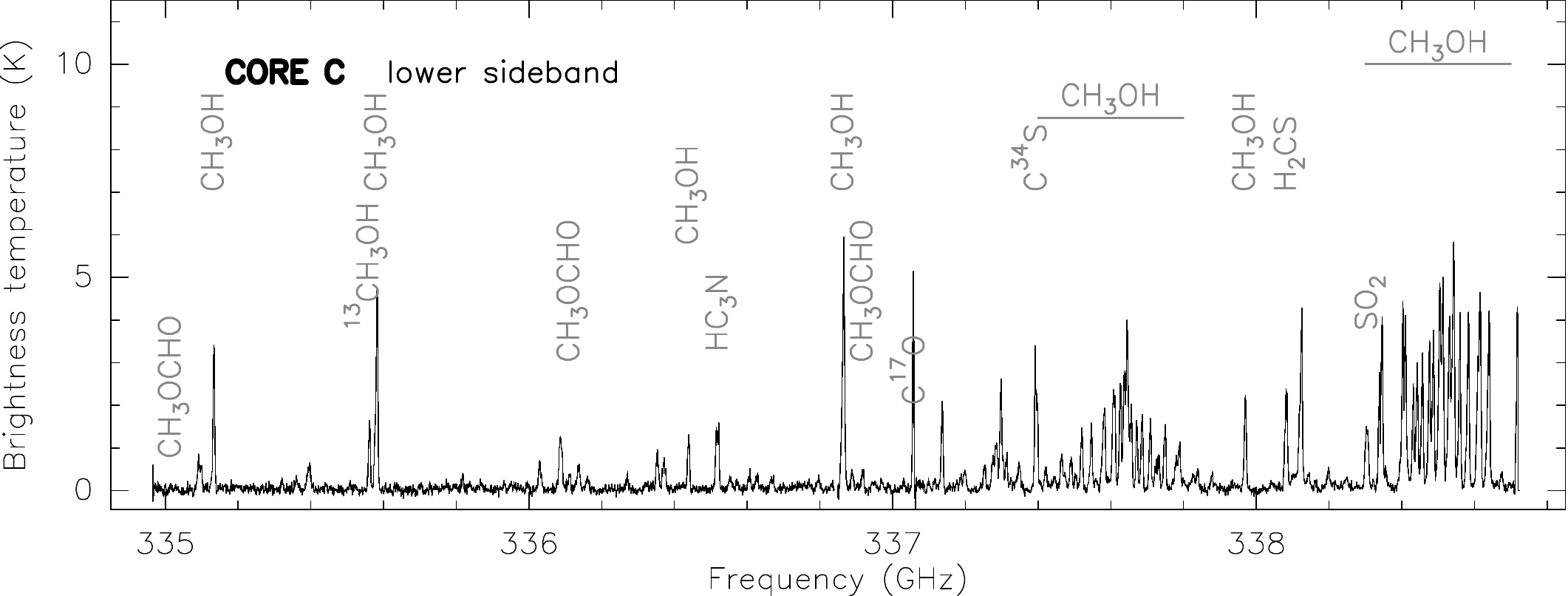, width=0.85\columnwidth, angle=0} &&
 \epsfig{file=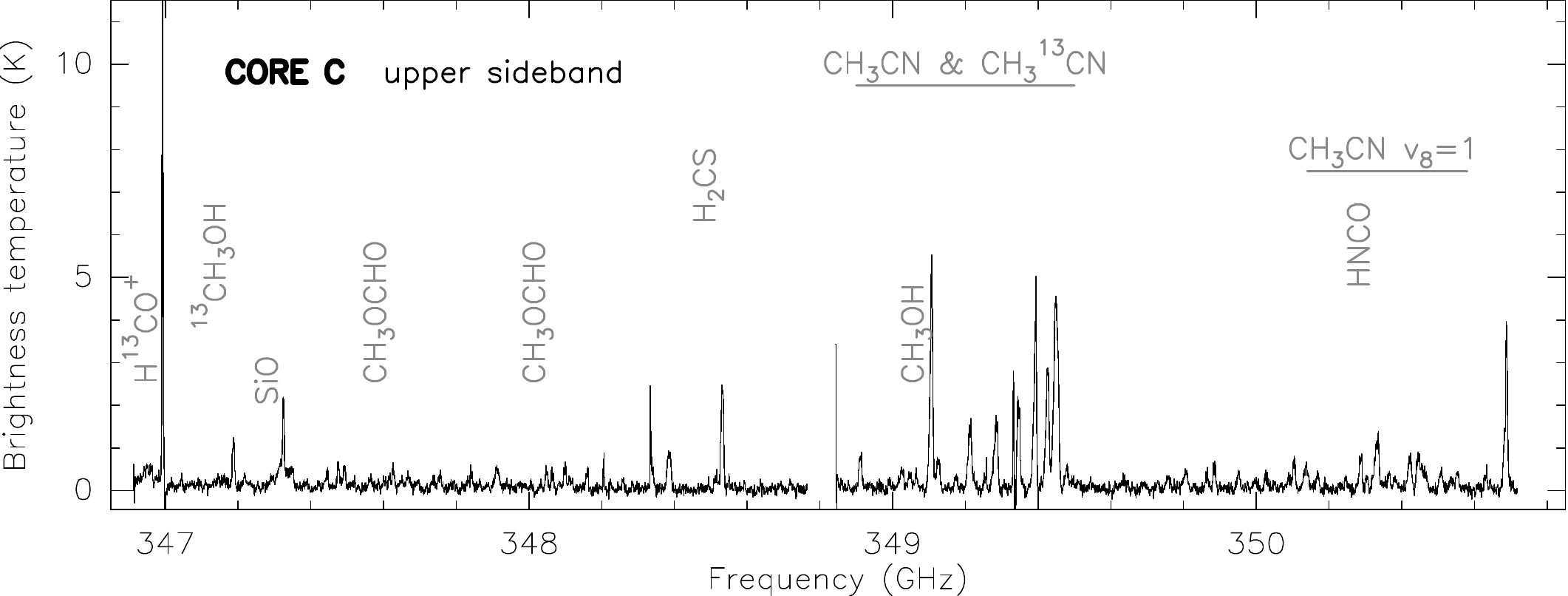, width=0.85\columnwidth, angle=0} \\
 \epsfig{file=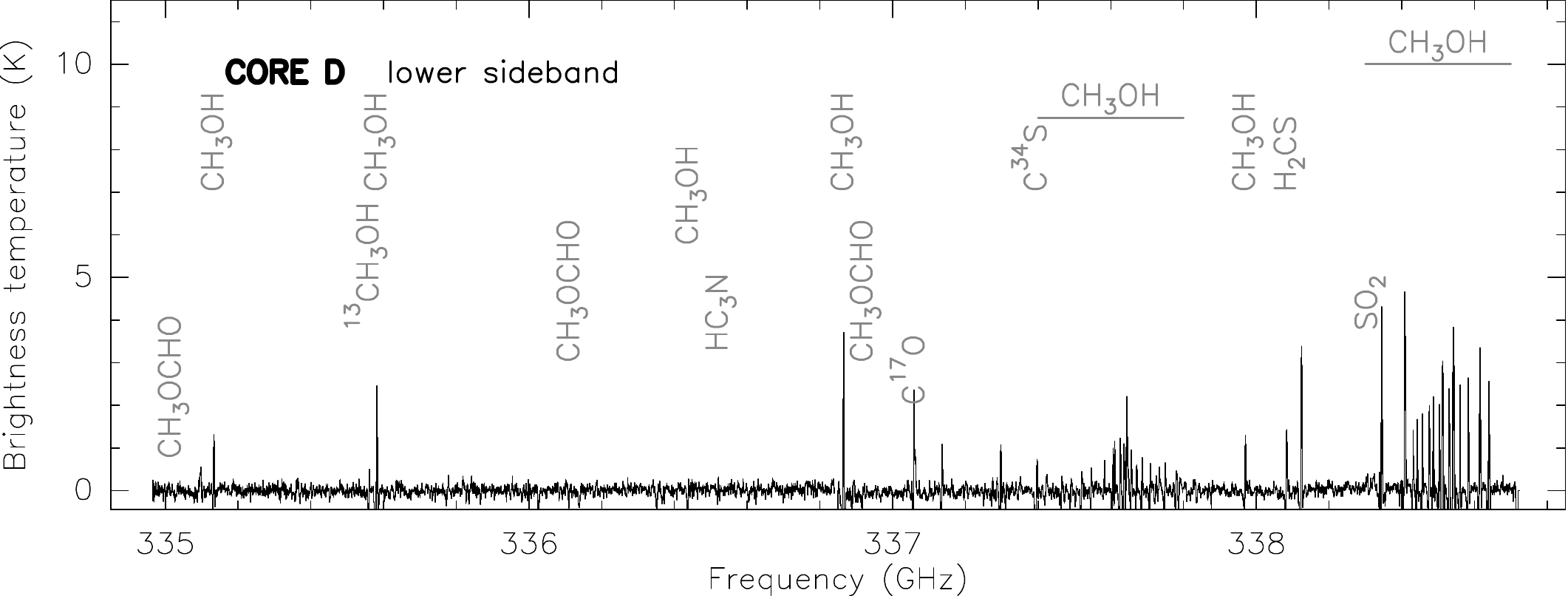, width=0.85\columnwidth, angle=0} &&
 \epsfig{file=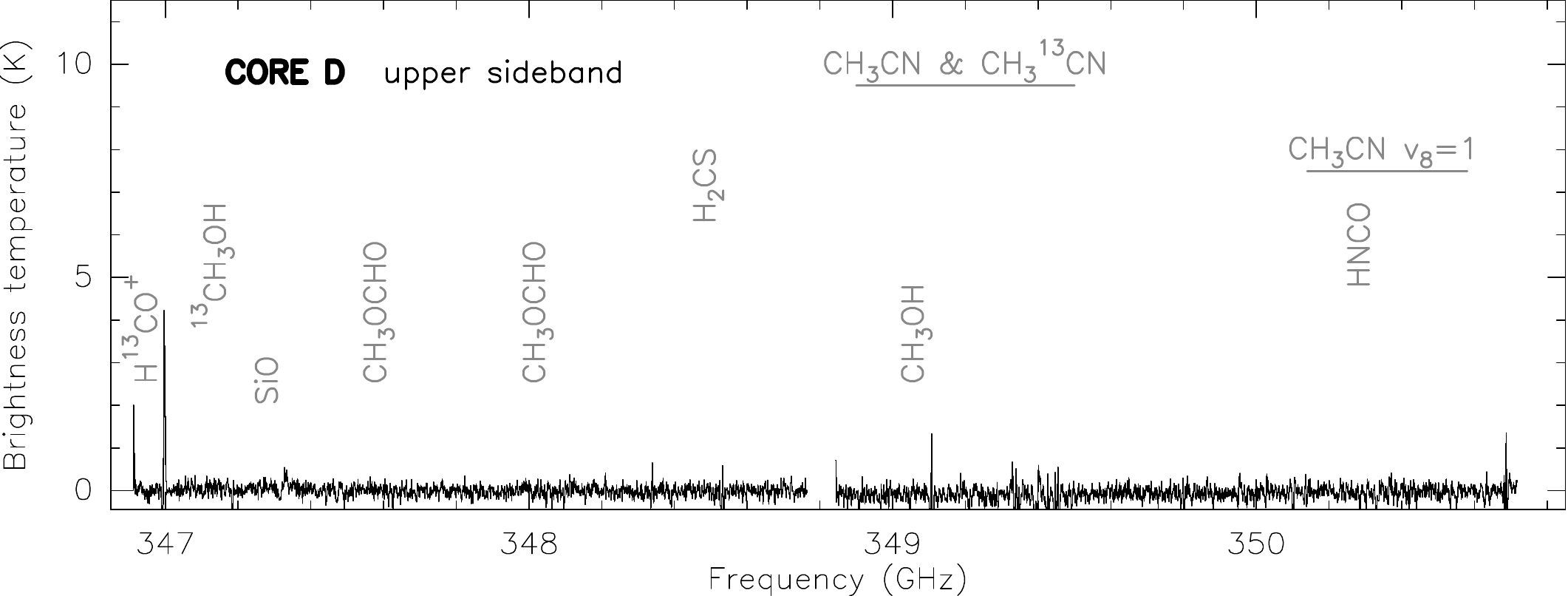, width=0.85\columnwidth, angle=0} \\
 \epsfig{file=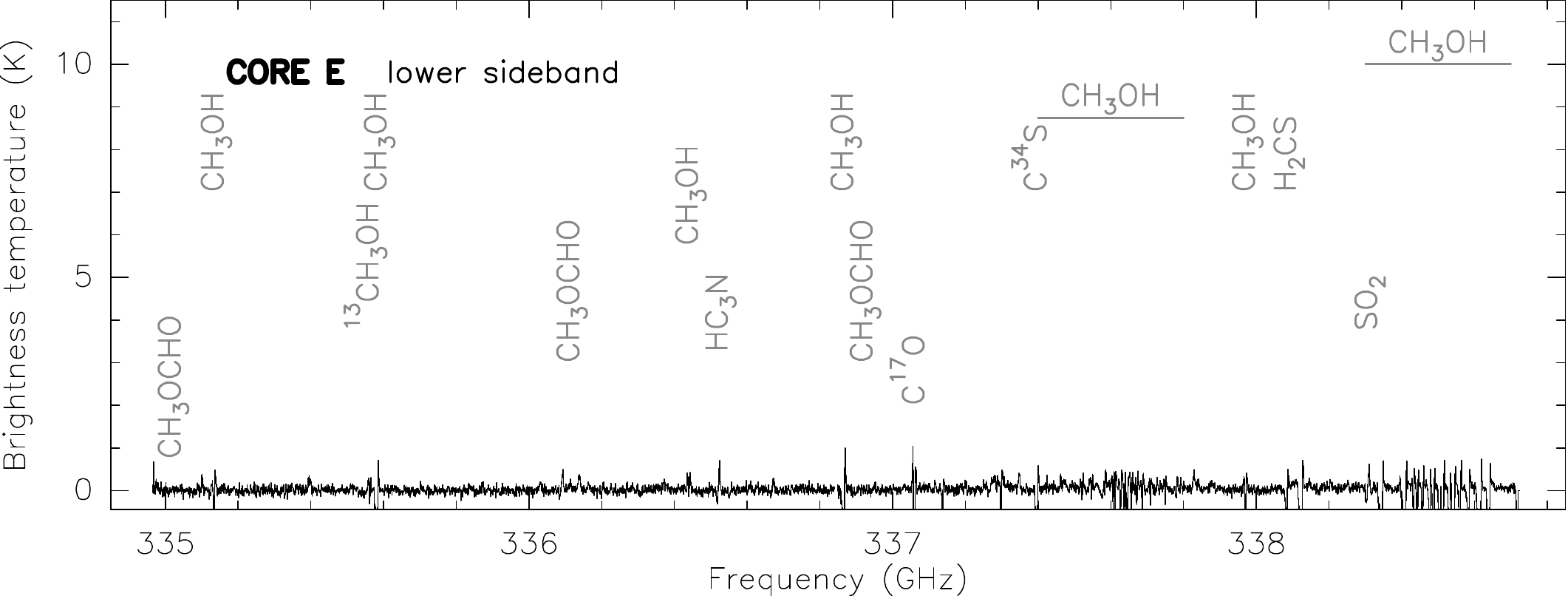, width=0.85\columnwidth, angle=0} &&
 \epsfig{file=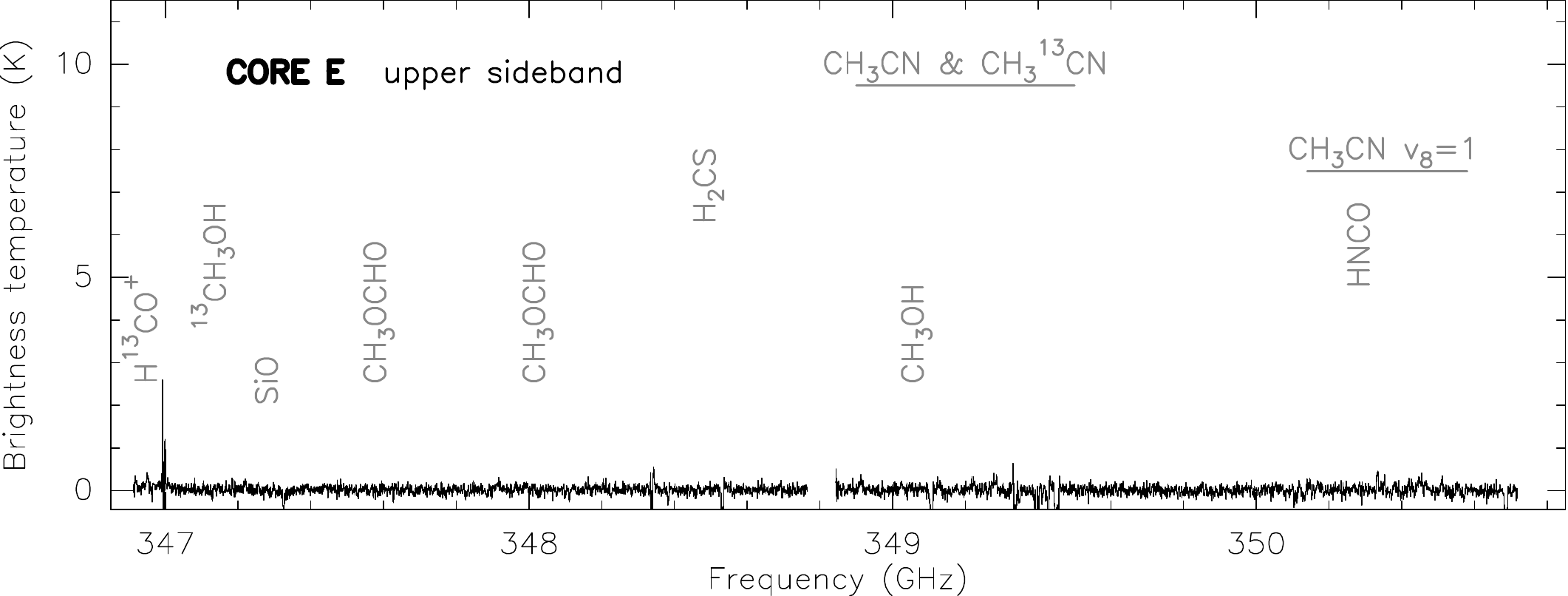, width=0.85\columnwidth, angle=0} \\
 \epsfig{file=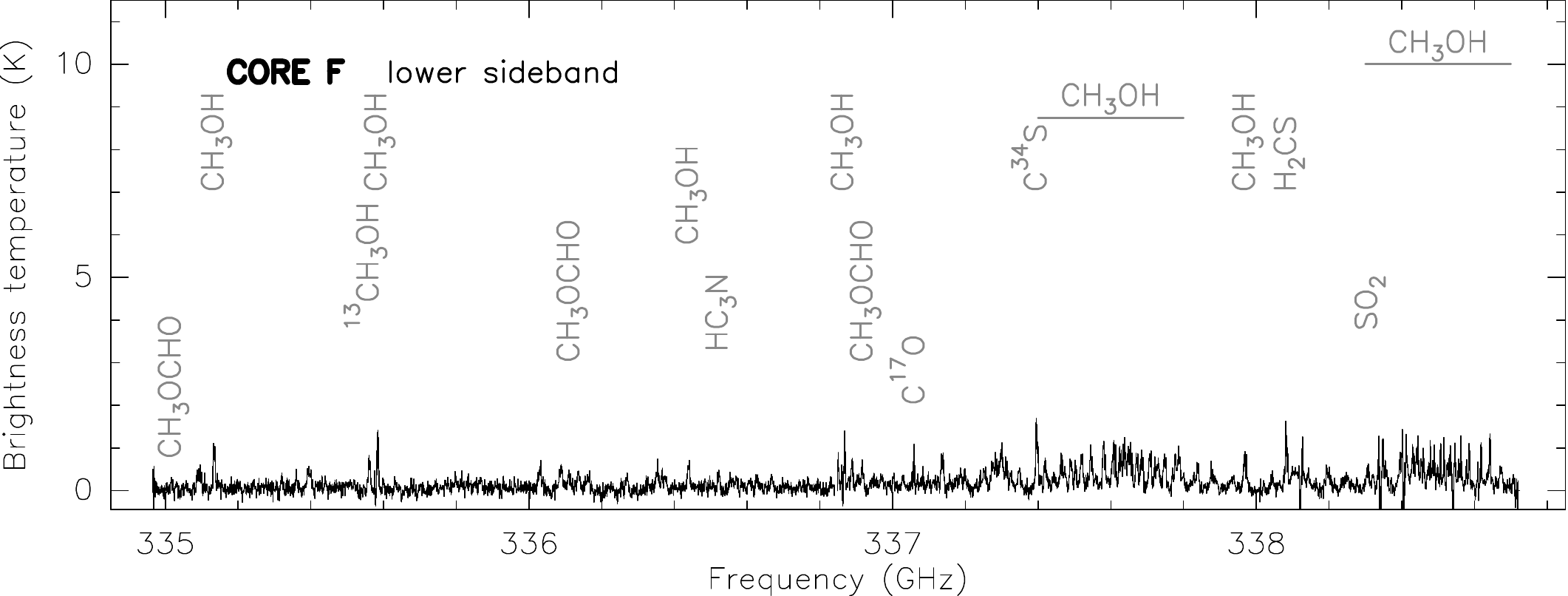, width=0.85\columnwidth, angle=0} &&
 \epsfig{file=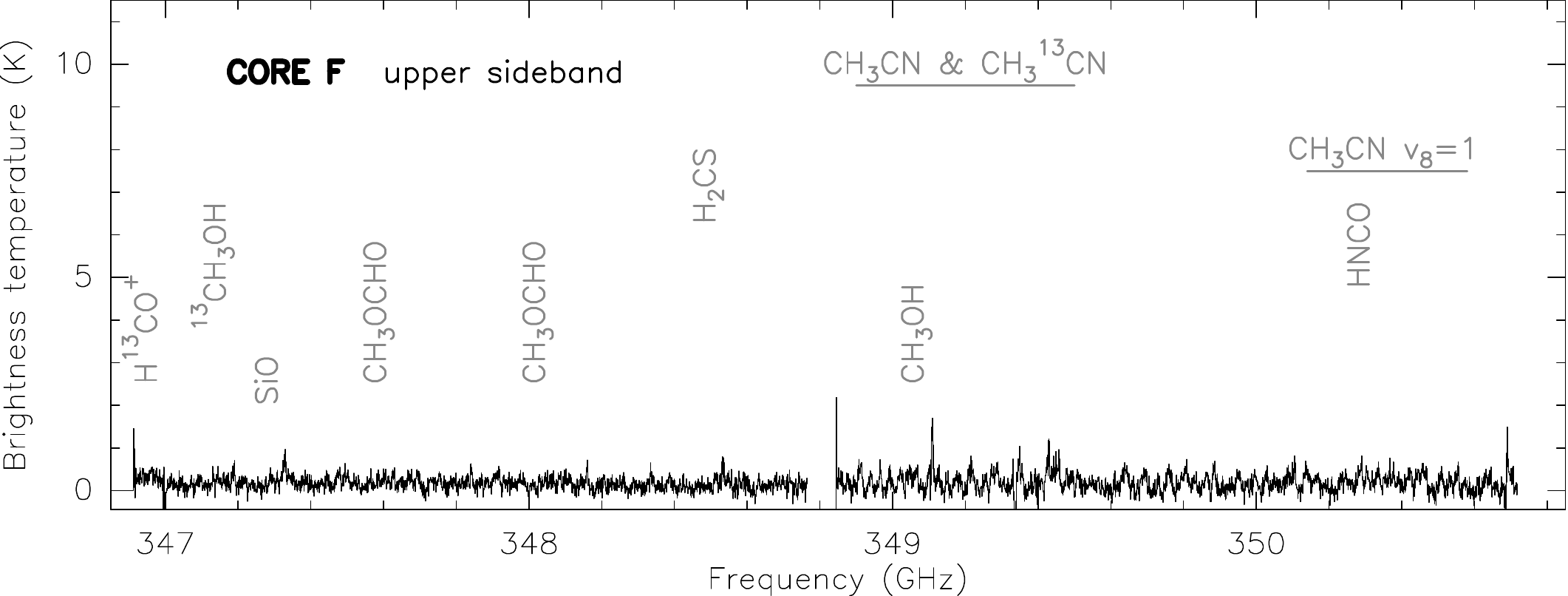, width=0.85\columnwidth, angle=0} \\
\end{tabular}
\caption{Spectra of the full frequency range surveyed by the ALMA observations, toward the six cores identified in the continuum image (see Fig.~\ref{f:continuum}). The spectra are extracted by averaging the intensity over an area of 1.37, 2.35, 1.04, 0.55, 0.75, 0.44~arcsec$^{2}$, yielding conversion factors of 8.9, 5.0, 11.3, 21.7, 15.7, 27.1~K~Jy$^{-1}$, for cores A to F respectively. The brightness temperature scale of cores~A and B is different from that of the rest of the cores. The absorption seen in some spectra is likely produced by the filtering of extended emission.}
\label{f:globspectra}
\end{center}
\end{figure*}
%----------------------------------------------------------------------
%----------------------------------------------------------------------
\begin{figure*}[t!]
\begin{center}
\begin{tabular}[b]{c}
 \epsfig{file=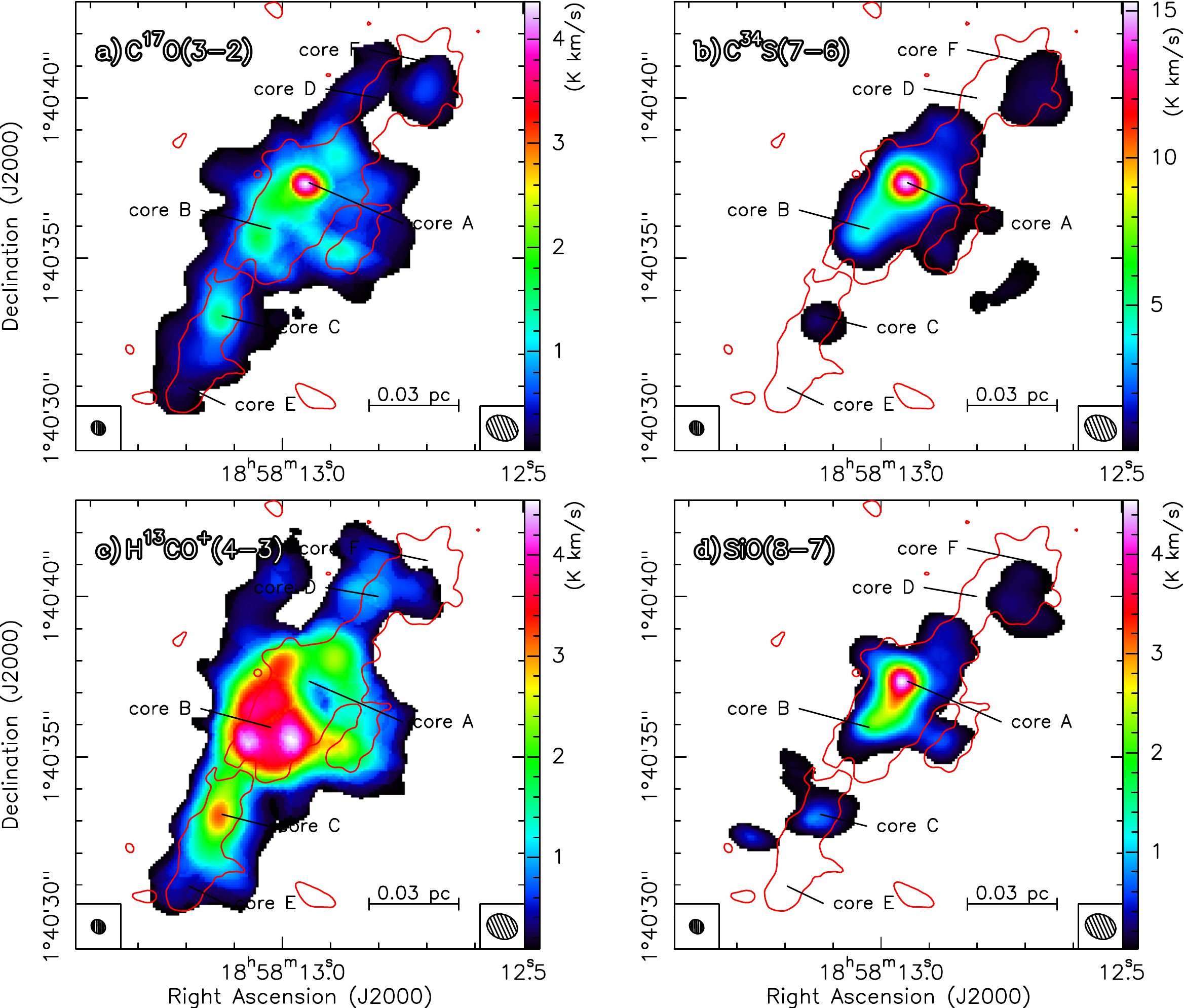, width=0.85\textwidth, angle=0} \\
\end{tabular}
\caption{Overlay of the 870~$\mu$m continuum emission (\emph{red contour}) on the integrated intensity maps (\emph{colors}) of C$^{17}$O, C$^{34}$S, H$^{13}$CO$^+$, and SiO. The contour level of the continuum emission corresponds to the 9.0~m\jpb\ intensity (see Fig.~\ref{f:continuum}). The intensity scale for the integrated emission for each molecule is shown in the color bar at the right of each panel, in K~\kms. The synthesized beam of the continuum emission ($0\farcs474\times0\farcs411$, PA=46\degr) is shown in the bottom-left corner of each panel. For each molecule, we show the synthesized beam of the tapered maps in the bottom-right corner: \textbf{a)} $1\farcs02\times0\farcs76$, PA=67\degr, \textbf{b)} $1\farcs02\times0\farcs76$, PA=67\degr, \textbf{c)} $0\farcs97\times0\farcs76$, PA=69\degr, and \textbf{d)} $0\farcs98\times0\farcs76$, PA=69\degr.}
\label{f:largescale}
\end{center}
\end{figure*}
%----------------------------------------------------------------------

%----------------------------------------------------------------------
\section{ALMA Observations\label{s:obs}}

G35.20N was observed with ALMA in Cycle 0 between May and June 2012. The source was observed in Band 7 ($\sim$350~GHz) with the 16 antennas of the array in the extended configuration, \ie\ baselines in the range 36--400~m, providing sensitivity to structures $\la$2\arcsec. The digital correlator was configured in four spectral windows (with dual polarization) of 1875~MHz and 3840~channels each, providing a resolution of $\sim$0.4~\kms. The four spectral windows covered the frequency ranges [336\,849.57--338\,723.83]~MHz, [334\,965.73--336\,839.99]~MHz, [348\,843.78--350\,718.05]~MHz, and [346\,891.29--348\,765.56]~MHz. The phase center of the observations is $\alpha$(J2000) = 18$^{\rm h}$ 58$^{\rm m}$ 13$\fs03$, $\delta$(J2000) = 01$\degr$ 40$'$ 36$\farcs$0. Flux, gain, and bandpass calibrations were obtained through observations of Neptune and J1751$+$096. The data were calibrated and imaged using CASA\footnote{The Common Astronomy Software Applications (CASA) software can be downloaded at http://casa.nrao.edu} \citep{mcmullin2007}. A continuum map was obtained from line-free channels and subtracted from the data, directly in the \emph{(u,v)}-domain. Channel maps were created for some selected molecular transitions (see Table~\ref{t:beams} and Section~\ref{s:res}). Continuum and channel maps were created with the robust parameter of \citet{briggs1995} set equal to 0.5, as a compromise between resolution and sensitivity to extended sources. The resulting continuum image has a synthesized CLEANed beam of $0\farcs47\times0\farcs41$, PA=46\degr\ and a rms noise of 1.8~m\jpb. The synthesized CLEANed beams of the spectral line maps vary from $0\farcs50\times0\farcs42$ to $0\farcs48\times0\farcs41$ with increasing frequency, with PA=40--45\degr. The rms noise of each spectral channel (of 0.6~\kms) varies between 5 and 20~m\jpb, with the largest values measured in channels with strong line emission. The spectral line images were analyzed using the CLASS and GREG programs of the GILDAS\footnote{http://www.iram.fr/IRAMFR/GILDAS} software package developed by the IRAM and Observatoire de Grenoble. All the resulting spectra have been smoothed to a resolution of 0.6~\kms.

%----------------------------------------------------------------------
\begin{figure}[t!]
\begin{center}
\begin{tabular}[b]{c c}
 \epsfig{file=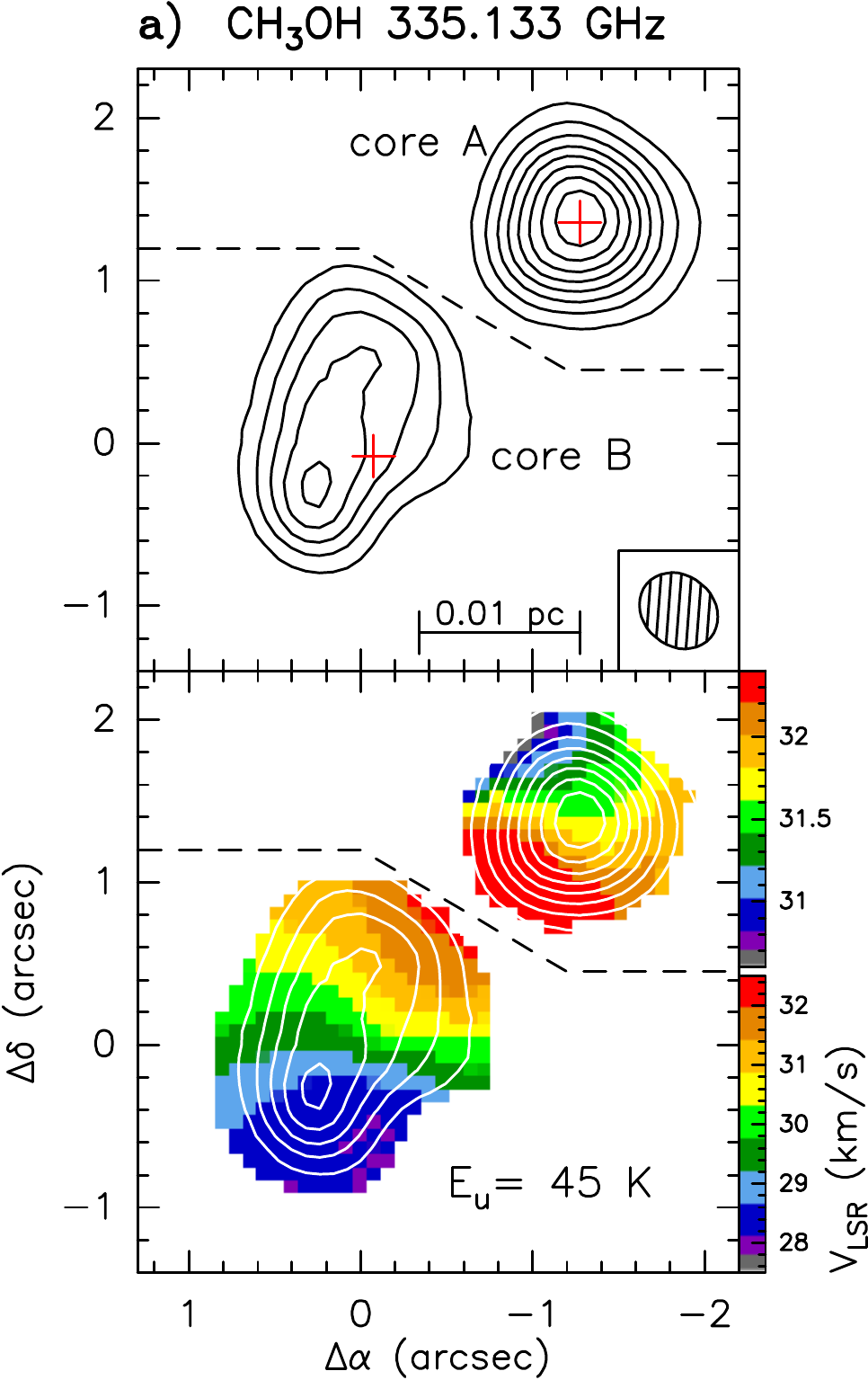, width=0.4\columnwidth, angle=0} &
 \epsfig{file=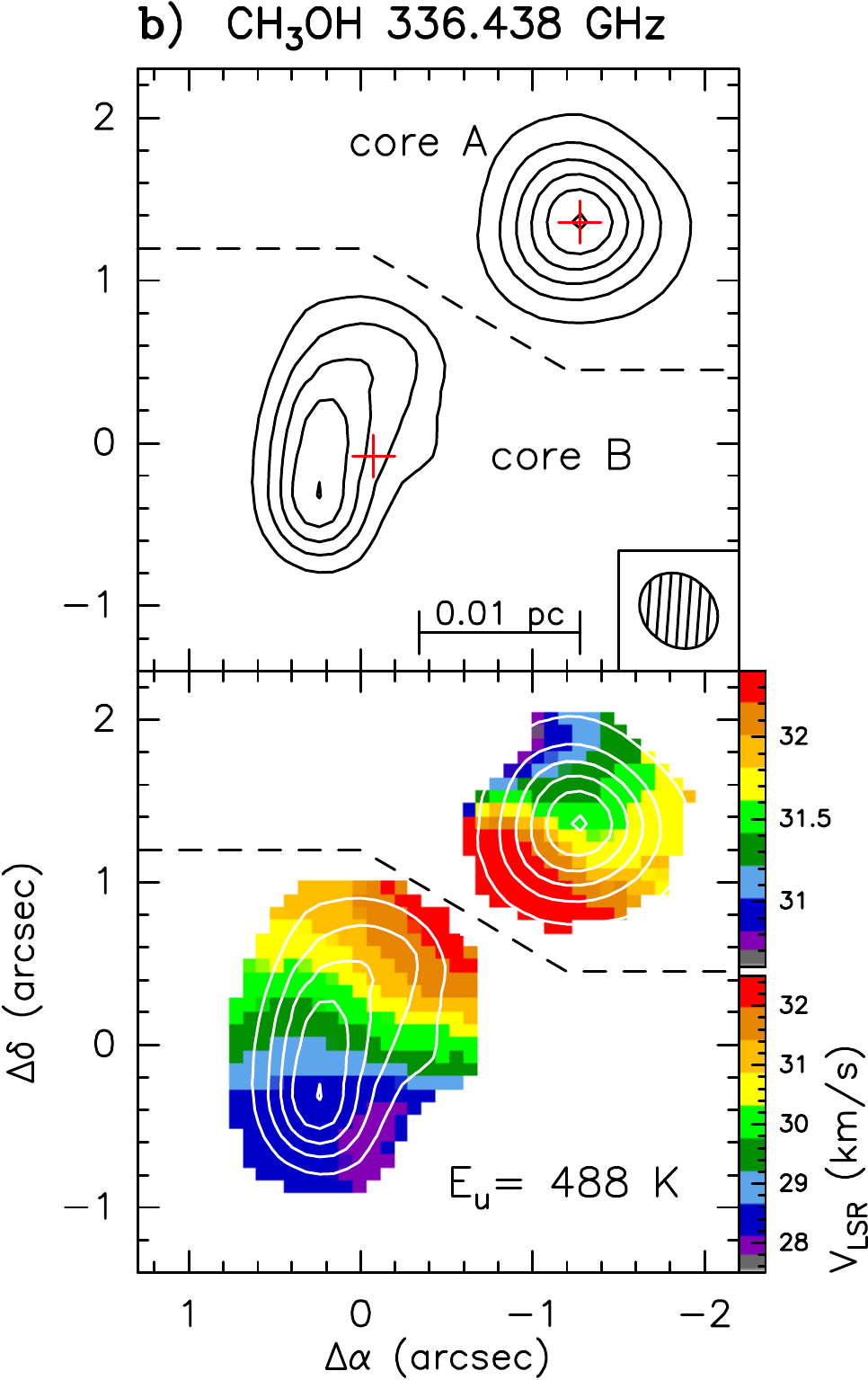, width=0.4\columnwidth, angle=0} \\
 \epsfig{file=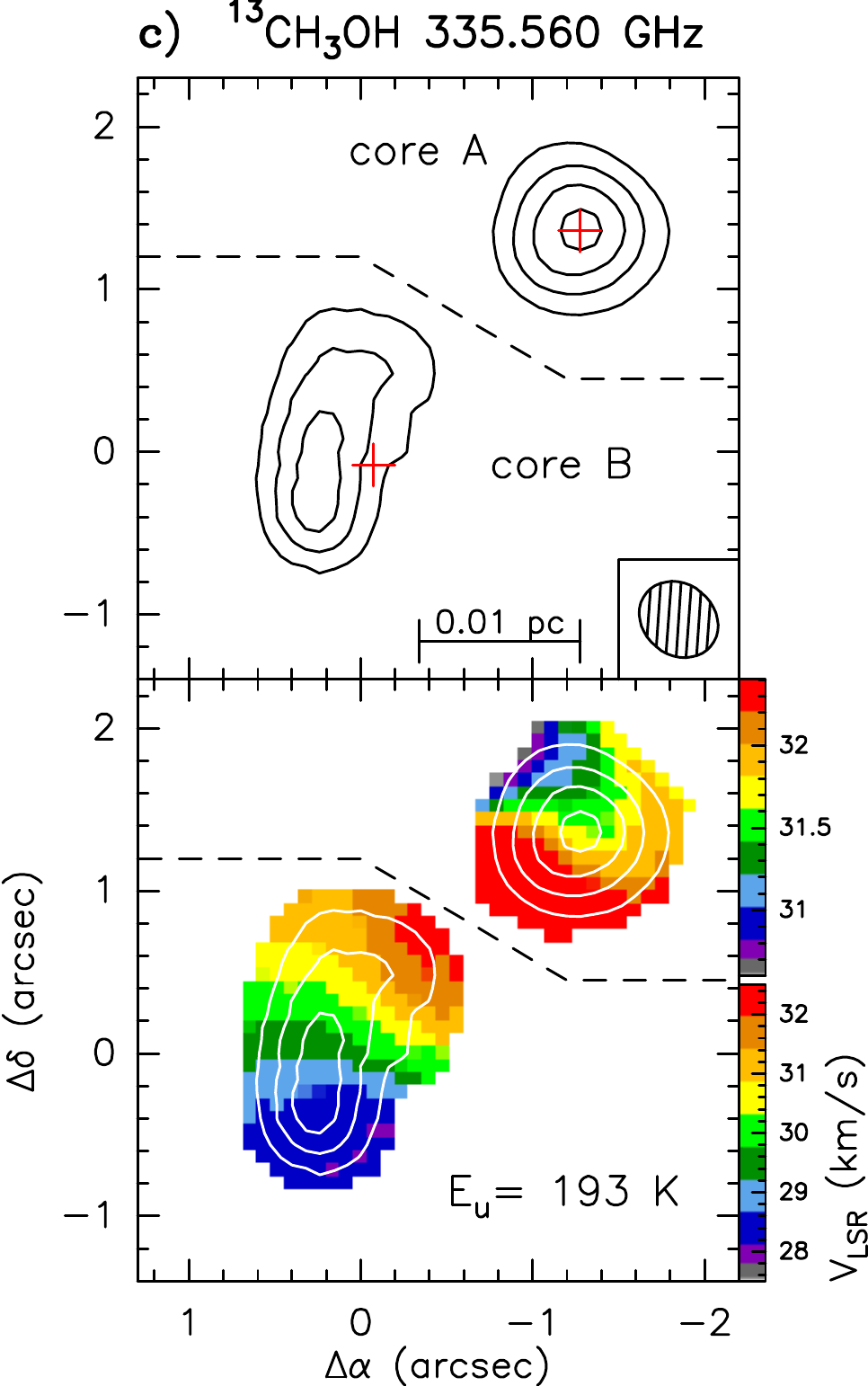, width=0.4\columnwidth, angle=0} &
 \epsfig{file=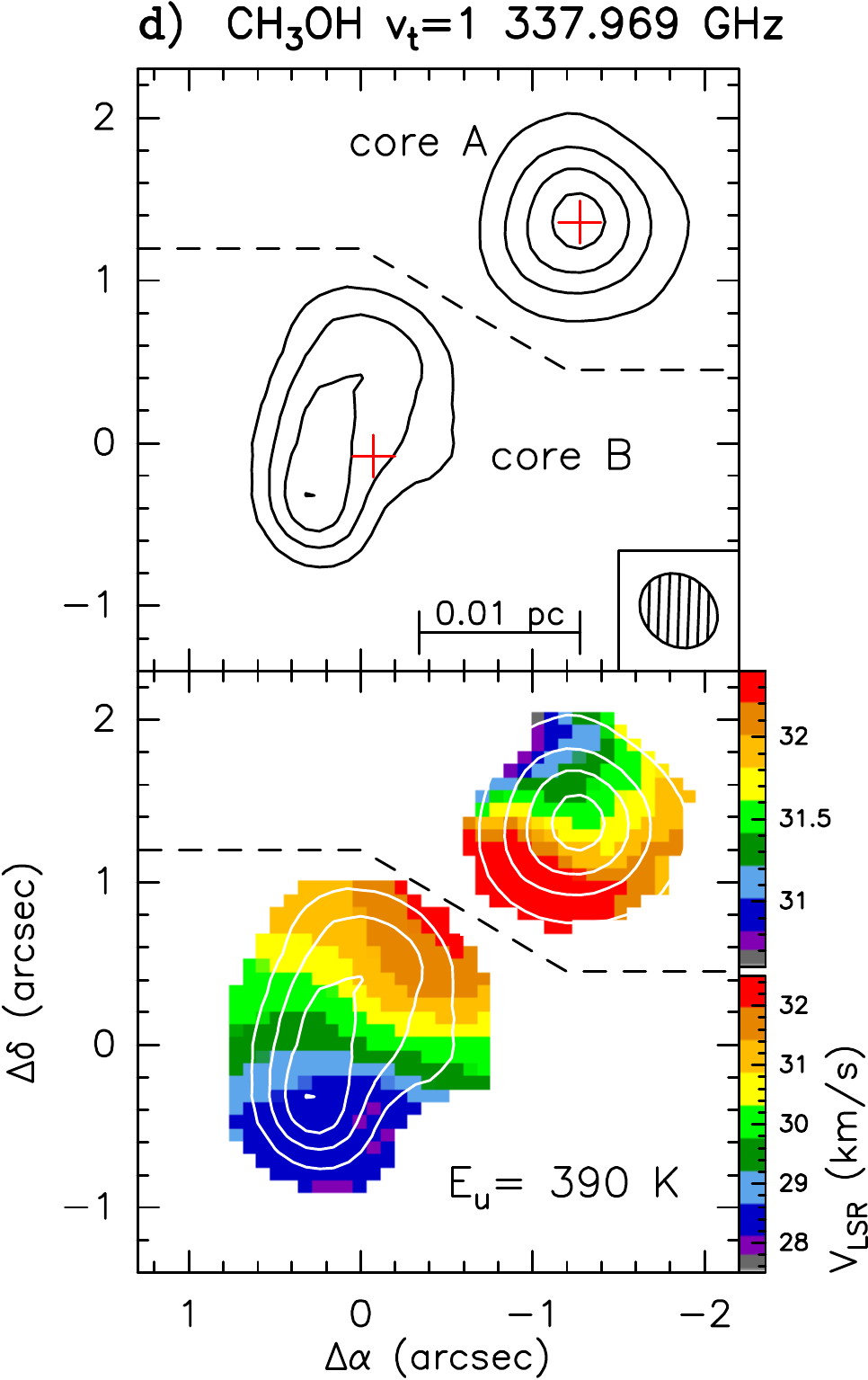, width=0.4\columnwidth, angle=0} \\
\end{tabular}
\caption{Maps of the integrated intensity (contours) and velocity field (color maps in the bottom panels) of \textbf{a)} CH$_3$OH\,(2$_{2,1}$--3$_{1,2}$), \textbf{b)} CH$_3$OH\,(14$_{7,8}$--15$_{6,9}$), \textbf{c)} $^{13}$CH$_3$OH\,(12$_{1,11}$--12$_{0,12}$), and \textbf{d)} CH$_3$OH $v_t$=1 (7$_{1,6}$--6$_{1,5}$) towards cores~A and B. For the different panels, the starting (and increasing contour levels in terms of $\sigma$, in \jpb~\kms) are 5$\sigma$ (5$\sigma$, with $\sigma$=0.18), 5$\sigma$ (10$\sigma$, with $\sigma$=0.11), 5$\sigma$ (5$\sigma$, with $\sigma$=0.25), and 5$\sigma$ (10$\sigma$, with $\sigma$=0.19) for core~A; and 5$\sigma$ (5$\sigma$, with $\sigma$=0.11), 5$\sigma$ (5$\sigma$, with $\sigma$=0.06), 3$\sigma$ (3$\sigma$, with $\sigma$=0.14), and 5$\sigma$ (5$\sigma$, with $\sigma$=0.11) for core~B. Offsets are measured with respect to the phase center (see Sect.~\ref{s:obs}). The red crosses indicate the position of the dust emission peaks of cores~A and B (see Table~\ref{t:continuum}). The synthesized beam and spatial scale are shown in the bottom-right corner (\emph{upper panels}).  The energy of the upper level of each transition is indicated in the bottom-right corner (\emph{lower panels}). The dashed line is intended to stress that the velocity intervals used to compute the moment maps are different for cores~A and B.}
\label{f:hotcore1}
\end{center}
\end{figure}
%----------------------------------------------------------------------
%----------------------------------------------------------------------
\begin{figure}[t!]
\begin{center}
\begin{tabular}[b]{c c}
 \epsfig{file=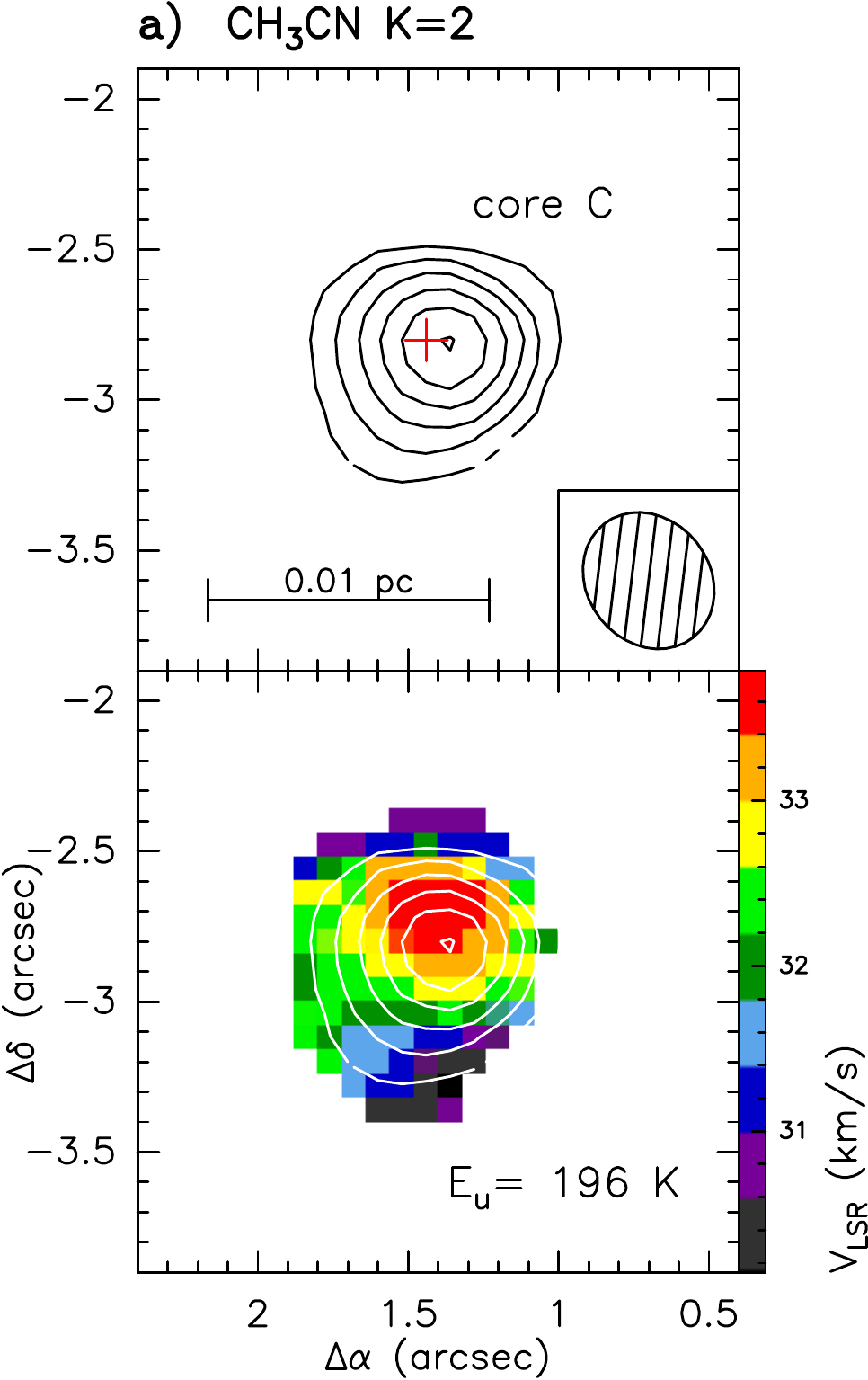, width=0.4\columnwidth, angle=0} &
 \epsfig{file=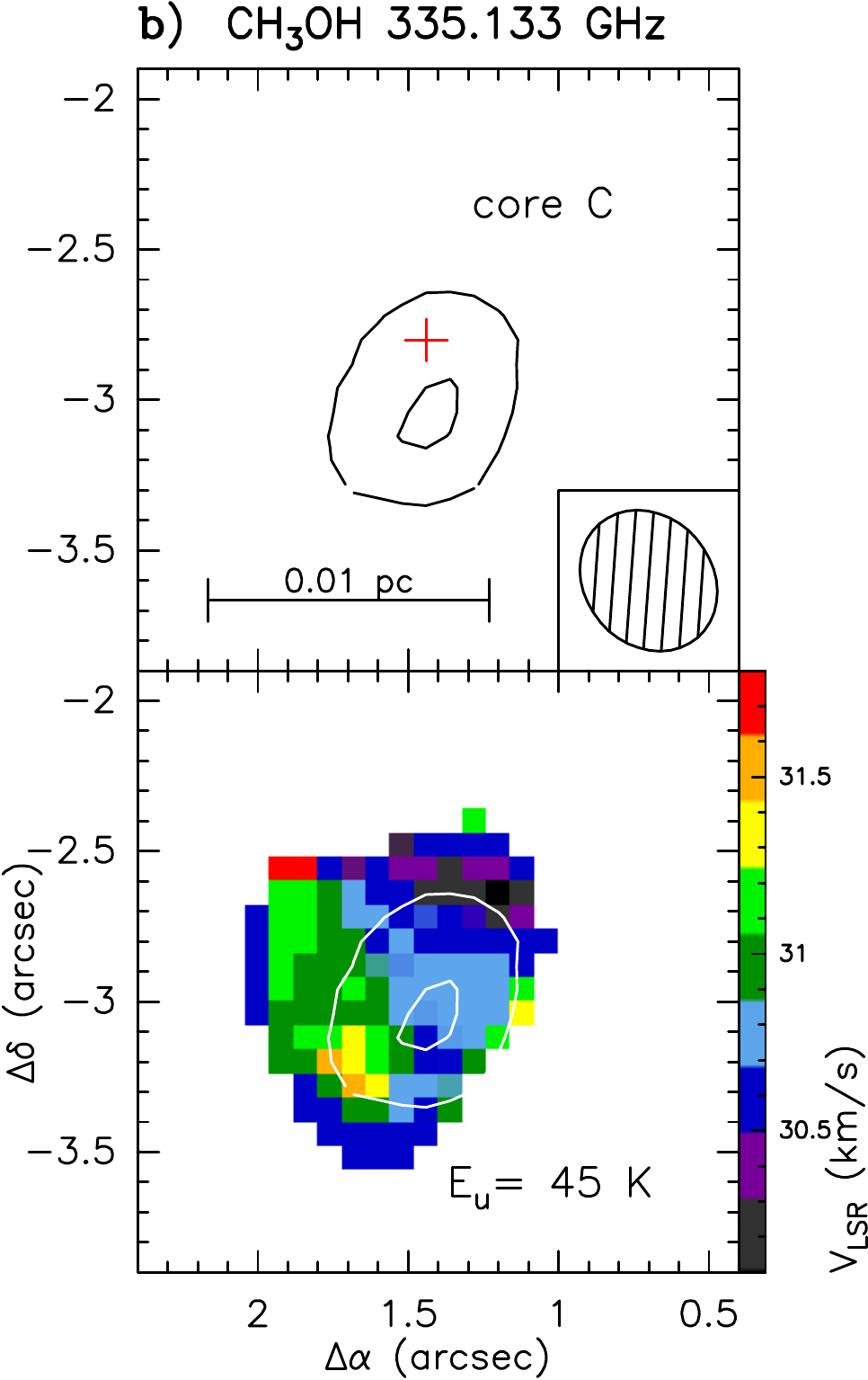, width=0.4\columnwidth, angle=0} \\
 \epsfig{file=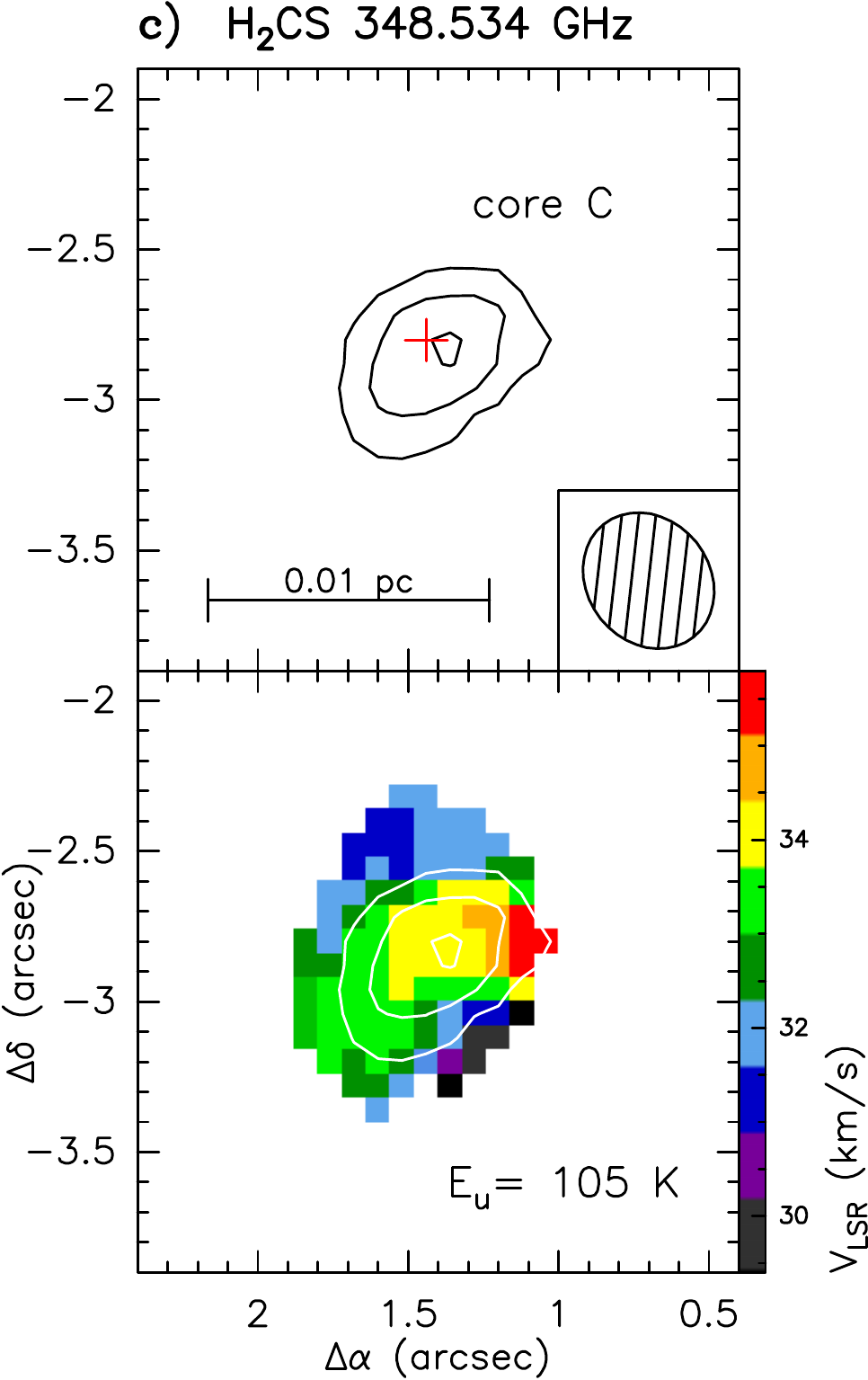, width=0.4\columnwidth, angle=0} &
 \epsfig{file=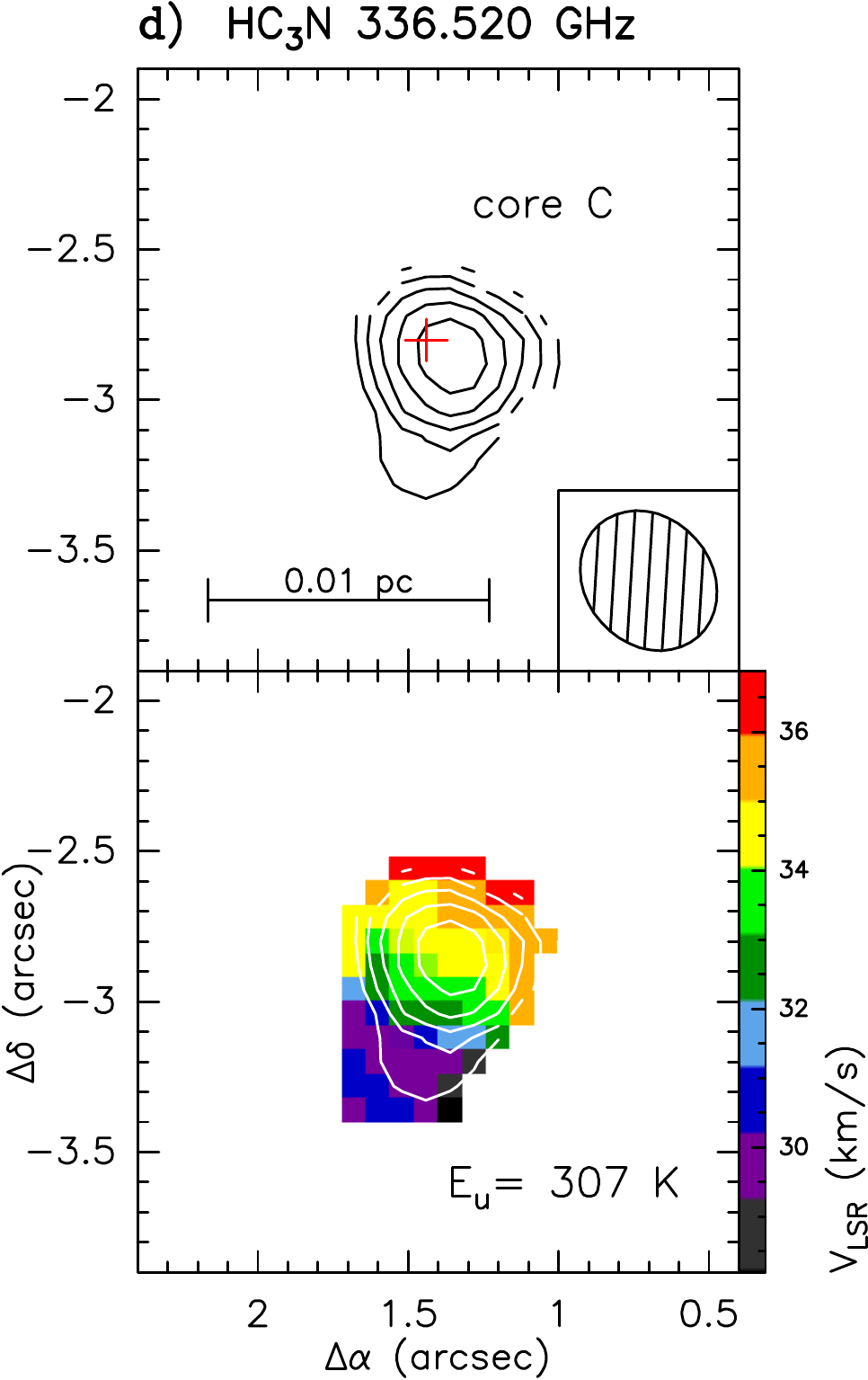, width=0.4\columnwidth, angle=0} \\
\end{tabular}
\caption{Maps of the integrated intensity (\emph{upper panels}) and velocity field (\emph{bottom panels}) of \textbf{a)} CH$_3$CN $K$=2 at 349.426~GHz, \textbf{b)} CH$_3$OH\,(2$_{2,1}$--3$_{1,2}$), \textbf{c)} H$_2$CS\,(10$_{1,9}$--9$_{1,8}$), and \textbf{d)} HC$_3$N\,(37--36) towards core~C. The starting (and increasing) contour levels for the four panels are 1$\sigma$ (1$\sigma$), with $\sigma$ equal to 0.18, 0.22, 0.20 and 0.14~\jpb~\kms, for the four panels. Other symbols as in Fig.~\ref{f:hotcore1}.}
\label{f:hotcore4}
\end{center}
\end{figure}
%----------------------------------------------------------------------

%----------------------------------------------------------------------
\section{Results\label{s:res}}

%----------------------------------------------------------------------
\subsection{Continuum emission \label{s:cont}}

In Fig.~\ref{f:continuum}a, we show the map of the ALMA 870~$\mu$m (350~GHz) continuum emission towards G35.20N overlaid on the \emph{Spitzer} 4.5~$\mu$m image, whose resolution has been enhanced with HiRes \citep{velusamy2008}. The submillimeter continuum emission appears to trace an elongated structure in the SE-NW direction, perpendicular to the 4.5~$\mu$m extended emission, and located across the waist of the butterfly-shaped nebula seen in the infrared. With an extension of $\sim$0.15~pc and a mean width of $\sim$0.013~pc (full width at half maximum), this is probably the densest part of the elongated structure observed at larger scales (0.21~pc long and 0.13~pc wide; \citealt{little1985, brebner1987, gibb2003}). This structure is not smooth but fragmented in several smaller condensations. The 2D CLUMPFIND algorithm \citep{williams1994} identifies six of these, which we name cores A--F (see Fig.~\ref{f:continuum}b), although additional substructure is seen in the map. The strongest of these, cores A and B, are located at the center of the elongated structure, and were already identified by \citet{sanchezmonge2013b}. These two cores are associated with faint centimeter continuum emission likely tracing young \hii\ regions or radiojets \citep[see][]{gibb2003, sanchezmonge2013b}, with a number of 1665-MHz OH maser features \citep{brebner1987, hutawarakorncohen1999}, and with probably\footnote{The 6.7~GHz CH$_3$OH maser observations carried out with the Japanese VLBI Network (JVN; \citealt{sugiyama2008}) and the European VLBI Neatwork (EVN; \citealt{surcis2012}) were performed without phase-referencing and it is thus impossible to establish the absolute position of the masers. However, the CH$_3$OH maser distribution seems to match the OH maser distribution \citep{brebner1987, hutawarakorncohen1999}; since the latter are associated with cores A and B, it seems likely that the methanol masers are also associated with these two cores.} 6.7-GHz CH$_3$OH maser emission. Core~C, located $\sim$3\arcsec\ to the SE of core~B, coincides with the millimeter source G35MM2 reported by \citet{gibb2003}. The ALMA 870~$\mu$m dust continuum emission shown in Fig.~\ref{f:continuum}a is consistent with the Submillimeter Array (SMA) 880~$\mu$m image obtained by \citet[][see their Fig.~1c]{qiu2013}. There are some morphological differences to the southeast of the elongated structure, probably due to the vicinity of the primary beam edge in our ALMA observations.

In Table~\ref{t:continuum}, we list for all the cores the coordinates of the peak position, the primary beam corrected fluxes, and the deconvolved sizes. The fluxes have been computed by integrating the emission within the 5$\sigma$ contour level. The peak intensities range from 40~m\jpb\ to 200~m\jpb, while the integrated fluxes are in the range 80--700~mJy. These fluxes are slightly lower (a factor of 2) than those measured by \citet{qiu2013} in their SMA observations. However, we note that Qiu \et\ determine the flux densities for larger dust entities containing at least two cores (\eg\ their object MM1, with a total flux of 3~Jy, is the sum of cores A and B). We have convolved the ALMA image to the beam of the SCUBA 850~$\mu$m image (15\arcsec; \citealt{qiu2013}), and compared the peak fluxes of both images ($\sim$2.5~\jpb\ versus $\sim$8.5~\jpb) to conclude that we are recovering around 30\% of the flux measured with SCUBA. The deconvolved diameters of the dust cores (see Table~\ref{t:continuum}) have been computed from the 50\% contour level applying Gaussian deconvolution (see procedure described in Table~3 of \citealt{sanchezmonge2013a}). The mean deconvolved diameter is $\sim$1600~AU, with core~A being the most compact (900~AU), and core~F the most extended (2600~AU). The different cores appear regularly spaced, with a mean (projected) separation of $\sim$$2\farcs1$ or $\sim$0.023~pc.

% SEPARATIONS BETWEEN CORES:
% core F - core D   1.90 arcsec; 0.020 pc
% core D - core A   3.41 arcsec; 0.036 pc
% core A - core B   1.87 arcsec; 0.020 pc
% core B - core C   3.11 arcsec; 0.033 pc
% core C - core E   2.47 arcsec; 0.026 pc
%
% JEANS ANALYSIS
% Mjeans = 5 Msun * [T/K]^(3/2) * [n/cm^-3]^(-1/2)
% Ljeans = 0.19 pc * [T/10K]^(1/2) * [n/10^4cm^-3]^(-1/2)
%
% for a T =  30K, and n = 10^6cm^-3 --> Mjeans =  0.8 Msun; Ljeans = 0.03 pc
% for a T =  50K, and n = 10^6cm^-3 --> Mjeans =  1.8 Msun; Ljeans = 0.04 pc
% for a T = 100K, and n = 10^6cm^-3 --> Mjeans =  5.0 Msun; Ljeans = 0.06 pc
%
% for a T =  30K, and n = 10^5cm^-3 --> Mjeans =  2.6 Msun; Ljeans = 0.10 pc
% for a T =  50K, and n = 10^5cm^-3 --> Mjeans =  5.6 Msun; Ljeans = 0.13 pc
% for a T = 100K, and n = 10^5cm^-3 --> Mjeans = 15.8 Msun; Ljeans = 0.19 pc
%
% for a T =  30K, and n = 10^4cm^-3 --> Mjeans =  8.2 Msun; Ljeans = 0.33 pc
% for a T =  50K, and n = 10^4cm^-3 --> Mjeans = 17.7 Msun; Ljeans = 0.42 pc
% for a T = 100K, and n = 10^4cm^-3 --> Mjeans = 50.0 Msun; Ljeans = 0.60 pc

%----------------------------------------------------------------------
\subsection{Line emission\label{s:line}}

In Fig.~\ref{f:globspectra}, we show the spectra obtained by averaging the line emission over the cores in G35.20N (lower sideband: 334\,965.73--338\,723.83~MHz, and upper sideband: 346\,891.29--350\,718.05~MHz). Cores~A and B show rich spectra, with emission of several complex organic molecules (\eg\ CH$_3$CN, CH$_3$OH, CH$_3$OCHO, NH$_2$CHO, HC$_3$N, C$_2$H$_5$OH, CH$_3$OCHO). Out of the rest of the cores, only core~C is likely associated with emission in complex molecules such as CH$_3$CN and CH$_3$OH. Core~C was found to be associated with molecular emission at millimeter wavelengths in previous studies \citep[\eg][]{gibb2003, birks2006}. The rest of the cores are only associated with faint emission of more abundant species such as C$^{17}$O, C$^{34}$S, H$^{13}$CO$^+$ (see Fig.~\ref{f:largescale}) and weak CH$_3$OH, suggesting that these cores could either host only low-mass star formation, or be in an evolutionary phase prior to the hot core stage.

%----------------------------------------------------------------------
\subsubsection{Large-scale molecular emission\label{s:largescale}}

Out of all the species detected in the region (mainly associated with cores~A and B; see Sect.~\ref{s:hotcores}), there are four molecules (C$^{17}$O, C$^{34}$S, H$^{13}$CO$^+$ and SiO) that show a complex (extended) spatial distribution. Their emission appears partially resolved out by the interferometer, which is sensitive to angular scales $\approx$2\arcsec\ (see Sect.~\ref{s:obs}). In order to improve the detection of the extended emission, we generated new CLEANed maps for these four species after tapering the data at 1\arcsec\ resolution. In Fig.~\ref{f:largescale}, we show the corresponding integrated emission (zeroth-order moment) maps, over the velocity range 23 to 42~\kms, and compare them with the dust continuum emission. For these molecules, the line emission coincides with the dust continuum emission, with C$^{17}$O and H$^{13}$CO$^+$ being detected toward all the cores and tracing the elongated structure, and SiO and C$^{34}$S mostly concentrated towards cores~A and B, consistent with the SMA images reported by \citet{qiu2013}. We estimate a missing flux of 50\% from the comparison of our ALMA H$^{13}$CO$^+$ image and the map obtained by Qiu \et\ using compact configurations of the SMA. Such an elongated structure has also been revealed at larger scales ($\sim$40\arcsec) in the H$^{13}$CO$^+$\,(1--0) and H$^{13}$CN\,(1--0) lines \citep[\eg][]{gibb2003}. In Sect.~\ref{s:kinematics}, we will analyze the kinematics of this extended molecular structure. 

%----------------------------------------------------------------------
\begin{figure}[t!]
\begin{center}
\begin{tabular}[b]{c}
 \epsfig{file=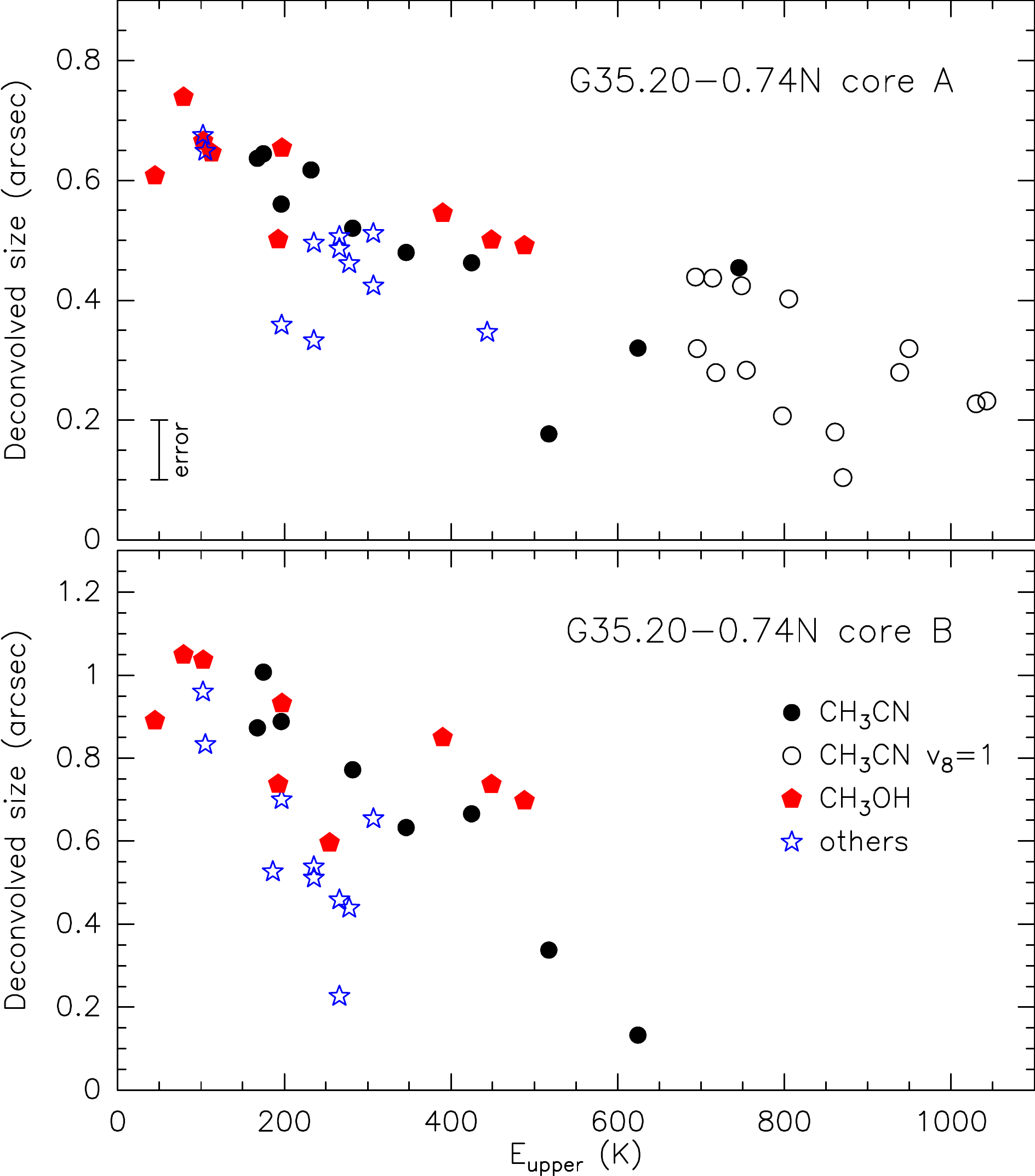, width=0.85\columnwidth, angle=0} \\
\end{tabular}
\caption{Deconvolved size (in arcsec) for different molecular transitions (listed in Table~\ref{t:beams}) versus the upper level energy of the transition for cores~A (\emph{top panel}) and B (\emph{bottom panel}). In both panels, black circles correspond to CH$_3$CN, open circles to CH$_3$CN vibrationally excited, red pentagons to CH$_3$OH, and blue stars to transitions of other molecules. The typical error in size is shown in the top panel.}
\label{f:sizeEu}
\end{center}
\end{figure}
%----------------------------------------------------------------------
%----------------------------------------------------------------------
\begin{figure*}[t!]
\begin{center}
\begin{tabular}[b]{c}
 \epsfig{file=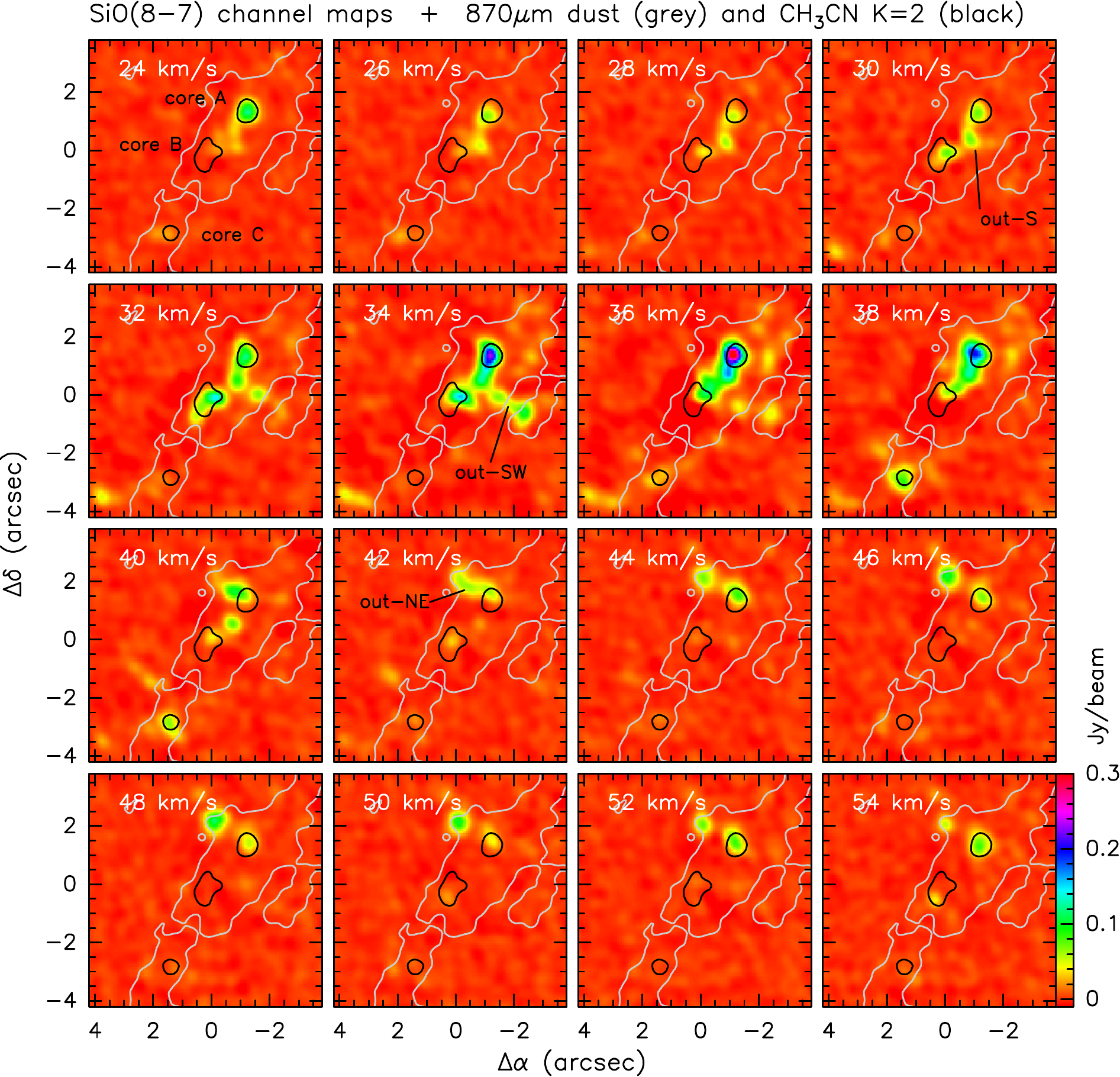, width=0.85\textwidth, angle=0} \\
\end{tabular}
\caption{SiO\,(8--7) channel map of the G35.20$-$0.74\,N region, averaged over 2~\kms\ wide velocity intervals. The central velocity of each channel is indicated in the upper left corner, and the velocity to take as reference is $+$32~\kms. Gray contour: 5$\sigma$ level of the 870~$\mu$m ALMA continuum emission shown in Fig.~\ref{f:continuum}. Black contour: 50\% intensity level of the CH$_3$CN\,(19--18) $K=2$ integrated emission as shown in Figs.~\ref{f:hotcore2} and \ref{f:hotcore4}. The structures named `out-S', `out-SW' and `out-NE' are discussed in Section~\ref{s:outflow}, while the spectra of these three regions are shown in Fig.~\ref{f:SiOspectra}.}
\label{f:SiOchannel}
\end{center}
\end{figure*}
%----------------------------------------------------------------------

%----------------------------------------------------------------------
\subsubsection{Compact molecular emission: hot cores\label{s:hotcores}}

As shown in Fig.~\ref{f:globspectra}, cores~A and B show a very rich chemistry with strong emission of many species typical of hot cores. Core~C also shows emission in complex molecules, although with an intensity several times fainter than the two cores located at the center of the elongated structure. A detailed analysis of the molecular content towards these three cores will be the subject of a forthcoming paper. In the present study, we will only focus on a few hot core tracers, basically those that are clearly detected, and less blended with other species: CH$_3$CN, CH$_3$$^{13}$CN, and vibrationally excited CH$_3$CN, as well as CH$_3$OH, CH$_3$OCHO, HC$_3$N, HNCO, H$_2$CS, and SO$_2$. Similar to the large-scale molecular and continuum emission, we have estimated the missing flux in our maps of hot-core tracers by comparing the ALMA CH$_3$CN\,$K$=0,1 lines with the same lines observed with the SMA in compact configurations \citep{qiu2013} and with the Atacama Submillimeter Telescope Experiment (ASTE; R.\ Furuya, private communication). The ALMA observations recover around 75\% of the flux measured with ASTE. The fluxes between the SMA and ALMA images (after convolution to the same beam) are equal within a 8\%.

Figures~\ref{f:coreAmyxclass}, \ref{f:coreBmyxclass} and \ref{f:coreCmyxclass} show a close-up view of the CH$_3$CN and CH$_3$OH transitions detected towards cores~A, B and C, respectively. The observational setup (see Sect.~\ref{s:obs}) permits to cover up to the $K$=9 transition of CH$_3$CN\,(19$_K$--18$_K$), and up to the $K$=5 line of CH$_3$$^{13}$CN\,(19$_K$--18$_K$), with upper level energies of 745~K and 517~K, respectively. All of these CH$_3$CN lines have been detected in the three cores. Regarding the CH$_3$$^{13}$CN, only a few lines are not blended (\eg\ $K$=2, $K$=4), while others clearly overlap with methyl cyanide or methanol lines (\eg\ $K$=3, $K$=5). Several vibrationally excited CH$_3$CN transitions, with upper level energies ranging from $\sim$600~K to $\sim$1100~K, have been detected. The intensity of the methyl cyanide lines in core~A is almost twice the intensity of the lines in core~B, and ten times stronger than in core~C. A large number of methanol (CH$_3$OH) and methyl formate (CH$_3$OCHO) lines have been detected in these three cores, but in the following, we have only considered those transitions that are not blended with other (strong) lines. The selected lines are listed in Table~\ref{t:beams}, and cover a range of upper level energies from 45~K to 488~K in the case of methanol, and from 236~K to 444~K for methyl formate. Finally, we have also selected some strong and isolated lines of different species such as H$_2$CS, HC$_3$N, SO$_2$ or HNCO.

In Figs.~\ref{f:hotcore1}, \ref{f:hotcore2} and \ref{f:hotcore3}, we present the zeroth (integrated intensity) and first (velocity field) moment maps of different hot core species detected towards cores~A and B: CH$_3$CN $K$=2 and 8, CH$_3$$^{13}$CN $K$=2, and CH$_3$CN $v_8$=1 $K,l$=3,1, as well as different transitions of CH$_3$OH, CH$_3$OCHO, H$_2$CS, HNCO, SO$_2$, and HC$_3$N. The moment maps have been computed following the same procedure as in \citet{sanchezmonge2013b}, using the velocity interval 25.9--38.0~\kms\ for core~A, and 25.9--33.0~\kms\ for core~B. As seen in the maps, the line emission peak in core~A is perfectly coincident with the position of the dust emission peak. In both the molecular and continuum images, the emission of core~A is barely resolved with a mean deconvolved diameter of $\sim$1100~AU ($\sim$$0\farcs52$), obtained from a 2D Gaussian fit to the zero-order moment maps, and applying Gaussian deconvolution. In core~B, the strongest peak of the molecular emission is typically located to the SE of the dust peak. For most of the transitions, the molecular emission of core~B is clearly resolved (deconvolved size $\sim$$0\farcs95$; $\sim$2000~AU) in an elongated structure oriented in the SE-NW direction, with the dust peak located close to the center of the extended line emission. For both cores, the emission is more compact in the maps of high excitation energy transitions (\eg\ CH$_3$CN $v_8$=1, or CH$_3$OCHO at 335.015~GHz) or less abundant species (\eg\ CH$_3$$^{13}$CN $K$=2), suggesting the existence of temperature and/or molecular abundance gradients within the cores. This is better seen in Fig.~\ref{f:sizeEu}, where we plot the deconvolved size for different transitions against the upper level energy of the transition. For both cores, the trend is such that the higher the excitation energy of the transition is, the more compact the emission is. From the velocity field maps, we see a clear velocity gradient in both cores for all the species. The velocity gradient of core~B is oriented in the (S)SE-(N)NW direction, while the velocity gradient in core~A is closer to the N--S direction. We also note that the velocity gradient in both cores is reversed, with redshifted emission to the north for core~B and to the south for core~A. A more detailed analysis of the kinematics of these two cores is presented in Sect.~\ref{s:disks}.

Finally, in Fig.~\ref{f:hotcore4}, we present the integrated intensity and velocity field maps of four transitions of hot-core species towards core~C: CH$_3$CN\,(19$_2$--18$_2$), CH$_3$OH\,(2$_{2,1}$--3$_{1,2}$), H$_2$CS\,(10$_{1,9}$--9$_{1,8}$), and HC$_3$N\,(37--36). The moments were computed using the velocity interval 26.1--41.0~\kms. Similar to core~A, the molecular emission in core~C is compact and coincident with the dust emission. Differently from cores~A and B, we do not find a clear velocity gradient in any of the molecules. 

%----------------------------------------------------------------------
\begin{figure}[t!]
\begin{center}
\begin{tabular}[b]{c}
 \epsfig{file=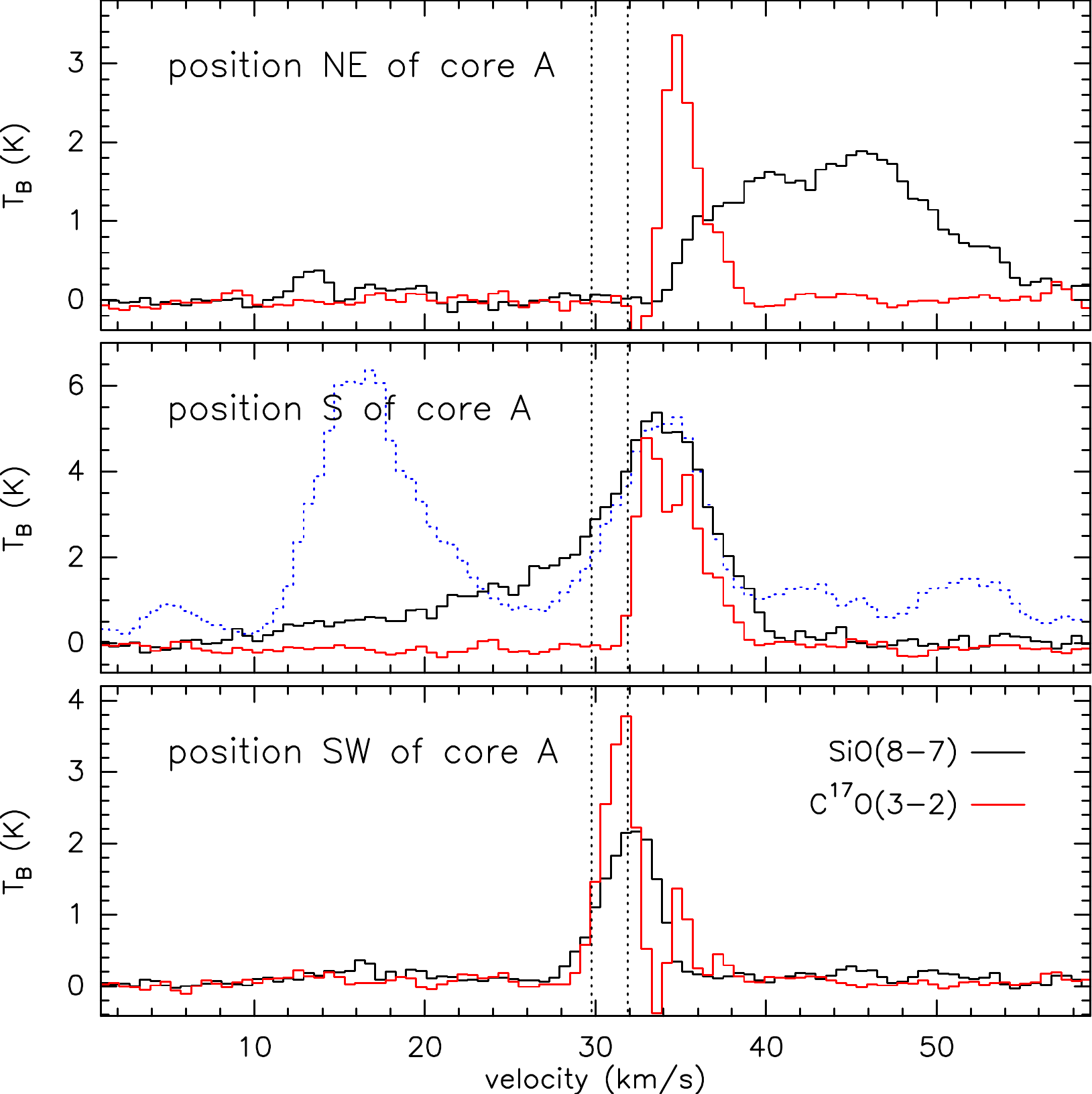, width=0.9\columnwidth, angle=0} \\
\end{tabular}
\caption{Spectra of the SiO\,(8--7) emission (black line) toward different positions in the G35.20N star forming region: positions `out-NE' (top), `out-S' (middle) and `out-SW' (bottom) as indicated in Fig.~\ref{f:SiOchannel}. The emission has been obtained by integrating over areas of 1.34, 0.45 and 1.19~arcsec$^{2}$, which yield to conversion factors of 8.5, 25.3 and 9.6~K~Jy$^{-1}$, respectively for the three panels. The red line corresponds to the C$^{17}$O\,(3--2) line emission toward the three positions. In middle panel, the blue dotted line corresponds to the SiO\,(8--7) line emission at the position of core~A (over an area of 1.37~arcsec$^{2}$, which yields to a conversion factor of 8.9~K~Jy$^{-1}$). The velocity component at $\sim$16~\kms\ corresponds to two blended CH$_3$CHO transitions. The two vertical dotted lines indicate the velocities of core~A (31.8~\kms) and core~B (29.6~\kms).}
\label{f:SiOspectra}
\end{center}
\end{figure}
%----------------------------------------------------------------------

%----------------------------------------------------------------------
\subsubsection{High-velocity SiO outflow emission\label{s:outflow}}

We have analyzed the SiO\,(8--7) line emission to search for outflow signatures, since this molecule is known to be one of the best jet/outflow tracers in star forming regions \citep[\eg][]{gusdorf2008a, gusdorf2008b, lopezsepulcre2011, sanchezmonge2013d}. Channel maps of the SiO\,(8--7) emission are displayed in Fig.~\ref{f:SiOchannel}, with the 870~$\mu$m continuum emission and hot core (CH$_3$CN) line emission overlaid. The SiO emission appears to span a wide range of velocities, however, line blending in hot cores A and B hinders the identification of high-velocity emission at the position of the two cores.

To overcome this problem, we identified three different regions to search for high-velocity SiO emission. These regions (named `out-S', `out-SW' and `out-NE') are located far from the cores identified in Fig.~\ref{f:continuum}, thus avoiding contamination from the chemically-rich hot cores (see Fig.~\ref{f:SiOchannel}). The spectra of the emission integrated over the three regions are shown in Fig.~\ref{f:SiOspectra}, together with the spectra of the C$^{17}$O\,(3--2) emission as a reference of the dense gas emission. The SiO emission at the `out-NE' position, spanning a range from 32 up to 60~\kms, is clearly red-shifted with respect to the velocities of cores~A and B (31.9 and 29.8~\kms, respectively), and to the C$^{17}$O line emission (with the peak at $\sim$32.3~\kms). In the `out-S' position, the SiO emission spans a range of velocities from 0 up to 32~\kms, \ie\ blue-shifted with respect to the C$^{17}$O line emission. Finally, at the `out-SW' position, the SiO line has a narrow profile similar to the C$^{17}$O line, both in velocity and linewidth. This suggests that the SiO emission at this position is associated with to the large-scale dense gas emission, rather than with high-velocity outflows.

%----------------------------------------------------------------------
\begin{figure}[t!]
\begin{center}
\begin{tabular}[b]{c}
 \epsfig{file=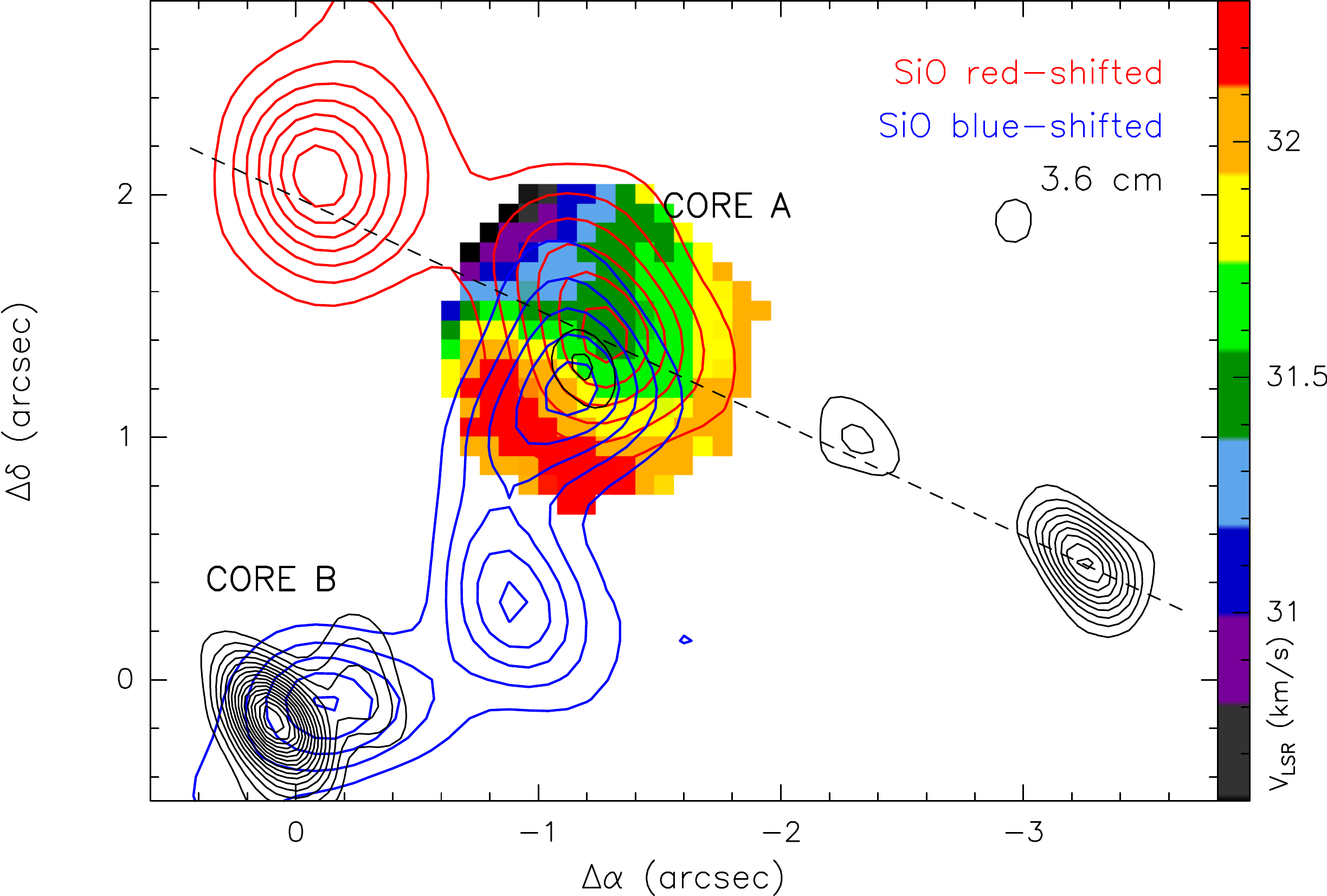, width=0.9\columnwidth, angle=0} \\
\end{tabular}
\caption{Maps of the blue- (\emph{blue contours}) and red-shifted (\emph{red contours}) SiO\,(8--7) emission overlaid on the CH$_3$CN\,(19--18) $K$=2 velocity field (\emph{color scale}) towards core~A. The maps have been obtained by averaging the emission over the velocity intervals 25.2--28.6~\kms\ and 40.8--57.7~\kms. Contour levels range from 0.05 to 0.27 in steps of 0.05~\jpb~\kms\ and from 0.15 to 1.19 in steps of 0.15~\jpb~\kms, for blue and red-shifted emission, respectively. Black contours are a map of the 3.6~cm continuum emission \citep{gibb2003}, with levels from 0.039 to 0.533 in steps of 0.026~m\jpb. The dashed line shows the direction of a possible jet associated with core~A.}
\label{f:sioout}
\end{center}
\end{figure}
%----------------------------------------------------------------------

In Fig.~\ref{f:sioout}, we show the zeroth-order moment maps of the SiO emission integrated over two different velocity ranges: (1) blue-shifted velocities in the range 25.2--28.6~\kms, and (2) red-shifted velocities in the range 40.8--57.7~\kms. These velocity ranges have been chosen to avoid contamination of the hot core species associated with cores~A and B (see middle panel in Fig.~\ref{f:SiOspectra}). The blue-shifted emission is located in between cores~A and B, and thus it is difficult to determine the powering source. On the other hand, SiO red-shifted emission seems to be clearly emanating from core~A. While the configuration of the SiO high-velocity emission does not resemble a bipolar outflow, it is worth comparing the SiO map with that of the free-free 3.6~cm continuum obtained by \citet{gibb2003} with the Very Large Array (see black contours in Fig.~\ref{f:sioout}). One sees two radio peaks that are lined with core~A and the red-shifted SiO emission, suggesting that both the radio continuum and SiO emission might be tracing a jet from the core, oriented in the SW-NE direction.

%PARAMETERS of the SiO (red) outflow and CM radiojet?
%
%PARAMETERS of SiO outflow
%red lobe size 1.8arcsec=3942AU for a range velo of 60-32=28 km/s result in time=668 years
%Mass = 0.054139271671395 Mo 
%Kinetic Energy = 909.87194620269 Lo yrs 
%Kinetic Energy = 1.10327359919E+44 erg 
%Momentum = 0.73449022491798 Mo km/s 
%blue lobe size 1.8arcsec=3942AU for a range velo of 32-1=31 km/s result in time=603 years
%Mass = 0.023080182578648 Mo 
%Kinetic Energy = 187.89981818562 Lo yrs 
%Kinetic Energy = 2.278396532193E+43 erg 
%Momentum = 0.19334537815754 Mo km/s 
%0.0003-0.0011 Mo/yr km/s
%
%PARAMETERS of CM radiojet
%flux 0.25-0.40 mJy
%Pdot = 10^-2.5*(Snu*d^2)^1.1 --> 0.0036-0.0060 Mo/yr km/s 

In the following we test if the SiO and centimeter continuum emission could be arising from the same outflow. \citet{anglada1996} found an observational relation between the outflow momentum rate, $\dot{P}$, and the radio continuum luminosity, $S_\nu d^{2}$, of the jet associated with the outflow. This observational relation is in agreement with the shock induced ionization mechanism proposed by \citet{curiel1989} to explain the radio emission of the jets. The relation reads as
\begin{equation}
\left[\frac{S_\nu d^{2}}{\mathrm{mJy~kpc}^{2}}\right]
=10^{3.5}\left[\frac{\Omega}{4\pi}\right]\,\left[\frac{\dot{P}}{M_\sun~\mathrm{yr}^{-1}~\mathrm{km~s}^{-1}}\right],
\end{equation}
where $\Omega/4\pi$ is an efficiency factor that indicates the fraction of the solid angle that is shocked ($\sim$0.1 for low-mass radiojets/outflows, \citealt{anglada1996}), $S_\nu$ is the flux density at 5~GHz, and $d$ the heliocentric distance. The flux density of the 3.6~cm continuum emission of the radiojet is $\sim$0.25--0.40~mJy (depending on the sources/knots considered), which corresponds to a flux density of $\sim$0.18--0.30~mJy at 5~GHz, assuming a typical spectral index of 0.6. This results in an outflow momentum rate of 2.7--4.5$\times10^{-3}$~\mo~yr$^{-1}$~\kms. We compared this number with the outflow momentum rate that can be computed from the SiO high-velocity emission. We followed the same procedure described in \citet{lopezsepulcre2009}, and derived an outflow momentum rate of 1.5$\times10^{-3}$ and 0.4$\times10^{-3}$~\mo~yr$^{-1}$~\kms, for the red and blue-shifted emission lobes shown in Fig.~\ref{f:sioout}, assuming a SiO abundance of $3\times10^{-9}$ \citep{sanchezmonge2013d}. The derived outflow momentum rate for the red-shifted lobe agrees, within a factor of 2--3, with that obtained from the radio continuum emission. The small discrepancy may be related to the uncertainties involved in these calculations (\eg\ SiO abundance, $\Omega/4\pi$ efficiency factor). Observations, sensitive to scales $\sim$2\arcsec, of different outflow tracers should be performed to confirm the scenario in which core~A is powering an outflow-jet directed approximately NE-SW.

%----------------------------------------------------------------------
\begin{figure}[t!]
\begin{center}
\begin{tabular}[b]{c}
 \epsfig{file=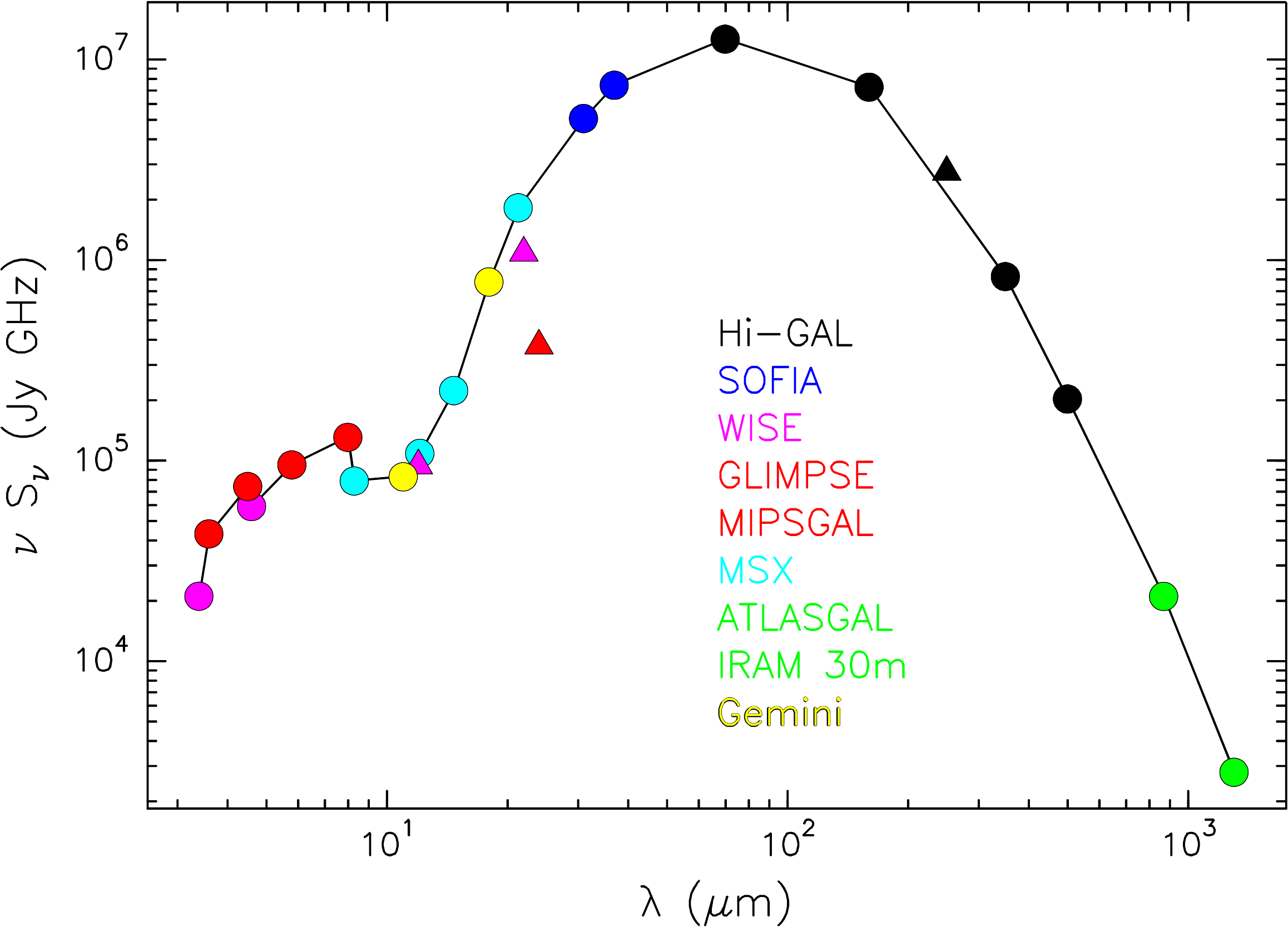, width=0.68\columnwidth, angle=-90} \\
\end{tabular}
\caption{Continuum spectrum of the G35.20N star forming region. Triangles denote lower limits. Different colors correspond to different surveys and instruments, as indicated in the figure.}
\label{f:sed}
\end{center}
\end{figure}
%----------------------------------------------------------------------
%----------------------------------------------------------------------
\begin{table*}[t!]
\caption{\label{t:myxclass}Best myXCLASS fit parameters for the CH$_3$CN and CH$_3$OH species (see Figs.~\ref{f:coreAmyxclass}--\ref{f:coreDmyxclass}).}
\begin{tabular}{c c c c c c c c c c c c  ccc}
\hline\hline
&\multicolumn{5}{c}{CH$_3$CN}
&
&\multicolumn{5}{c}{CH$_3$OH}
\\
\cline{2-6}\cline{8-12}

&v
&$\Delta$v
&size
&$T_\mathrm{CH_3CN}$
&$N_\mathrm{CH_3CN}$
&
&v
&$\Delta$v
&size
&$T_\mathrm{CH_3OH}$
&$N_\mathrm{CH_3OH}$
\\
Core
&(\kms)
&(\kms)
&(arcsec)
&(K)
&(cm$^{-2}$)
&
&(\kms)
&(\kms)
&(arcsec)
&(K)
&(cm$^{-2}$)
\\
\hline
A	&31.8	&5.2		&0.24	&309		&$3.8\times10^{16}$	&&31.8	&4.0		&0.34	&192		&$6.0\times10^{18}$	\\
B	&29.6	&2.8		&0.31	&148		&$1.0\times10^{16}$	&&29.9	&2.9		&0.36	&177		&$2.1\times10^{18}$	\\
C	&32.4	&8.1		&0.12	&280		&$1.0\times10^{16}$	&&32.0	&5.4		&0.09	&176		&$4.9\times10^{18}$	\\
D	&\ldots	&\ldots	&\ldots	&\ldots	&\ldots				&&31.2	&1.5		&0.07	&120		&$0.8\times10^{18}$	\\
\hline
\end{tabular}
\end{table*}
%----------------------------------------------------------------------
%----------------------------------------------------------------------
\begin{table*}[t!]
\caption{\label{t:physical}Physical parameters of the cores detected at 870~$\mu$m (see Fig.~\ref{f:continuum}).}
\begin{tabular}{c c c c c c c c c c c c c c c c}
\hline\hline

&$T_\mathrm{rot}$\supa
&$M_\mathrm{core}$\supb
&$N_\mathrm{core}$\supb
&$n_\mathrm{core}$\supb
&$\Sigma_\mathrm{core}$\supb
&$\Delta \mathrm{v}^\mathrm{hot-core}$
&$M_\mathrm{virial}^\mathrm{hot-core}$
&$\Delta \mathrm{v}^\mathrm{H^{13}CO^+}$
&$M_\mathrm{virial}^\mathrm{H^{13}CO^+}$
\\
Core
&(K)
&(\mo)
&($10^{24}$~cm$^{-2}$)
&($10^{8}$~cm$^{-3}$)
&(g~cm$^{-2}$)
&(\kms)
&(\mo)
&(\kms)
&(\mo)
\\
\hline
A	&\phn30--250	&11.2--1.0		&40.2--3.6		&44.3--4.0		&156.0--14.0\phn	&4.6		&\phn9.7	&$>$3.2	&$>$4.7	\\ % 900
B	&\phn30--150	&\phn5.4--0.9	&\phn4.7--0.8	&\phn2.6--0.4	&18.9--3.1\phn	&2.9		&\phn7.7	&$>$2.9	&$>$7.7	\\ %1800
C	&\phn30--200	&\phn3.1--0.4	&\phn3.7--0.5	&\phn2.4--0.3	&13.7--1.8\phn	&6.8		&37.7	&$>$2.6	&$>$5.5	\\ %1600
D	&\phn30--100	&\phn3.7--0.9	&\phn7.0--1.7	&\phn5.6--1.4	&24.8--6.0\phn	&1.5		&\phn1.5	&$>$2.9	&$>$5.6	\\ %1300
E	&30--50		&\phn2.6--1.4	&\phn2.3--1.2	&\phn1.2--0.7	&8.2--4.4		&\ldots	&\ldots	&$>$1.8	&$>$3.1	\\ %1900
F	&30--50		&\phn4.5--2.4	&\phn2.0--1.1	&\phn0.8--0.4	&7.5--4.0		&\ldots	&\ldots	&$>$1.0	&$>$1.3	\\ %2600
\hline
\end{tabular}

\supa\ Temperature derived from the rotational diagram method when possible (see Sect.~\ref{s:temperature}).\\
\supb\ Mass, $M_\mathrm{core}$, column density, $N_\mathrm{core}$, volume density, $n_\mathrm{core}$, and surface density, $\Sigma_\mathrm{core}$, estimated using a dust temperature equal to $T_\mathrm{rot}$, and a dust opacity of 1.75~cm$^2$~g$^{-1}$ at 870~$\mu$m (see Sect.~\ref{s:temperature}).\\
\end{table*}
%----------------------------------------------------------------------------

%----------------------------------------------------------------------
\subsection{Spectral energy distribution\label{s:sed}}

Figure~\ref{f:sed} shows the spectral energy distribution (SED) of G35.20N, obtained from the \emph{Herschel} infrared Galactic Plane Survey (Hi-GAL; \citealt{molinari2010a, molinari2010b}) images complemented with literature and archival data at wavelengths ranging from mid-infrared to millimeter. In particular, we used the following data: Galactic Legacy Infrared Mid-Plane Survey Extraordinaire (GLIMPSE; \citealt{benjamin2003}); Midcourse Space Experiment (MSX; \citealt{price1999}); MIPSGAL/\emph{Spitzer} \citep{carey2009}; Wide-field Infrared Survey Explorer (WISE; \citealt{wright2010}); APEX Telescope Large Area Survey of the Galaxy (ATLASGAL; \citealt{schuller2009}); \citet{mooney1995}; \citet{debuizer2006}; \citet{zhang2013}. We stress that inclusion of the Hi-GAL data at 70 and 160~$\mu$m, represents an important improvement with respect to previous estimates, as it allows us to sample the SED around the peak, \ie\ the spectral region responsible for most of the bolometric luminosity.

The bolometric luminosity has been derived by integrating the SED over the full observed spectral distribution. We note that different spatial apertures have been used at each wavelength due to the different angular resolution of the data. We obtain $\sim$$3\times10^{4}$~\lo, thus confirming the value quoted in the RMS database\footnote{The RMS database is available at http://www.ast.leeds.ac.uk/cgi-bin/RMS/RMS\_DATABASE.cgi}, but less than the luminosity ($\sim$0.7--$2.2\times10^5$~\lo) derived by \citet{zhang2013} from a model taking into account the photons which may escape through the outflow cavity (the so-called flashlight effect), but considering a single high-mass protostar forming within a massive core of radius 0.1~pc (which seems to be inconsistent with the fragmentation seen in our high-angular resolution observations).

%----------------------------------------------------------------------
\section{Analysis and discussion\label{s:analysis}}

%----------------------------------------------------------------------
\subsection{Temperature and mass estimates\label{s:temperature}}

The large number of methyl cyanide (CH$_3$CN) transitions detected towards cores~A, B and C, together with the methanol (CH$_3$OH) transitions found in almost all the cores, can be used to obtain an estimate of the temperature of the cores in G35.20N. We fitted the observed methyl cyanide and methanol spectra for all the cores using the CASA interface of the XCLASS software called myXCLASS\footnote{myXCLASS can be downloaded from https://www.astro.uni-koeln.de/projects/schilke/myXCLASSInterface} described in \citet{schilke2001}, \citet{comito2005} and \citet{zernickel2012}. The myXCLASS software includes entries from the CDMS (Cologne Database for Molecular Spectroscopy; \citealt{muller2001}) and JPL (Jet Propulsion Laboratory; \citealt{pickett1998}) catalogs. With this information, it is possible to simulate a spectrum of a molecule, by providing parameters such as the excitation temperature, abundance, size of the source, linewidth and velocity. Once an initial guess is set up, the parameters can be fitted using the program MAGIX (Modeling and Analysis Generic Interface for eXternal numerical codes; \citealt{bernst2011, moller2013}). Figures~\ref{f:coreAmyxclass} to \ref{f:coreDmyxclass} show the observed and myXCLASS fitted CH$_3$CN and CH$_3$OH spectra towards the cores in G35.20N. We fitted each molecule with a single component, \ie\ the five free parameters are the same for all the lines of a given species. A more complex model, taking into account temperature and density gradients, might be necessary to explain the variations in size shown in Fig.~\ref{f:sizeEu}. Finally, we note that the myXCLASS software takes the line opacity into account. The fits do not include only ground state transitions of the two species, but also vibrationally and torsionally excited as well as transitions from isotopologues such as CH$_3^{13}$CN and $^{13}$CH$_3$OH. We assumed the relative abundance [$^{12}$C/$^{13}$C] to be 50 for a galactocentric distance of 6.5~kpc \citep{wilsonrood1994}. The parameters of the best fits are listed in Table~\ref{t:myxclass}. The temperature ranges typically between 100--300~K for cores~A--D, in both methanol and methyl cyanide species. For cores~A and C, the estimated CH$_3$CN temperature is large (close to 300~K) but with a large uncertainty as shown in the $\chi^2$ plots (see Figs.~\ref{f:coreAmyxclass}--\ref{f:coreDmyxclass}). For core~D, the CH$_3$CN emission is too weak and cannot be fitted, while from the CH$_3$OH line fitting we measure a temperature of $\sim$120~K (cooler than the three strongest cores). The column densities range between 1--$4\times10^{16}$~cm$^{-2}$ for CH$_3$CN, and 1--$6\times10^{18}$~cm$^{-2}$ for CH$_3$OH.

The mass of each core has been computed from
\begin{equation}
\Bigg[\frac{M_\mathrm{core}}{M_{\sun}}\Bigg]=4.597\,
\Bigg[\frac{S_{870\mu\mathrm{m}}}{\mathrm{Jy}}\Bigg]
\Bigg[\frac{D}{\mathrm{kpc}}\Bigg]^{2}
\Bigg(\exp{\left(\frac{16.46}{[T/\mathrm{K}]}\right)}-1\Bigg),
\end{equation}
with $S_{870\mu\mathrm{m}}$ the dust continuum flux density at 870~$\mu$m (see Table~\ref{t:continuum}), and $D$ the distance to the source. We have adopted a dust opacity of 1.75~cm$^2$~g$^{-1}$ at 870~$\mu$m \citep{ossenkopfhenning1994}, with a gas-to-dust ratio of 100. We assume $T=T_\mathrm{ex}$, where $T_\mathrm{ex}$ is obtained from the myXCLASS analysis. Since CH$_3$CN and CH$_3$OH may be biased towards warm gas ($>$100~K), but the dust may also be tracing cooler material, we have set a lower limit for the temperature of 30~K. In Table~\ref{t:physical}, we give the range of temperatures used and the derived masses. For cores~E and F, without detection of hot core molecules, we assume a temperature range of 30--50~K. The masses of the cores in G35.20N are in the range 1--10~\mo.

We have also calculated the source-averaged column density, $N_\mathrm{core}$, and volume density, $n_\mathrm{core}$, under the assumption that the cores are spherically symmetric and homogeneous:
\begin{equation}
\Bigg[\frac{N_\mathrm{core}}{\mathrm{cm}^{-2}}\Bigg]=
2.95\times10^{24}\,
\Bigg[\frac{M_\mathrm{core}}{M_{\sun}}\Bigg]
\Bigg[\frac{\mu}{2.3}\Bigg]^{-1}
\Bigg[\frac{\theta_\mathrm{S}}{\mathrm{arcsec}}\Bigg]^{-2}
\Bigg[\frac{D}{\mathrm{kpc}}\Bigg]^{-2}
\end{equation}
\begin{equation}
\Bigg[\frac{n_\mathrm{core}}{\mathrm{cm}^{-3}}\Bigg]=
2.95\times10^{8}\,
\Bigg[\frac{M_\mathrm{core}}{M_{\sun}}\Bigg]
\Bigg[\frac{\mu}{2.3}\Bigg]^{-1}
\Bigg[\frac{\theta_\mathrm{S}}{\mathrm{arcsec}}\Bigg]^{-3}
\Bigg[\frac{D}{\mathrm{kpc}}\Bigg]^{-3},
\end{equation}
where $\mu=2.3$ is the mean molecular weight, and $\theta_\mathrm{S}$ is the deconvolved angular diameter (see Table~\ref{t:continuum}). The results are given in Table~\ref{t:physical}. The mean column and volume densities are 1--$10\times10^{24}$~cm$^{-2}$ and 2--9$\times10^{8}$~cm$^{-3}$, with core~A being the densest. From these values and the CH$_3$CN and CH$_3$OH column densities (see Table~\ref{t:myxclass}), we derive a CH$_3$CN fractional abundance of 0.10--1.1$\times10^{-8}$, 0.21--1.3$\times10^{-8}$ and 0.27--2.0$\times10^{-8}$, for cores~A, B, and C, respectively. For CH$_3$OH, we obtain 0.15--1.7$\times10^{-6}$, 0.45--2.6$\times10^{-6}$, 1.3--9.8$\times10^{-6}$ and 0.11--0.47$\times10^{-6}$, for cores~A, B, C, and D. These abundances\footnote{We note that missing flux may affect differently the spatial distributions of gas and dust, introducing some uncertainties in the reported abundances.} are similar to the values found toward other hot cores associated with high-mass and intermediate-mass YSOs \citep[\eg][]{charnley1992, fuente2005, fuente2009, sanchezmonge2010}, and only slightly lower than the abundances ($\ga$$10^{-8}$ for CH$_3$CN, and $\ga$$10^{-6}$ for CH$_3$OH) measured in the strongest hot cores \citep[\eg][]{remijan2004, bisschop2007, mookerjea2007, wang2010}. We also computed the source-averaged surface density as $\Sigma_\mathrm{core}=4\,M_\mathrm{core}/(\pi D_{\rm S}^2)$, where the source diameter, $D_{\rm S}$, is taken from Table~\ref{t:continuum}.

\begin{figure}[t!]
\begin{center}
\begin{tabular}[b]{c}
 \epsfig{file=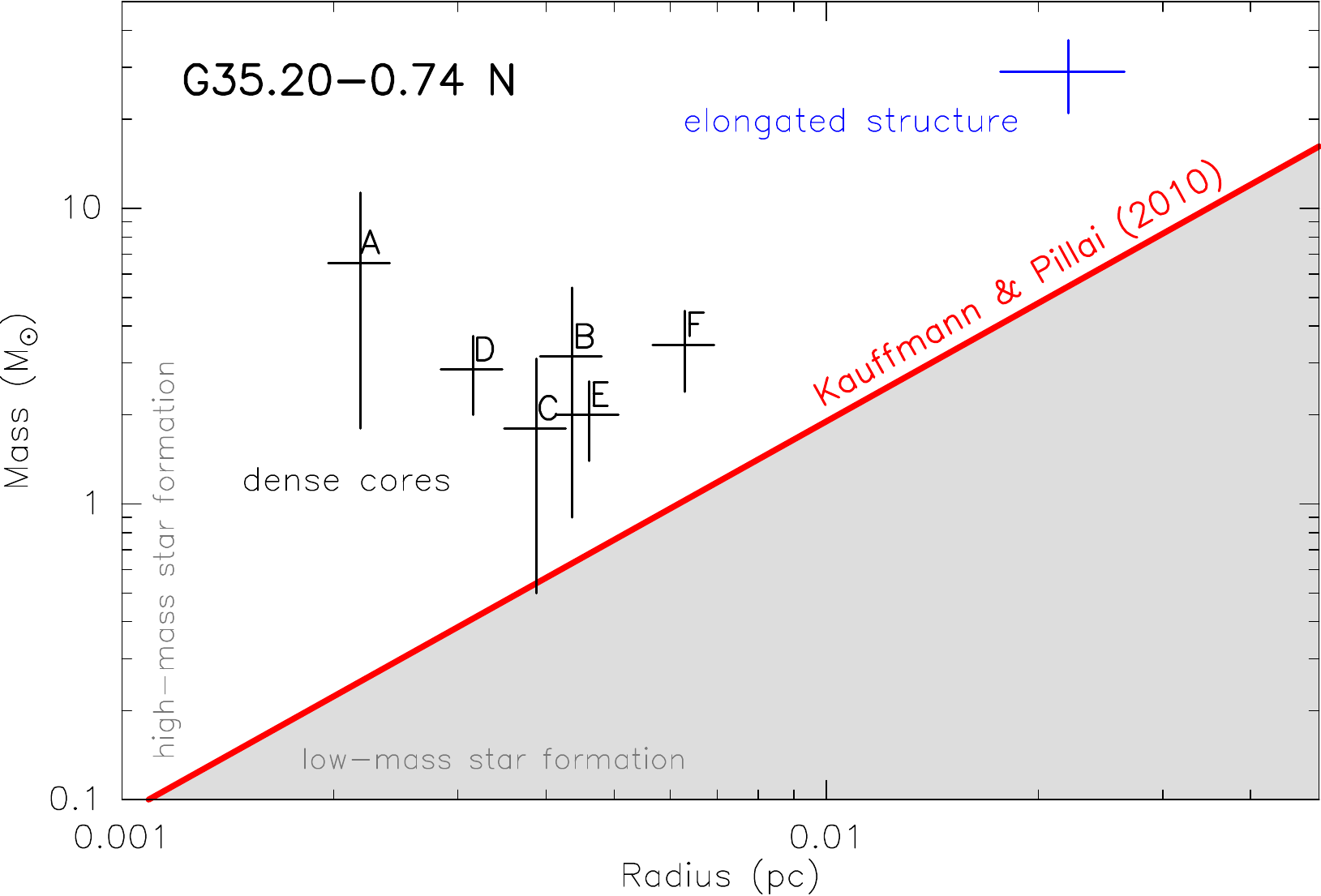, width=0.9\columnwidth, angle=0} \\
\end{tabular}
\caption{Mass-radius relation for the cores (\emph{black crosses}) and the elongated structure (\emph{blue cross}) in G35.20N. The vertical length of the crosses denotes the uncertainty in the mass (see Table~\ref{t:physical}), while the horizontal length denotes a 10\% error in the radius of the core (Table~\ref{t:continuum}). The red solid line indicate the criteria for massive star formation by \citet{kauffmannpillai2010}, with the gray area corresponding to the low-mass star formation region.}
\label{f:massradius}
\end{center}
\end{figure}

The G35.20N region is believed to be forming high-mass stars, based on the large luminosity (see Section~\ref{s:sed}), association with maser emission, and measurements of the mass of the embedded stars \citep{sanchezmonge2013b}. However, from Table~\ref{t:physical}, one sees that the values of the core masses are typically small, so one may wonder whether such a limited amount of gas is consistent with the formation of a massive star. We note that, as a matter of fact, there are quite a few B-type protostars associated with cores of a few \mo, such as IRAS\,20126$+$4104 \citep{cesaroni2005, beltran2011}, W3\,IRS5 \citep{wang2013} and G35.03$+$0.35 (Beltr\'an \et\ A\&A, submitted). It is possible that cores that ``light'' are only the remnants of the original envelopes enshrouding the protostars, while most of the material has already been accreted and/or dispersed. As suggested by \citet{cesaroni2005iau}, the ``light'' cores might be forming only one or very few massive stars, unlike the ``heavy'' cores $>$100~\mo, which likely contain multiple high-mass stars. The surface densities measured toward the cores in G35.20N (see Table~\ref{t:physical}) are significantly higher than 0.15--1~g~cm$^{-2}$, which is the theoretical minimum value needed to form massive stars \citep{krumholzmckee2008, butlertan2012}. We note, however, that the theoretical minimum surface density usually refers to clumps or clouds larger than the cores detected in G35.20N. We also compared the mass and size of the six cores identified in G35.20N with the empirical relation for a dense core to form a massive star as derived by \citet[][see also \citealt{elia2013}]{kauffmannpillai2010}. In Fig.~\ref{f:massradius}, we see that the six cores as well as the whole elongated structure\footnote{The mass of the whole elongated structure imaged with ALMA has been computed by adding the masses of the cores to the mass of the extended emission. The flux density of the whole structure is $\sim$3.3~Jy, with $\sim$2~Jy coming from the cores (see Table~\ref{t:continuum}). We assume that the remaining flux ($\sim$1.3~Jy) comes from gas with a temperature of 30--50~K to derive a total mass of the structure (extended emission plus dense cores) of 21--37~\mo.} lay above the mass-size relation. However, we note that this relation is derived for clumps/clouds with sizes 0.1--10~pc, and it might be no longer a valid description for cores with sizes $\sim$0.01~pc. Further observations are thus required to confirm if cores~D to F may form high-mass stars, or only intermediate/low-mass counterparts.

Finally, we note that the mass of the whole elongated structure derived by us is several times less than that of $\sim$130~\mo\ estimated by \citet{gibb2003} and \citet{birks2006} at lower angular resolution, assuming a dust temperature of 33~K and the opacities of \citet{ossenkopfhenning1994}. This indicates that our ALMA observations resolve out part of the large-scale, lower-density envelope, and are only sensitive to more compact and dense material, as suggested from the missing flux estimation done in Section~\ref{s:cont}. We also note that a fraction of the mass available in the elongated structure might be eventually accreted onto the different dense cores, and increase the final mass of the stars forming there.

%----------------------------------------------------------------------
%----------------------------------------------------------------------

%----------------------------------------------------------------------
\subsection{Dynamical status of the cores\label{s:virial}}

We estimated the virial masses of the cores identified in Fig.~\ref{f:continuum} as $M_\mathrm{vir}=105\,[D_{\rm S}/\mathrm{pc}][\Delta {\rm v}/\mathrm{km~s}^{-1}]^2$~\mo\ \citep[\eg][]{maclaren1988, bertoldimckee1992}, where $D_{\rm S}$ is the diameter of the core from Table~\ref{t:continuum}, $\Delta {\rm v}$ is the line full width at half maximum (FWHM), and the numerical value accounts for a uniform density distribution across the core. Different numerical factors have to be considered for non-uniform density distributions, \eg\ 95 for $n\propto r^{-1}$, and 63 for $n\propto r^{-2}$ \citep{maclaren1988}. This results in a virial mass smaller by a factor 1.1 and 1.7, respectively. The value of $\Delta {\rm v}$ is obtained from an average of the CH$_3$CN and CH$_3$OH spectral line fits (see Table~\ref{t:myxclass}). The virial mass derived from these hot core tracers is listed in column~8 of Table~\ref{t:physical}. An estimate of the virial mass is also obtained by using the H$^{13}$CO$^+$ line, which is detected throughout the whole elongated structure (see Fig.~\ref{f:largescale}) and hence towards all the cores. We note, however, that extended structures are heavily resolved out in our observations. This results in fake absorption features in the spectra of the cores, affecting our estimates of the line widths, which might be underestimated. In columns~9 and 10 of Table~\ref{t:physical}, we list the lower limits obtained for the H$^{13}$CO$^+$ line FWHM and the corresponding virial masses.

From Table~\ref{t:physical} one sees that the virial masses are typically greater than (or similar to) the core masses. Since the former are lower limits (as explained above), we conclude that the virial parameter is $M_\mathrm{virial}/M_\mathrm{core}\ge$1, with the cores possibly confined by the external pressure. It is worth noting, that other studies \citep[\eg][]{sanchezmonge2013c, kauffmann2013} have found similar results for the virial parameters when studying cores at scales $\sim$0.5~pc. There is a global trend in which virial parameters close to unity are found at scales $\sim$1--10~pc, whereas greater values of the virial parameter are obtained on smaller scales ($\sim$0.01--0.5~pc), suggesting that external pressure effects might be important mainly at small scales. These results provide some evidence in favor of the theoretical idea that on small scales, feedback from \eg\ protostellar winds is sufficiently effective to keep regions out of virial equilibrium at scales below 0.5~pc.

%----------------------------------------------------------------------
\subsection{Velocity field of cores~A and B\label{s:disks}}

\begin{figure}[t!]
\begin{center}
\begin{tabular}[b]{c}
 \epsfig{file=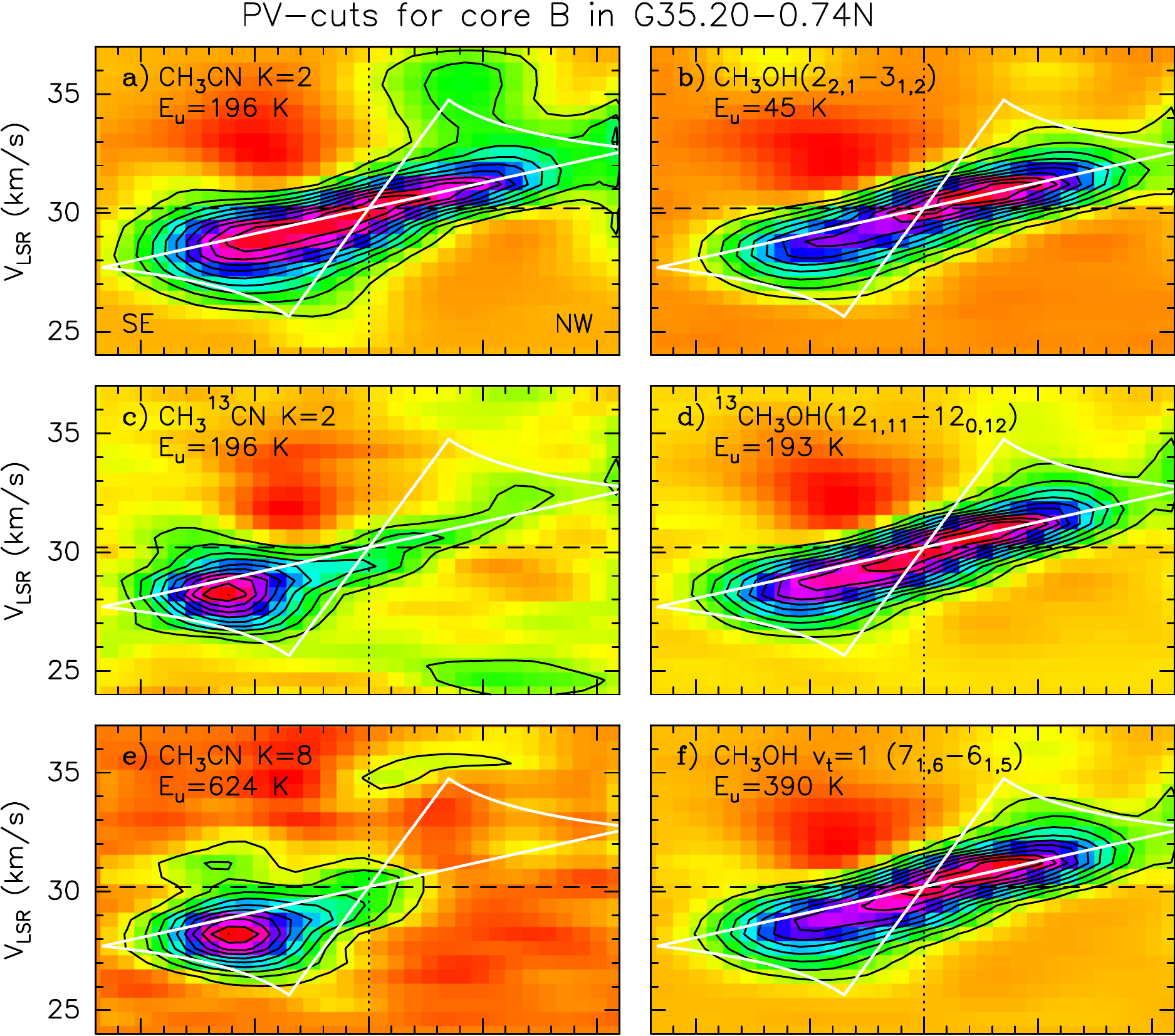, width=0.88\columnwidth, angle=0} \\
 \epsfig{file=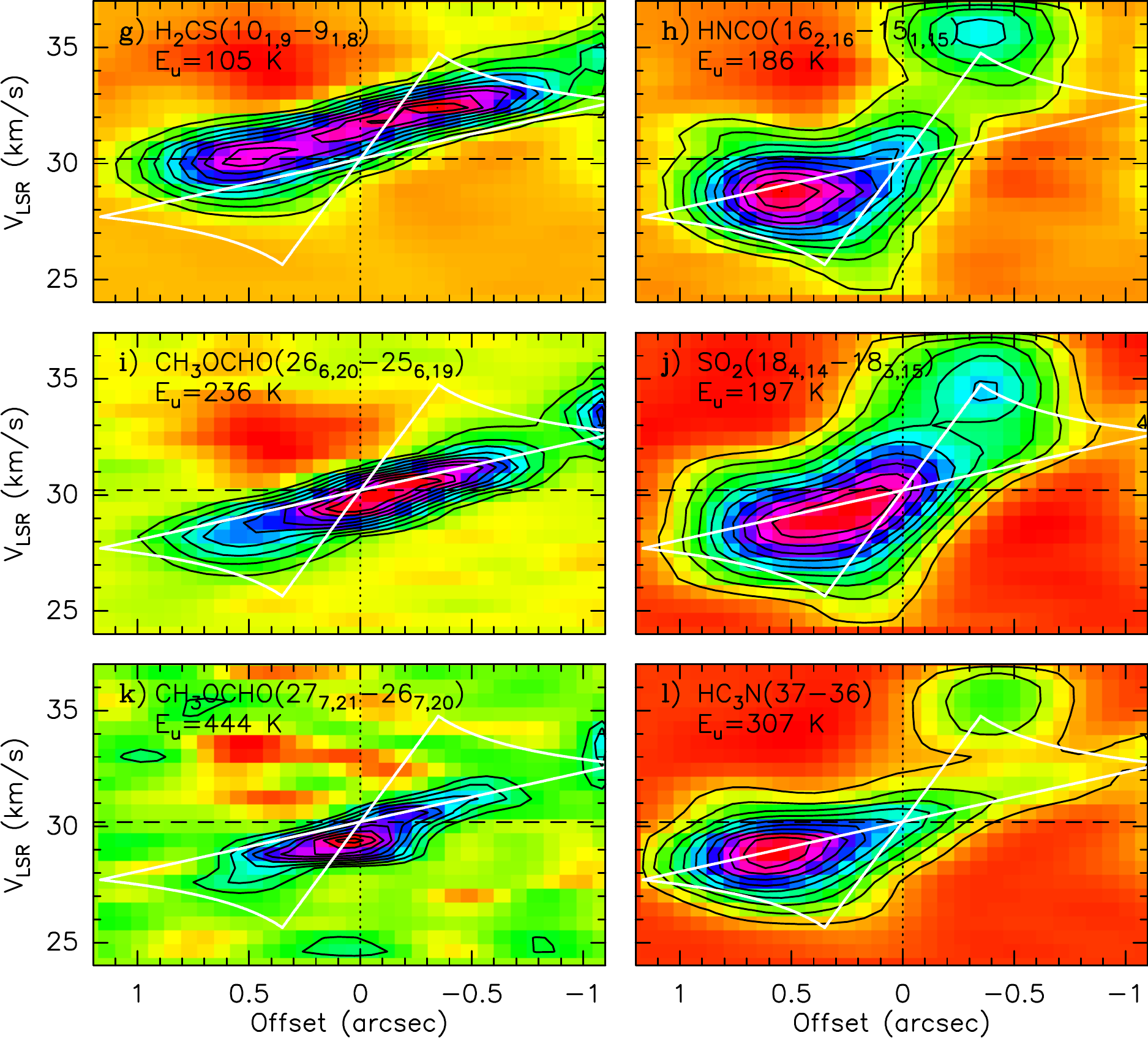, width=0.88\columnwidth, angle=0} \\
\end{tabular}
\caption{Position-velocity plots along the direction with P.A.=157\degr\ towards core~B for different species and transitions. The offsets are measured from the dust peak position of core~B (see Table~\ref{t:continuum}), positive towards southeast.  Contour levels start at 10\%, increasing in steps of 10\% of the peak: (a) 0.53, (b) 0.72, (c) 0.08, (d) 0.33, (e) 0.10, (f) 0.51, (g) 0.56, (h) 0.27, (i) 0.09, (j) 0.37, (k) 0.02 and (l) 0.52~Jy~\kms. The white solid line marks the border of the region where emission is expected for an edge-on Keplerian disk of radius $1\farcs2$ (2600~AU) rotating about a 18~\mo\ central mass.}
\label{f:pvplotsB}
\end{center}
\end{figure}

\begin{figure}[t!]
\begin{center}
\begin{tabular}[b]{c}
 \epsfig{file=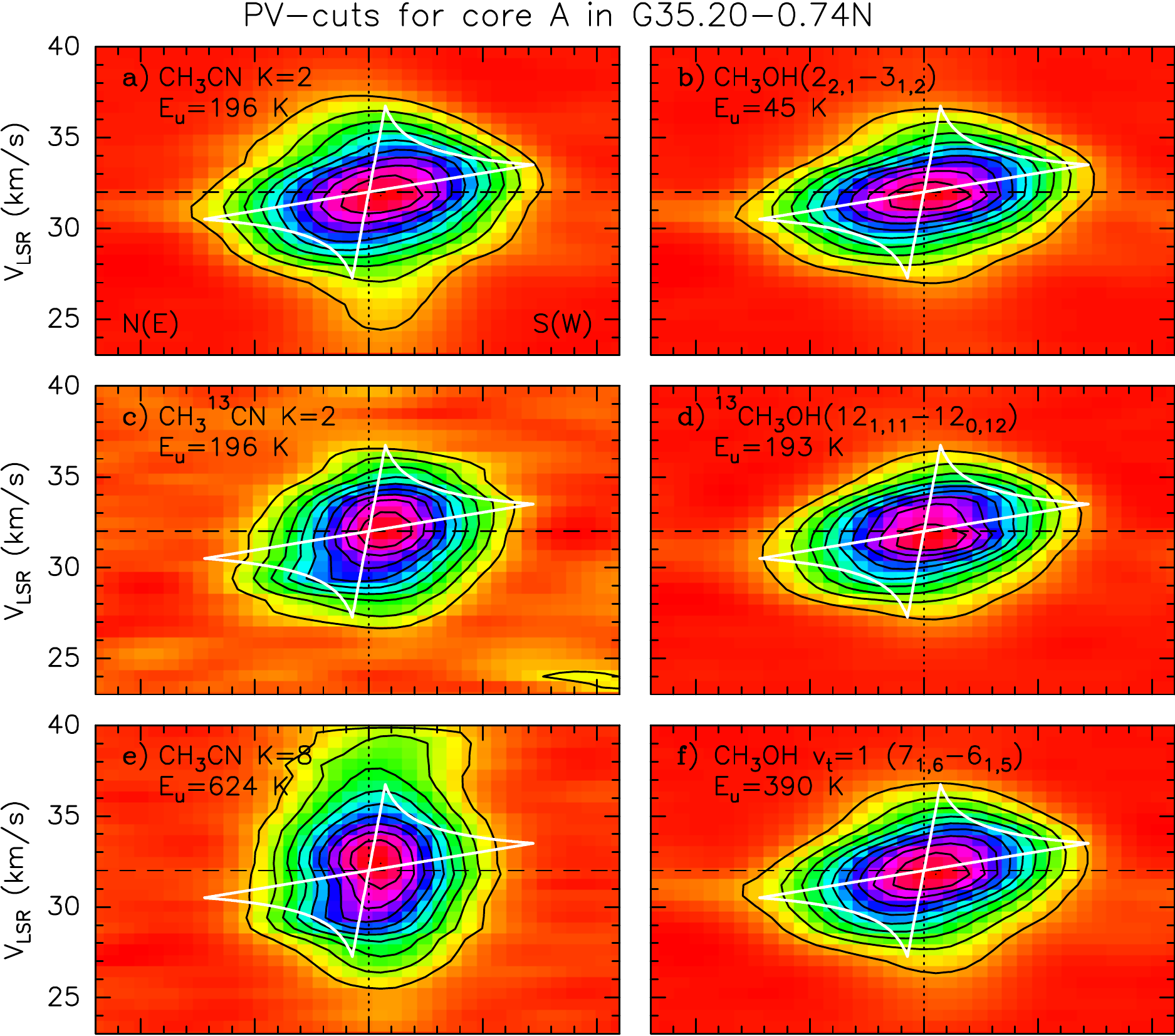, width=0.88\columnwidth, angle=0} \\
 \epsfig{file=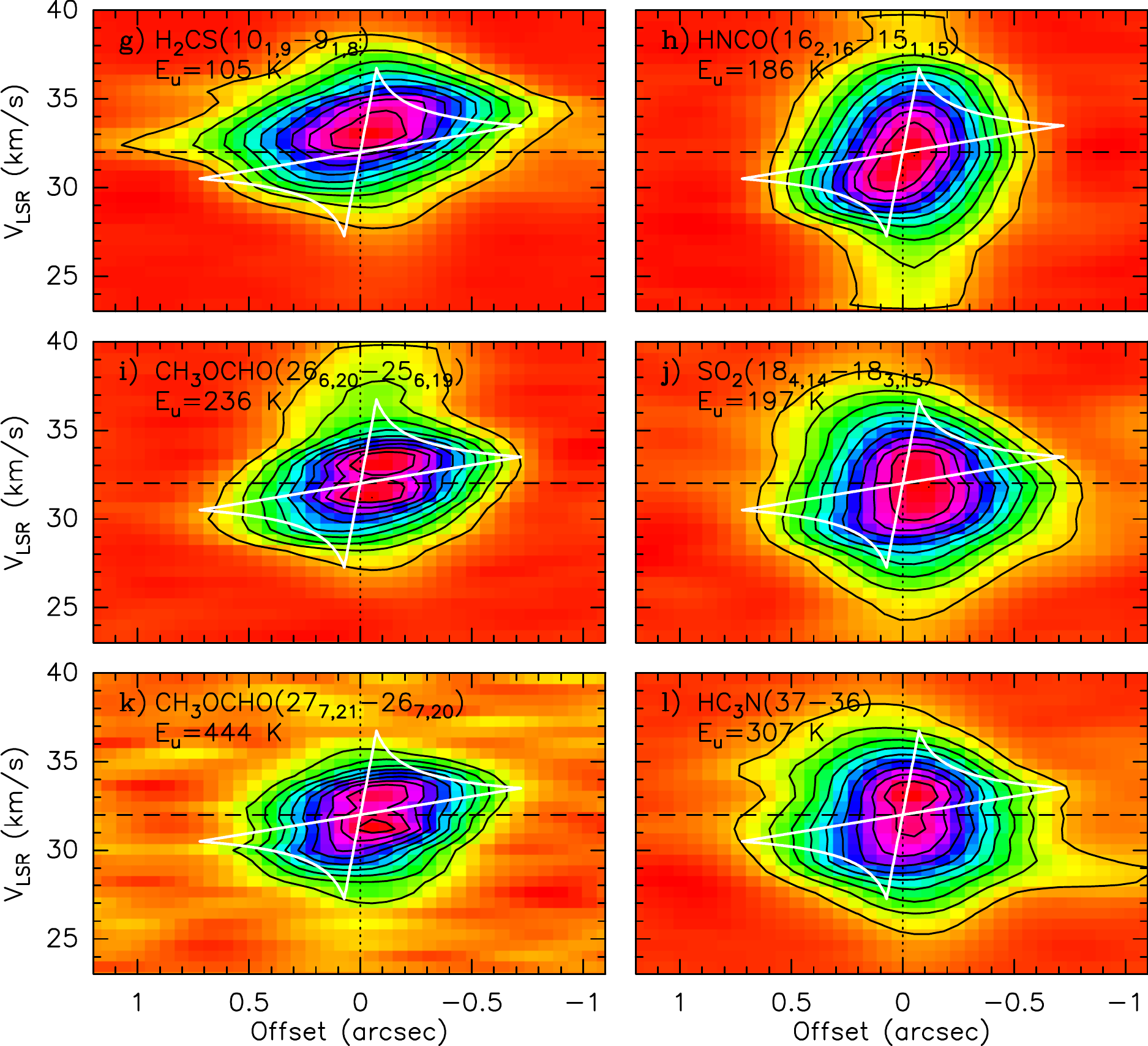, width=0.88\columnwidth, angle=0} \\
\end{tabular}
\caption{Position-velocity plots along the direction with P.A.=10\degr\ towards core~A for different species and transitions. The offsets are measured from the dust peak position of core~A (see Table~\ref{t:continuum}), positive towards northeast. Contour levels start at 10\%, increasing in steps of 10\% of the peak: (a) 0.87, (b) 0.99, (c) 0.13, (d) 0.64, (e) 0.23, (f) 0.81, (g) 0.76, (h) 0.47, (i) 0.27, (j) 0.30, (k) 0.06 and (l) 0.34~Jy~\kms. The white solid line marks the border of the region where emission is expected for an edge-on Keplerian disk of radius $0\farcs7$ (1500~AU) rotating about a 4~\mo\ central mass.}
\label{f:pvplotsA}
\end{center}
\end{figure}

As shown in Figs.~\ref{f:hotcore1}, \ref{f:hotcore2} and \ref{f:hotcore3}, the emission of almost all molecular species reveals a velocity gradient in cores~A and B, oriented N--S in core~A and SE--NW (\ie\ along the major axis of the hot core) in core~B. The velocity shift is approximately the same ($\sim$3--4~\kms) for all species and transitions. In Figs.~\ref{f:pvplotsB} and \ref{f:pvplotsA}, we present position-velocity (PV) cuts along the direction where the velocity gradient is maximum: P.A.=157\degr\ for core~B and P.A.=10\degr\ for core~A. In the following we analyze these plots separately for the two cores.

%----------------------------------------------------------------------
%----------------------------------------------------------------------
%----------------------------------------------------------------------
%----------------------------------------------------------------------

%----------------------------------------------------------------------
\subsubsection{Core B}

\citet{sanchezmonge2013b} studied the velocity pattern of core~B. The authors fitted the emission of several molecules in each velocity channel with a 2D Gaussian. The positions of the Gaussian peaks, which provide a picture of the velocity field (see Fig.~4 of \citealt{sanchezmonge2013b}), were fitted with a simple Keplerian disk model. From the best fit, the authors concluded that the velocity gradient in core~B traces a Keplerian disk rotating about a central mass of 18~\mo, and inclined 19\degr\ with respect to the plane of the sky. The position angle of the disk is 157\degr. Comparing the molecular disk detected with ALMA with the centimeter continuum emission \citep{gibb2003} and the 4.5~$\mu$m \emph{Spitzer}/IRAC emission (see Fig.~\ref{f:continuum}a), the authors concluded that the disk in core~B rotates around a central object which is the powering source of the NE--SW outflow/jet seen in the centimeter \citep{gibb2003}, in the infrared \citep{dent1985b, fuller2001, debuizer2006, zhang2013} and in large-scale molecular outflow tracers \citep{dent1985a, gibb2003, birks2006}.

The PV-plots towards core~B (see Fig.~\ref{f:pvplotsB}) are consistent with the interpretation of \citet{sanchezmonge2013b}, as they present a ``butterfly'' pattern typical of Keplerian rotation \citep[see \eg][]{cesaroni2005}, with the emission extending to high velocities ($\pm$5~\kms\ with respect to the systemic velocity of $\sim$30~\kms) at small offsets from the dust continuum peak (expected to trace the star position), and small velocities ($\pm$1~\kms) at large offsets. The white pattern shown in the figure, encompasses the region where emission is expected for an edge-on Keplerian disk of radius 1\farcs2 (or 2600~AU), rotating about an 18~\mo\ central mass. This pattern is roughly consistent with the data, taking into account that it is obtained for zero line width and infinite angular resolution.

The PV-plots appear to change depending on the molecule. Species such as CH$_3$CN, HC$_3$N or SO$_2$ have the emission peak to the southeast at blue-shifted velocities ($\sim$28~\kms), whereas species such as CH$_3$OH, CH$_3$OCHO, and H$_2$CS peak to the northwest at red-shifted velocities ($\sim$31~\kms). We find that N-bearing molecules are typically stronger to the southeast of the continuum peak, while O-bearing molecules are stronger to the northwest. These differences could be a consequence of chemical variations across the disk, perhaps due to the interaction with nearby sources. An asymmetry was also observable in the Keplerian disk model fit presented in \citet{sanchezmonge2013b}. 

The Keplerian pattern is less evident in transitions like CH$_3$CN $K$=8, CH$_3$$^{13}$CN $K$=2 or CH$_3$OCHO\,(27$_{7,21}$--26$_{7,20}$) (see Figs.~\ref{f:pvplotsB}c, e, and k), because these are detected only on one side of the disk. This suggests that not only the molecular abundance, but also the density distribution is not axially symmetric, similar to the case of the disk around the B-type protostar IRAS~20126$+$4104 (\eg\ Cesaroni \et\ A\&A, submitted).

%----------------------------------------------------------------------
\begin{figure}[t!]
\begin{center}
\begin{tabular}[b]{c}
 \epsfig{file=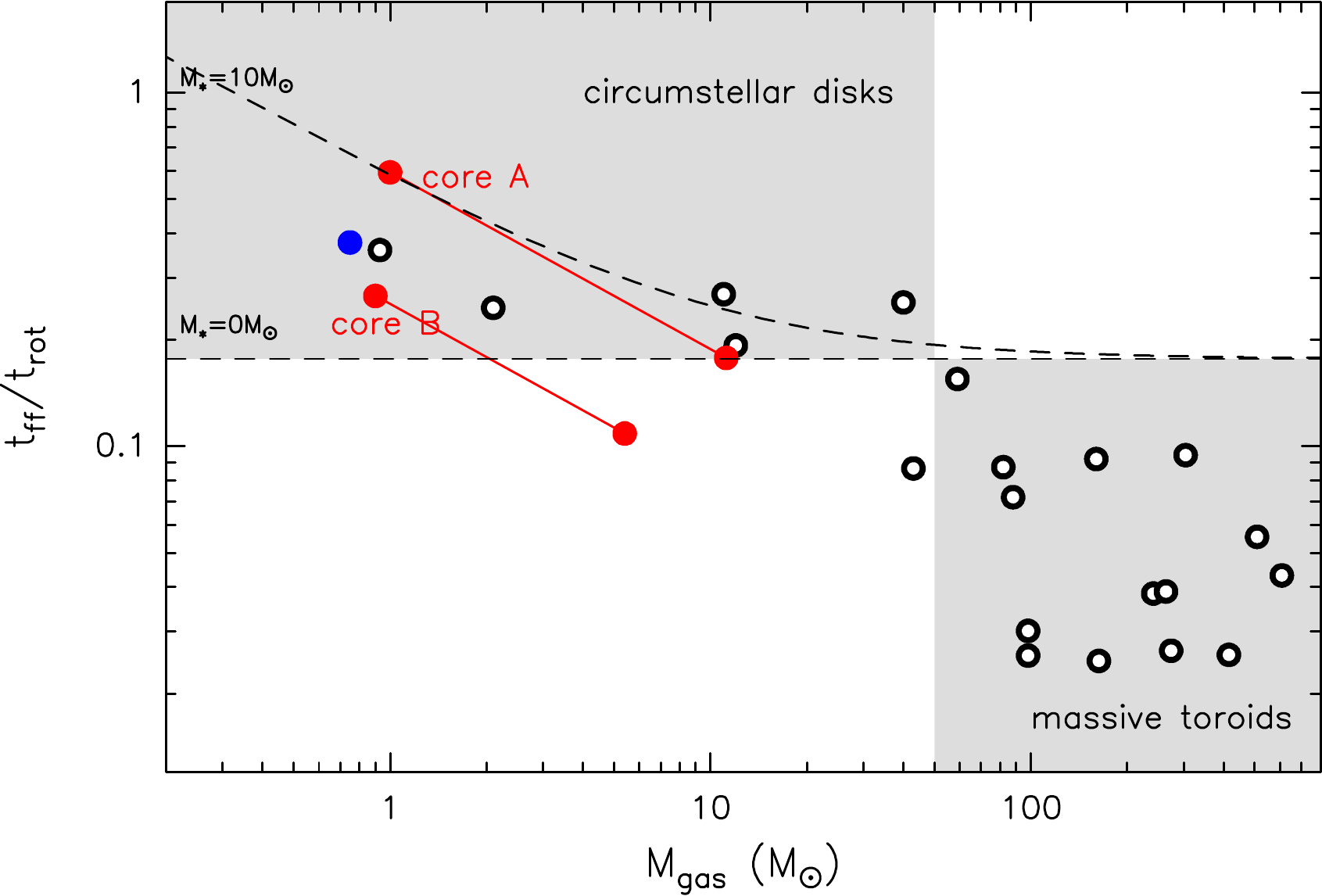, width=0.9\columnwidth, angle=0} \\
\end{tabular}
\caption{Free-fall timescale to rotational period ratio ($t_\mathrm{ff}/t_\mathrm{rot}$) versus gas mass ($M_\mathrm{gas}$) of known rotating disks or toroids. Open circles correspond to the objects studied by \citet{beltran2011}, blue dot corresponds to source G35.03$+$0.35 (Beltr\'an \et\ A\&A, submitted), and the red dots correspond to cores~A and B of G35.20N (this work). The connecting line corresponds to the different gas mass estimates for different temperatures (see Table~\ref{t:physical}). The gray areas show the regions where circumstellar disks and toroids are found. The dashed lines correspond to spheres of mass $M_\mathrm{gas}$, containing a star of mass $M_{*}$ (\eg\ 0~\mo\ and 10~\mo; see Section~\ref{s:otherdisks}).}
\label{f:tfftrotmass}
\end{center}
\end{figure}
%----------------------------------------------------------------------

%----------------------------------------------------------------------
\subsubsection{Core A}

Core~A is less resolved than core~B, which hinders an analysis like that carried out by \citet{sanchezmonge2013b}. The velocity gradient in this case is almost N--S. If this is tracing a rotating disk, one should see a jet/outflow perpendicular to it, \ie\ in the E--W direction. The only clearly confirmed outflow with a similar orientation, is the one in the NE--SW direction observed with an angular resolution $\ga5$\arcsec\ \citep[\eg][]{gibb2003, birks2006}. As discussed later in Sect.~\ref{s:jet}, this outflow is likely powered by a precessing jet arising from core~B \citep[see also][]{sanchezmonge2013b}. In Section~\ref{s:outflow}, we analyzed the SiO\,(8--7) transition covered in our ALMA observations. The combination of the SiO emission together with the radio continuum emission (see Fig.~\ref{f:sioout}) suggests the presence of a jet/outflow centered in core~A and oriented approximately ENE-WSW, \ie\ roughly perpendicular to the velocity gradient in the core. This lends support to the hypothesis that core~A might also be a rotating disk.

The PV-plots of core~A shown in Fig.~\ref{f:pvplotsA}, reveal a compact source, with a small velocity gradient. Although the evidence of Keplerian rotation for core~A is less clear than that in core~B, in a few species the pattern looks slightly elongated (see panels a, b, f, and g), with the emission close to the center extending up to velocities as high as $\pm$5~\kms\ (with respect to the systemic velocity of $\sim$32~\kms) and the low-velocity emission ($\pm$1.5~\kms) detected at relatively large offsets ($\pm$$0\farcs7$ with respect to the continuum peak).  Assuming that one is indeed observing a Keplerian disk with a rotation velocity of 1.5~\kms\ at a radius of 0\farcs7 (or 1500~AU), we derive a mass of $\sim$4~\mo\ for the putative central star. Higher angular resolution data are needed to confirm whether core A is undergoing Keplerian rotation and derive a reliable estimate of the central mass.

%----------------------------------------------------------------------
\subsubsection{Comparison with other disk candidates\label{s:otherdisks}}

In Fig.~\ref{f:tfftrotmass}, we follow the approach of \citet{beltran2011} to compare the rotating structures found in G35.20N with those found around B-type and O-type stars. These authors plot the ratio of the free-fall timescale to the rotational period ($t_\mathrm{ff}/t_\mathrm{rot}$) versus the gas mass ($M_\mathrm{gas}$), for a sample of 20 high-mass YSOs. In this expression, $t_\mathrm{ff}$ is the free-fall timescale of the gas mass calculated from (sub)mm continuum emission, given by the expression $t_\mathrm{ff}=\sqrt{3\pi/(32G\rho)}$ ($=1.83\times10^{7}\,[M_\mathrm{gas}/M_{\sun}]^{-0.5}\,[R_\mathrm{S}/\mathrm{pc}]^{1.5}$~yr). This quantity is an estimate of the timescale of collapse of the gas under its own gravity. The rotational period, $t_\mathrm{rot}=2\pi R_\mathrm{S}/V_\mathrm{rot}$ ($=6.15\times10^{6}\,[R_\mathrm{S}/\mathrm{pc}]\,[V_\mathrm{rot}/\mathrm{km~s^{-1}}]^{-1}$~yr), is instead obtained from the observed velocity gradient and is basically the timescale needed for the disk to re-adjust its internal structure to the newly accreted material. \citet{beltran2011} find that rotating structures around B-type stars likely tracing circumstellar disks have higher $t_\mathrm{ff}/t_\mathrm{rot}$ values than the massive toroids. In Fig.~\ref{f:tfftrotmass}, we have added the results of cores~A and B in G35.20N to the plot of \citet{beltran2011}. In our calculations we consider a rotation velocity $V_\mathrm{rot}$ of 2~\kms\ and 1.2~\kms, and a radius of 0.01~pc and 0.005~pc for cores~B and A, respectively. The two cores of G35.20N are located in the region of Keplerian disks, which lends additional support to our hypotheses. In Fig.~\ref{f:tfftrotmass}, we also show the theoretical curves corresponding to the value of $t_\mathrm{ff}/t_\mathrm{rot}$ for a disk of mass $M_\mathrm{gas}$ rotating about a star of mass $M_*$. This can be obtained from the previous expressions of $t_\mathrm{ff}$ and $t_\mathrm{rot}$, by substituting the rotation velocity with $V_\mathrm{rot}=\sqrt{G(M_\mathrm{gas}+M_*)/R_\mathrm{S}}$:
\begin{equation}
\frac{t_\mathrm{ff}}{t_\mathrm{rot}}=\sqrt{\frac{M_\mathrm{gas}+M_{*}}{32\,M_\mathrm{gas}}} \label{etrat}
\end{equation}
The dashed line shown in the figure correspond to $M_*$=0~\mo\ and $M_*$=10~\mo. From these it is clear that the objects lying in the `circumstellar disks' region are characterized by dynamically important levels of rotation, and have rotation timescales of the order (or shorter) than the dynamical timescale of collapse of the gas under its own self-gravity. Conversely, objects in the region labeled `massive toroids' are gas masses possibly supported by a combination of thermal pressure, turbulence, and magnetic fields with a little amount of rotation that however plays no role in their support. Thus, for these objects, the ratio $t_\mathrm{ff}/t_\mathrm{rot}$t can take arbitrarily small values below the horizontal dashed line that corresponds to a homogeneous gas sphere in rotational equilibrium.

\begin{figure}[t!]
\begin{center}
\begin{tabular}[b]{c}
 \epsfig{file=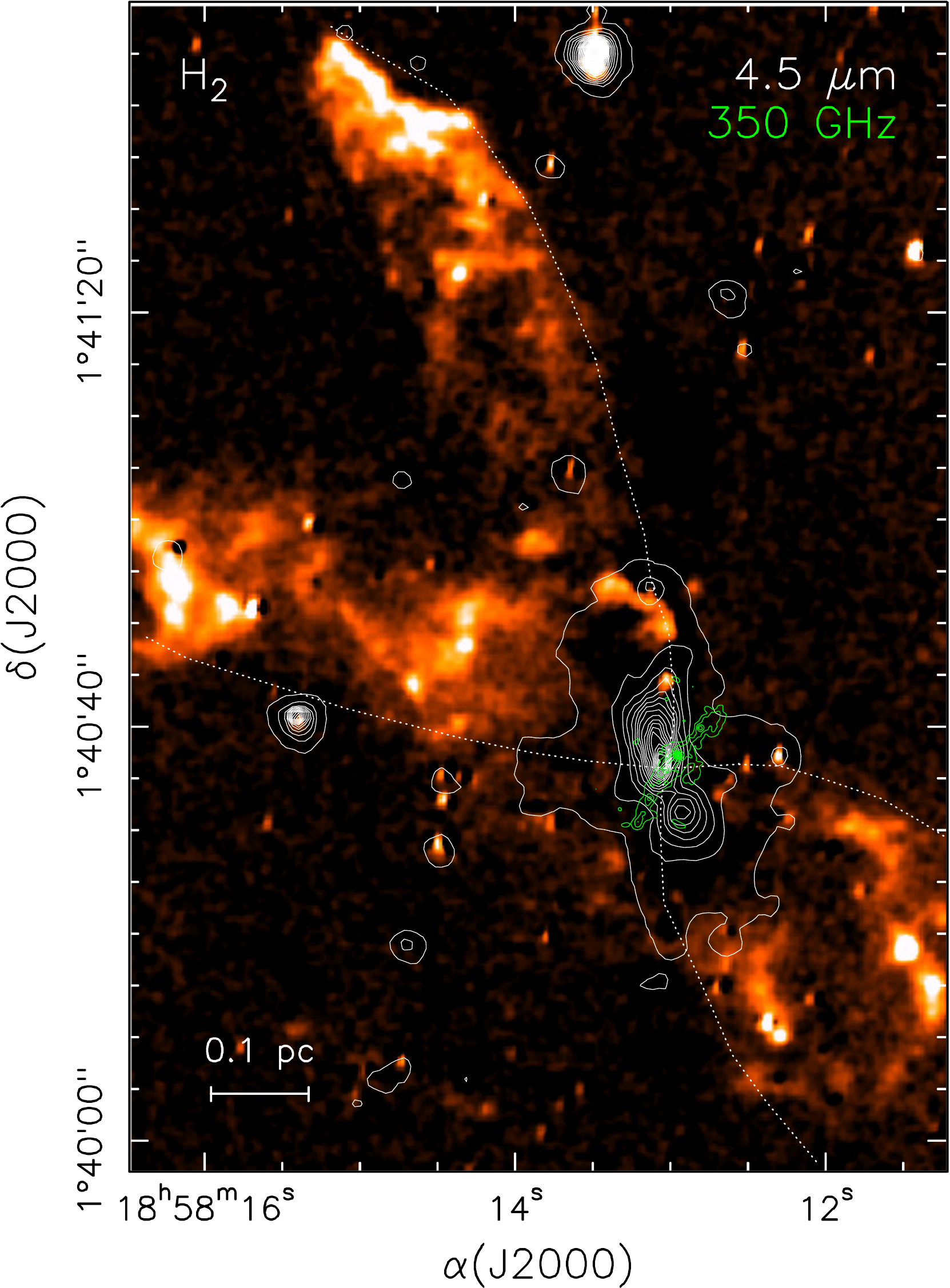, width=0.9\columnwidth, angle=0} \\
\end{tabular}
\caption{Image of the H$_2$ 2.12~$\mu$m line emission towards G35.20. The white and green contours are, respectively, maps of the 4.5~$\mu$m emission from the \emph{Spitzer}/GLIMPSE survey \citep{benjamin2003} and 350~GHz continuum emission from our observations. The dotted curves outline the bipolar pattern.}
\label{f:jet}
\end{center}
\end{figure}

\begin{figure}[t!]
\begin{center}
\begin{tabular}[b]{c}
 \epsfig{file=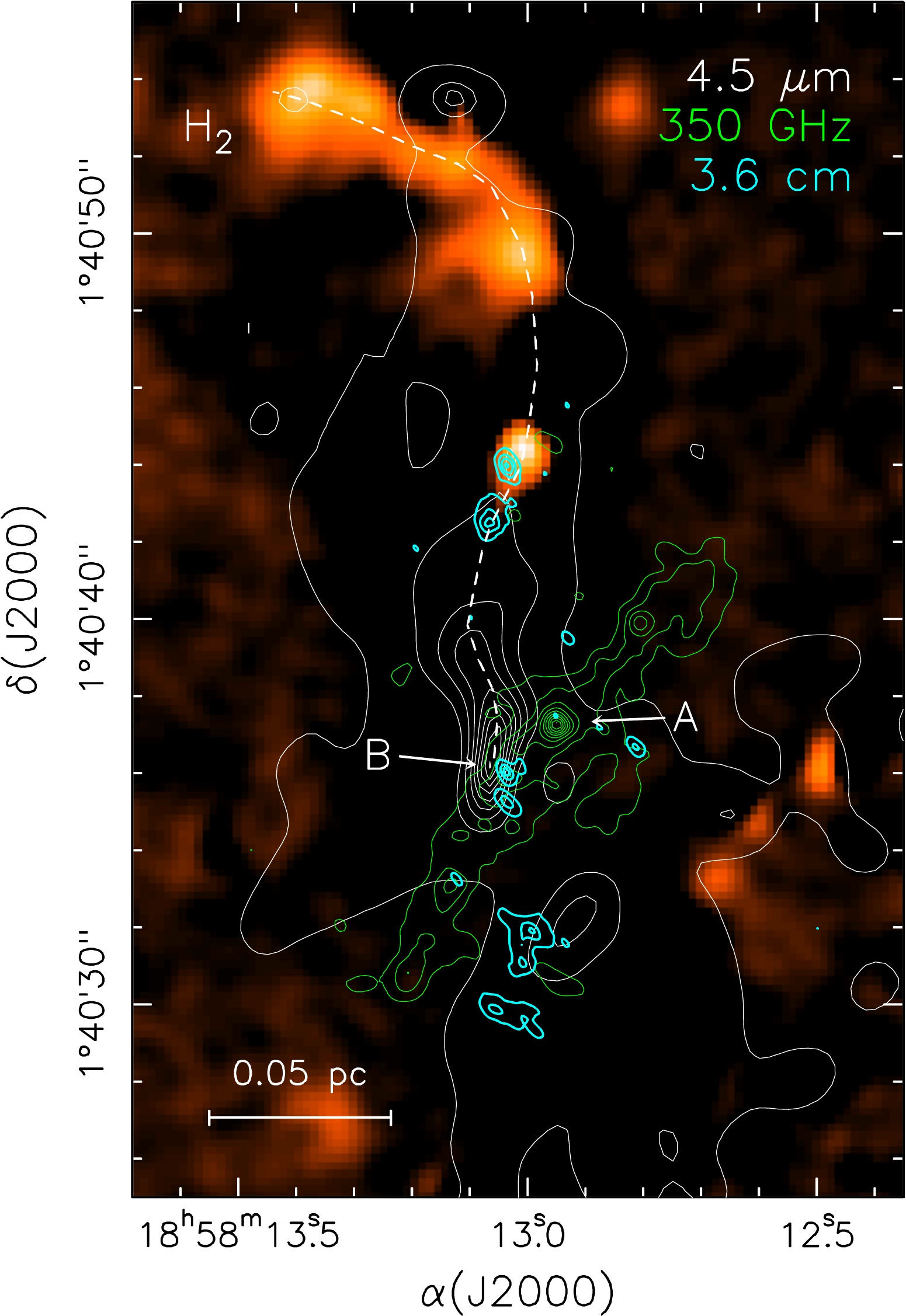, width=0.9\columnwidth, angle=0} \\
\end{tabular}
\caption{Same as Fig.~\ref{f:jet}, with overlaid also the 3.6~cm continuum map of \citet{gibb2003} (cyan contours). The resolution of the IRAC image (white contours) has been enhanced by HiRes deconvolution \citep{velusamy2008}. The dashed curve outlines the shape of the jet, which is bending by almost $\sim$90\degr\ at an offset of $\sim$11\arcsec\ to the north of core~B.}
\label{f:jetzoom}
\end{center}
\end{figure}

\begin{figure}[t!]
\begin{center}
\begin{tabular}[b]{c}
 \epsfig{file=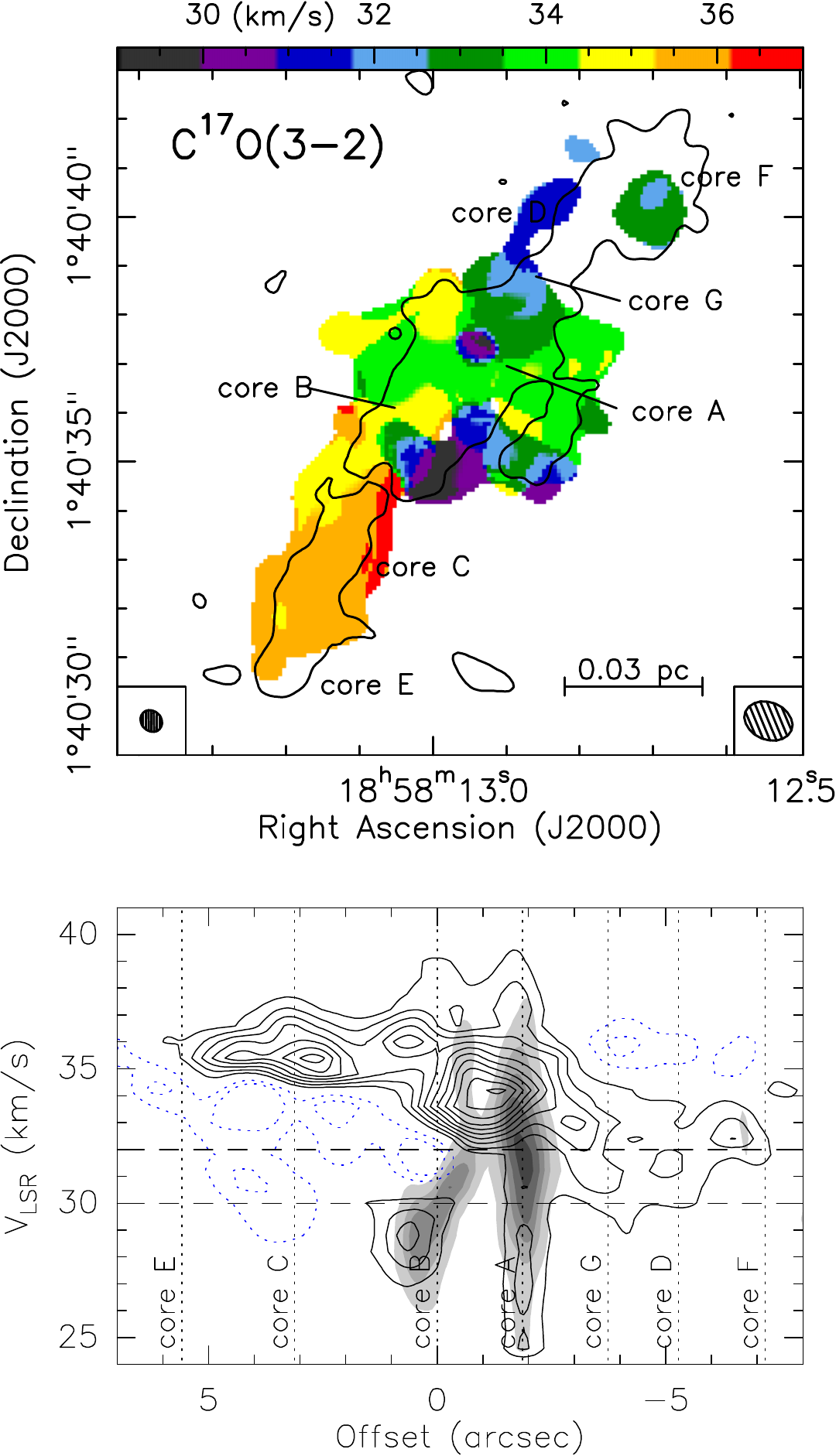, width=0.88\columnwidth, angle=0} \\
\end{tabular}
\caption{{\bf Top:} Overlay of the 870~$\mu$m continuum emission (\emph{black contour}) on the velocity field (\emph{colors}) of the C$^{17}$O\,(3--2) line, after tapering the line emission at 1\arcsec. The contour level of the continuum emission is the 9.0~m\jpb\ level (see Fig.~\ref{f:continuum}). The velocity scale is shown in the color bar at the top of the panel, in \kms. Other symbols as in Fig.~\ref{f:largescale} {\bf Bottom:} Position-velocity plot along the major axis of the elongated structure (P.A.=$-$60\degr), centered at the position of core~B, for C$^{17}$O (contours) and CH$_3$CN (gray scale). Vertical dotted lines mark the position of the cores identified in the continuum. The two horizontal dashed lines mark the velocities 30~\kms\ and 32~\kms, associated with the hot core emission of cores~B and A, respectively.}
\label{f:largescaleC17O}
\end{center}
\end{figure}

\begin{figure}[t!]
\begin{center}
\begin{tabular}[b]{c}
 \epsfig{file=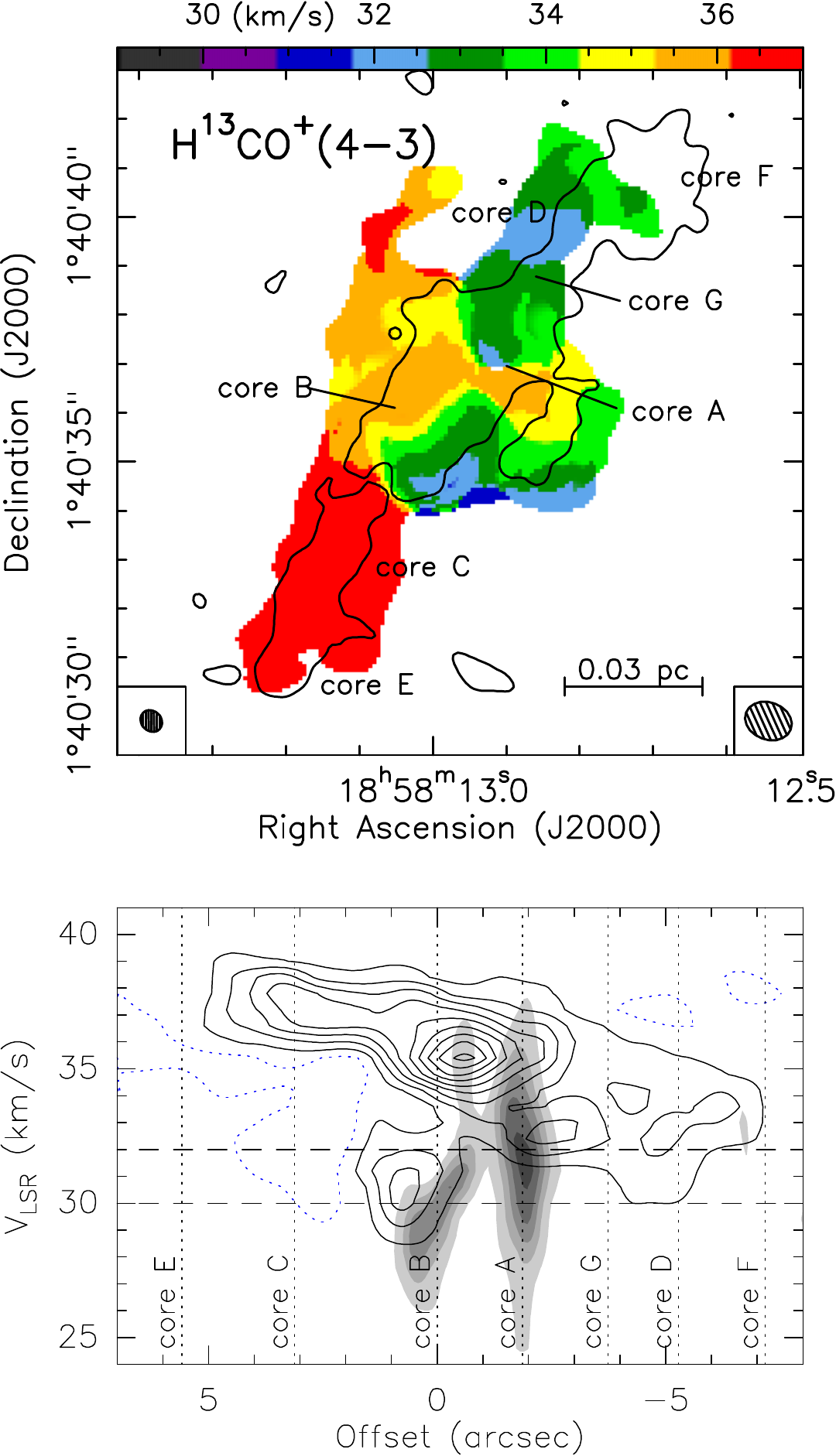, width=0.88\columnwidth, angle=0} \\
\end{tabular}
\caption{As Figure~\ref{f:largescaleC17O}, but for H$^{13}$CO$^{+}$\,(4--3) line.}
\label{f:largescaleH13CO+}
\end{center}
\end{figure}

Finally, we estimated the velocity gradients of cores~A and B, from the velocity shifts measured for the best resolved transitions (CH$_3$CN $K$=2, CH$_3$OH\,(7$_{1,6}$--6$_{1,5}$) H$_2$CS\,(10$_{1,9}$--9$_{1,8}$) and HC$_3$N\,(37--36)). The shifts are $\sim$4~\kms\ over an extent of $\sim$0.02~pc for core~B, and $\sim$2.5~\kms\ over $\sim$0.01~pc for core~A (see Figs.~\ref{f:hotcore1}--\ref{f:hotcore3} and Figs.~\ref{f:pvplotsA} and \ref{f:pvplotsB}), which correspond to a velocity gradient at scales $\sim$0.01~pc of about 250~\kms~pc$^{-1}$ and 200~\kms~pc$^{-1}$ for cores~A and B, respectively. These values are consistent with the expected Keplerian velocity at a radius of 0.01~pc (0.005~pc) and around a central mass of 16~\mo\ (4~\mo), for core~B (core~A): $\sim$2.5~\kms\ ($\sim$1.8~\kms).

%----------------------------------------------------------------------
%----------------------------------------------------------------------
%----------------------------------------------------------------------
%----------------------------------------------------------------------

%----------------------------------------------------------------------
\subsection{Precessing jet\label{s:jet}}

As previously discussed, the velocity fields of cores A and B are suggestive of rotation about an axis oriented E--W and NE--SW, respectively. The latter direction is consistent with the axis of the bipolar nebula in Fig.~\ref{f:continuum}a and corresponding outflow (see Sect.~\ref{s:intro}). This seems to confirm that the velocity gradient in core~B is indeed due to rotation about a NE--SW axis. However, we know that several starforming cores are present in G35.20N, and consequently the butterfly-shaped nebula in Fig.~\ref{f:continuum}a might be the result of the superposition of multiple outflows arising from these cores.  Indeed, the morphology of both the mid-IR emission (see Fig.~2 of \citealt{sanchezmonge2013b}; \citealt{zhang2013}) and 3.6~cm free-free continuum emission \citep{gibb2003} appears to indicate the existence of a jet from core~B directed N--S, and hence {\it not} perpendicular to the velocity gradient in core~B. Here, we argue that the non orthogonality between this jet and the core velocity gradient can be explained if the jet is precessing about an axis directed NE--SW. \citet{little1998} were the first to suggest that such a bipolar structure could be created by a precessing jet, and \citet{sanchezmonge2013b} have proposed that a binary system at the center of the Keplerian disk in core~B could be causing the precession.

To confirm this scenario, we have inspected the morphology of the H$_2$ emission (a typical jet tracer) at different scales: from the scale of the core to that of the bipolar nebula, and beyond. For this purpose, we retrieved the UKIDSS image of the H$_2$ 2.12~$\mu$m line, shown in Fig.~\ref{f:jet}. One clearly sees that the line emission outlines a bipolar structure (dotted curves) that is the ideal extrapolation to a larger scale, of the butterfly-shaped nebula seen in the IRAC 4.5~$\mu$m image (white contours). Even more instructive is the small-scale structure of the H$_2$ jet illustrated in Fig.~\ref{f:jetzoom}. Putting together the information provided by the free-free 3.6~cm continuum emission (cyan contours), and the 2.12~$\mu$m and 4.5~$\mu$m images, one sees that the jet from core~B is directed N--S, up to a distance of 8\arcsec\ (0.085~pc) from the origin, but at that point it sharply bends to the east (see dashed curve). This behavior is strongly suggestive of precession about an axis significantly inclined with respect to the N--S direction.

From the previous evidence we conclude that the morphology of the jet/outflow from core~B is compatible with precession of the jet itself about an axis roughly perpendicular to the plane of the circumbinary disk in core~B, as hypothesized by \citet{sanchezmonge2013b}. We note that this interpretation does not rule out the possibility that other outflows are also present in this star forming region, as suggested by the complex structure of the SiO emission (see Sect.~\ref{s:outflow}). Moreover, the morphology of the large-scale H$_2$ emission seen in Fig.~\ref{f:jet} is difficult to explain with a single precessing outflow. In particular, the branch of H$_2$ emission extending to the east might be associated with the red lobe of the putative jet arising from core~A (see Fig.~\ref{f:sioout}).

%----------------------------------------------------------------------
%----------------------------------------------------------------------
%----------------------------------------------------------------------
%----------------------------------------------------------------------

%----------------------------------------------------------------------
\subsection{Nature of the elongated structure\label{s:kinematics}}

The shape and orientation of the elongated structure in Fig.~\ref{f:continuum} suggest that this might be a 2D flattened structure seen edge-on, lying along the waist of the bipolar nebula observed at IR wavelengths. The existence of a large-scale, rotating disk-like structure has been indeed proposed by several authors. In particular, \citet{little1985} and \citet{brebner1987} revealed a large ($\sim$1\arcmin\ or 0.6~pc) elongated structure in NH$_3$ and CS, oriented SE--NW, \ie\ perpendicular to the large-scale outflow, and with a velocity gradient, which they interpret as a large rotating disk/toroid. More recently, \citet{gibb2003} studied the G35.20N star forming complex with the BIMA interferometer in the H$^{13}$CO$^+$\,(1--0) and H$^{13}$CN\,(1--0) transitions, with an angular resolution of $\sim$10\arcsec. These authors find evidence of fragmentation across the same structure that they interpret as a fragmented, rotating envelope containing multiple young stellar objects, rather than a smooth rotating disk.

An alternative to the previous hypothesis, one should also consider the trivial possibility that what looks like a filament on the plane of the sky, is indeed a filament also in space. Recent continuum \emph{Herschel}/Hi-GAL and spectral line observations of the interstellar medium have revealed that molecular clouds are highly filamentary \citep[\eg][]{andre2010, molinari2010b, busquet2013}, which makes the filament hypothesis very plausible.

In the following we will discuss the nature of the elongated structure, confronting the two scenarios depicted above: 2D disk-like {\it ``pancake''} seen edge-on, versus 1D {\it filament}.

First of all, it is important to determine the velocity field of the elongated structure, and verify whether this could be rotating about the outflow axis. For this purpose we need a suitable tracer, detected all over the structure. As already shown in Fig.~\ref{f:largescale} and discussed in Sect.~\ref{s:largescale}, a few molecules show extended emission which is spatially coincident with the dust 870~$\mu$m continuum emission in Fig.~\ref{f:continuum}. Despite the limited sensitivity of our observations to angular scales $\ga$2\arcsec, the C$^{17}$O\,(3--2) and H$^{13}$CO$^+$\,(4--3) transitions can be imaged toward most of the elongated structure, and may hence be used to characterize its kinematics. In the top panels of Figs.~\ref{f:largescaleC17O} and~\ref{f:largescaleH13CO+}, we present the 870~$\mu$m continuum emission map overlaid on the first order moment (velocity field) maps of the two lines. The southeastern part of the filament is associated with red-shifted emission ($\sim$36--37~\kms), while the central and northwestern regions have emission at velocities $\sim$31--32~\kms. This velocity shift is more visible in the bottom panels of the two figures, which show the PV-plots of the two species along the major axis of the elongated structure (P.A.=$-$60\degr). In these plots, the velocity appears to shift gradually from $\sim$37~\kms\ close to cores~C and E (southeast), to $\sim$31~\kms\ close to cores~D and F (northwest).

\begin{figure}[t!]
\begin{center}
\begin{tabular}[b]{c}
 \epsfig{file=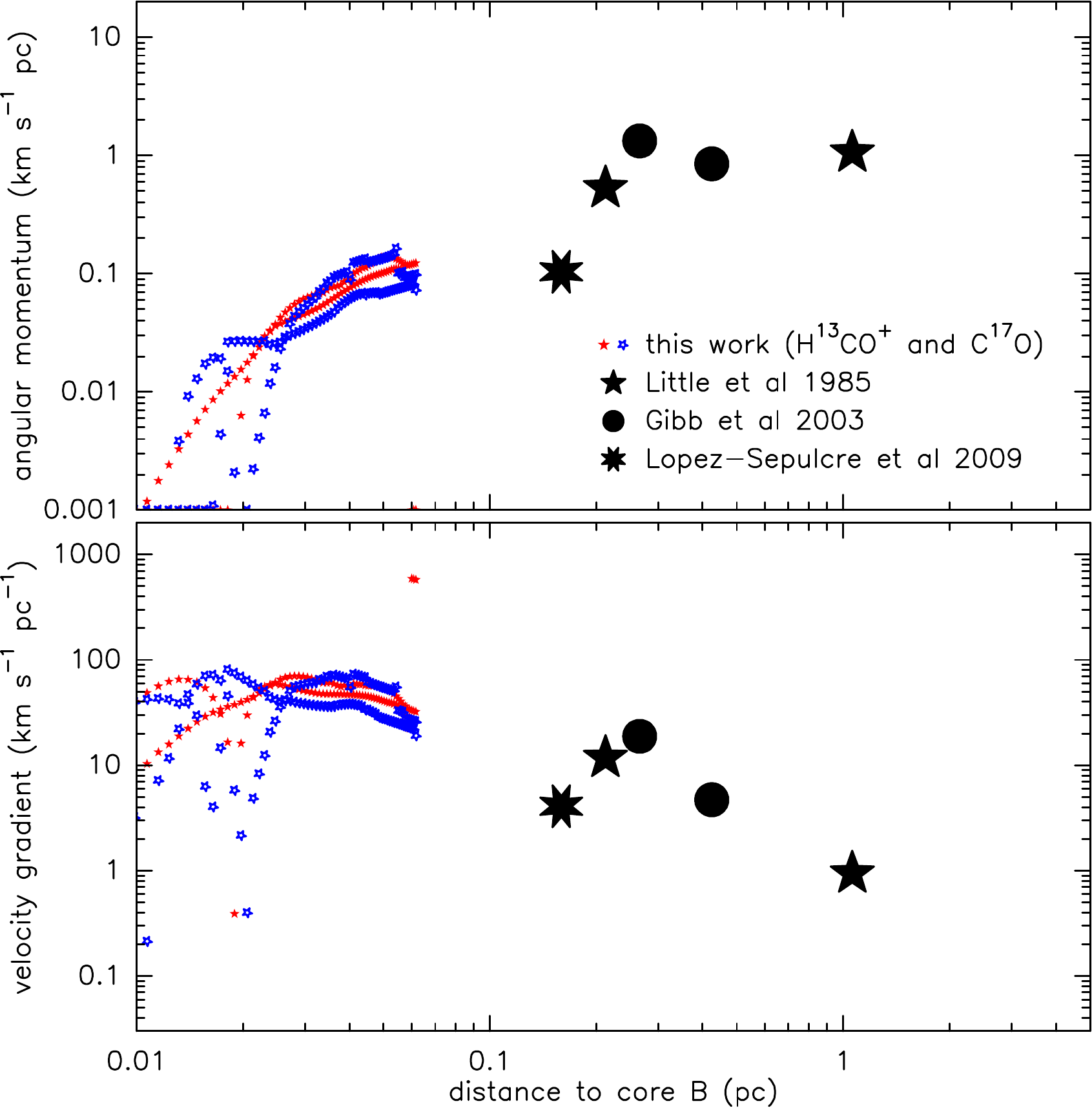, width=0.88\columnwidth, angle=0} \\
\end{tabular}
\caption{Specific angular momentum (\emph{top panel}) and velocity gradient (\emph{bottom panel}) at different radii traced by the available molecular line data. Small red and blue stars correspond to the H$^{13}$CO$^+$ and C$^{17}$O ALMA line data presented in this work, respectively. Black five-point stars: NH$_3$\,(1,1) and CS\,(2--1) line data from \citet{little1985}. Black filled circles: H$^{13}$CN\,(1--0) and H$^{13}$CO$^+$\,(1--0) line data from \citet{gibb2003}. Black eight-point star: C$^{18}$O\,(2--1) line data from \citet{lopezsepulcre2009}.}
\label{f:velocitygradient}
\end{center}
\end{figure}

The existence of a smooth velocity gradient seems to suggest that we are observing an edge-on, large-scale, rotating ``pancake''. Under this hypothesis one can plot the specific angular momentum as a function of distance from the center (assumed to be the position of core~B), complementing our data with data from the literature \citep{little1985, gibb2003, lopezsepulcre2009} that allow us to sample the putative rotating ``pancake'' on scales as large as $\sim$2~pc. In Fig.~\ref{f:velocitygradient} we show this plot, which clearly indicates that angular momentum is dissipated from the large to the small scale. Magnetic braking might be a plausible explanation of this effect, but this requires a magnetic field ($B$) perpendicular to the disk, whereas recent polarimetric observations \citep{qiu2013} suggest that the $B$ field is mainly aligned with the elongated structure. We will come back to the role of the $B$ field later on.

Another problem with the ``pancake'' scenario is that core~B would be counter-rotating with respect to the large-scale velocity field. Even more striking is that cores A and B appear to be kinematically inconsistent with respect to the putative ``pancake'', as one can see by comparing the gray-scale map (hot-core tracers) to the contour map (large-scale molecular emission) in Figs.~\ref{f:largescaleC17O} and~\ref{f:largescaleH13CO+}. A discrepancy between the velocities of the cores and that of the structure enshrouding them is difficult to justify in a rotating ``pancake'', but can be explained if the two cores are created by the dynamical interaction of different filaments. Is there any evidence of such a dynamical environment around G35.20?

\begin{figure*}[t!]
\begin{center}
\begin{tabular}[b]{c}
 \epsfig{file=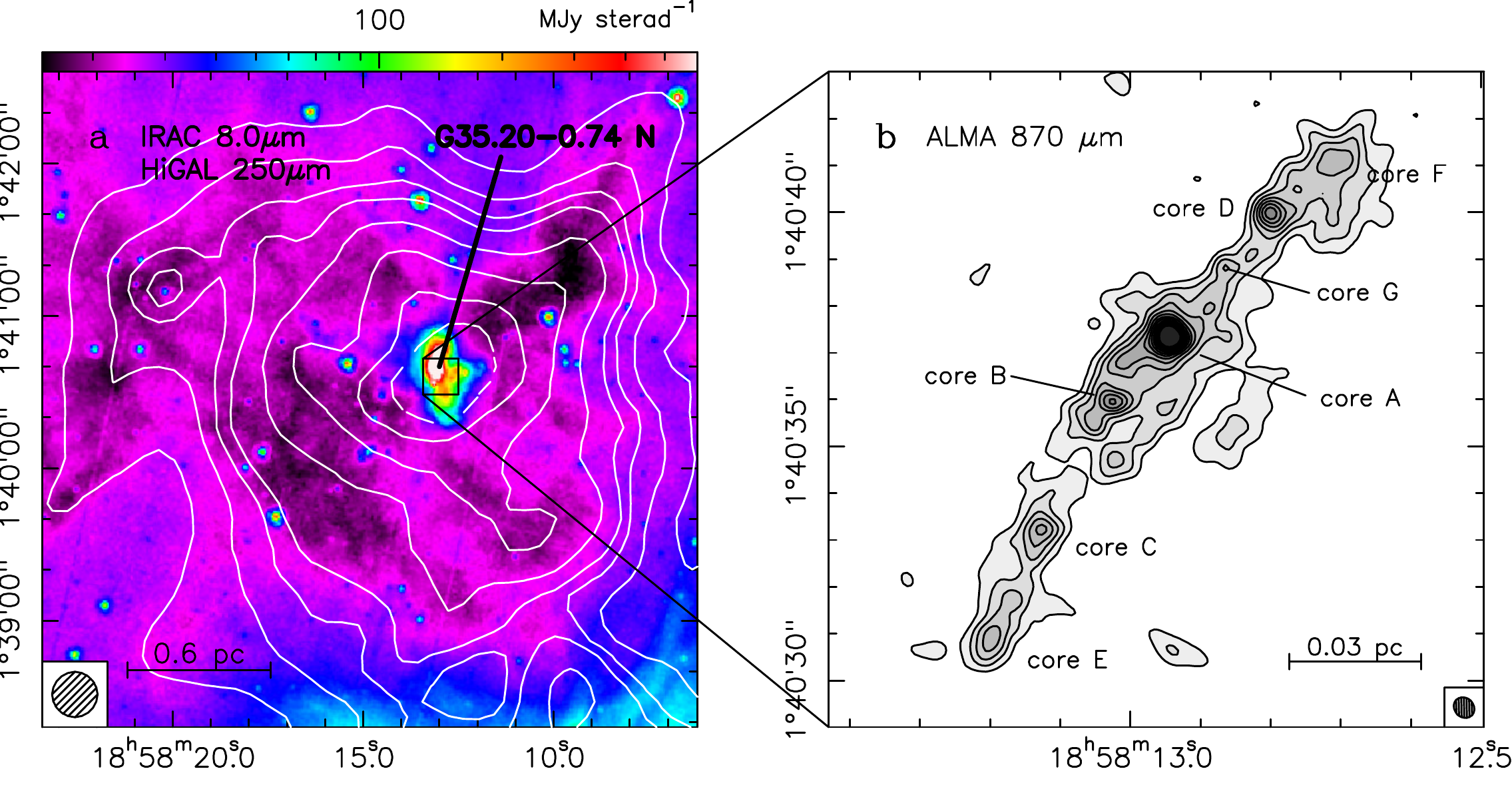, width=0.88\textwidth, angle=0} \\
\end{tabular}
\caption{{\bf (a)} Overlay of the Hi-GAL 250~$\mu$m continuum emission (\emph{white contours}) on the \emph{Spitzer}/IRAC 8.0~$\mu$m image (\emph{colors}) of the G35.20$-$0.74 region. Contour levels start at 65~m\jpb, increasing in steps of 0.39~\jpb\ until 1.885~\jpb, then levels increase in steps of 1.98~\jpb\ until the peak: 12~\jpb. {\bf (b)} Same as Fig.~\ref{f:continuum}b.}
\label{f:iracatlasgal}
\end{center}
\end{figure*}

Figure~\ref{f:iracatlasgal} presents a large-scale ($\sim$5\arcmin\ or 3~pc) view of the G35.20N star forming region. The \emph{Spitzer}/IRAC 8.0~$\mu$m image \citep{benjamin2003} reveals a series of elongated dark features converging towards the position of G35.20N, \eg\ the darkest feature, likely associated with the densest filament, towards the northwest, a curved arc-like structure towards the northeast, and two elongated structures to the south perpendicular to the main dark feature. The dust emission at 250~$\mu$m (from Hi-GAL; \citealt{molinari2010a}) matches the infrared dark features quite well, with the strongest emission associated with G35.20N, where the 8.0~$\mu$m emission also peaks. The overall structure (panel a) resembles that of the SDC335.579$-$0.272 infrared dark cloud \citep{peretto2013}, with multiple filaments converging and accreting matter towards the center, where a number of massive stars are forming. Our ALMA observations might be imaging the most prominent filament, where the accreting material is converging.

As previously noted, to discriminate an edge-on ``pancake'' from a filament, one could use the direction of the magnetic field, which is expected to be perpendicular to the disk plane and parallel to the filament. Recent SMA polarization measurements of the submillimeter continuum emission from G35.20N \citep{qiu2013} have found that the magnetic field on scales of $\sim$15\arcsec\ (0.16~pc) is almost parallel to the elongated structure, a result that supports the filament hypothesis. However, in this case the rotation axes of cores A and B should also be parallel to the filament, whereas they are significantly inclined with respect to it. The direction of the rotation axes is instead consistent with the {\it small-scale} direction of the magnetic field, obtained from observations of OH and CH$_3$OH masers \citep{hutawarakorncohen1999, surcis2012}, which indicates that the $B$ field is oriented in the (N)E-(S)W direction (see Fig.~5 of \citealt{surcis2012}) over $\sim$0\farcs3 (660~AU). The change of the direction of $B$ from the large to the small scale suggests that turbulence could play an important role in regulating the fragmentation of G35.20N at scales $\sim$0.1~pc. Turbulence might also explain the drastic change in the magnetic field orientation (by approximately 90\degr) toward the southeast of the elongated structure (see Fig.~2 by \citealt{qiu2013}).

One should also establish whether a big ``pancake'' could be unstable against perturbations and hence produce the observed fragments. This can be investigated by estimating the Toomre stability parameter for disks, $Q$. Assuming solid-body rotation (consistent with the PV-plots in Figs.~\ref{f:largescaleC17O} and~\ref{f:largescaleH13CO+}), this can be expressed as $Q=2\Omega\Delta V/(\sqrt{8\ln 2}\pi G \Sigma)$, with $\Omega$ angular velocity, $\Sigma$ surface density, and $\Delta V$ line FWHM. $\Omega\simeq40$~\kms\,pc$^{-1}$ is estimated from the slope of the PV-plots in Figs.~\ref{f:largescaleC17O} and~\ref{f:largescaleH13CO+}, while $\Sigma$ is given by the ratio between the mass of the elongated structure (30~\mo) and the disk surface ($\pi R^2$, where $2\,R=0.15$~pc is the length of the elongated structure). In order to obtain $Q\le1$, the condition for instability, the line width must be $\Delta V\le0.7$~\kms, a reasonable value for the initial conditions of the unperturbed gas, prior to star formation. We hence conclude that the ``pancake'' should break up into fragments.

%----------------------------------------------------------------------
%----------------------------------------------------------------------
%----------------------------------------------------------------------
%----------------------------------------------------------------------

The filament hypothesis may also explain the observed distribution of cores A--F. A filament (or isothermal gas cylinder) supported by `turbulence pressure' against its self-gravity has a maximum mass per unit length equal to
\begin{equation}
\left(M/l\right)_\mathrm{crit} = 2\sigma^2/G = 465\,\left(\sigma/1~\mathrm{km~s}^{-1}\right)^2\,M_\sun~\mathrm{pc}^{-1}
\end{equation}
\citep{nagasawa1987, wang2014}, over which it starts to fragment. The theoretical separation of the fragments is given by the expression
\begin{equation}
 \lambda = 9.39 \frac{\Delta V}{\sqrt{4\pi G\rho}}
\end{equation}
where $\Delta V$ is the line FWHM due to both turbulent at the thermal broadening, and relates to the velocity dispersion by $\Delta V=\sigma\,\sqrt{8\ln{2}}$. The line width to be used in these equations should be the initial value, before fragmentation, and not the current value, measured by us, which is likely to be broadened by additional turbulence injected by the newly formed stars. In practice, one sees that a value of $\Delta V=0.7$~\kms\ is required for the theoretical expression to match the observed spacing between the cores ($\sim$0.023~pc), after considering a filament with a mass $\sim$30~\mo, a size 0.15~pc$\times$0.013~pc, and a gas density of $\rho\simeq10^{-16}$~g~cm$^{-3}$. In addition, this line width of 0.7~\kms\ results in a critical mass per unit length of $\sim$40~\mo~pc$^{-1}$, while the measured mass per unit length in our filament is $\sim$200~\mo~pc$^{-1}$. Such an initial value of $\Delta V$ is plausible and we conclude that the filament should undergo fragmentation. Similar structures undergoing fragmentation have been found in other star forming regions such as G11.11$-$0.12 and G28.34$+$0.06 \citep{wang2011, wang2014}.

Based on these results, both a rotating ``pancake' and a filament appear to be unstable and hence consistent with the observed clumpiness. However, one should take into account also an intriguing feature of the elongated structure, namely the regular spacing between the cores. The distance between two adjacent cores lies between 0.020~pc and 0.036~pc, quite a narrow range of values. It seems unlikely that a random 2D distribution of fragments over a ``pancake'' can produce such a regular distribution, when seen in an edge-on projection on the plane of the sky. In conclusion, for this reason and the orientation of the magnetic field, we tend to favor the filament interpretation of the elongated structure.

%----------------------------------------------------------------------
\section{Summary\label{s:conclusions}}

We have observed the G35.20N star forming region with ALMA at 870~$\mu$m (350~GHz), achieving an angular resolution $\sim$$0\farcs4$ ($\sim$900~AU at 2.19~kpc), and covering a broad frequency range including dense gas (\eg\ H$^{13}$CO$^+$, C$^{17}$O), outflow (\eg\ SiO) and hot core (\eg\ CH$_3$CN, CH$_3$OH) tracers. Our conclusions can be summarized as follows:

\begin{itemize}

\item[-] The continuum emission shows an elongated dust structure (length $\sim$0.15~pc and width $\sim$0.013~pc), likely tracing the densest part of the elongated structure observed at larger scales, and perpendicular to the infrared and molecular large-scale outflow  \citep[\eg][]{gibb2003}. This structure is fragmented in six identified dense cores, although additional substructure is seen. The masses of fragments are between 1-10~\mo, and have sizes of around 1600~AU. The cores appear regularly spaced with a mean separation of $\sim$0.023~pc.

\item[-] Three out of the six dense cores show strong emission in complex organic molecules typical of hot cores. We fitted a large number of methyl cyanide (CH$_3$CN) and methanol (CH$_3$OH) spectral lines and derived temperatures of around 150--250~K for the three cores and relative abundances of 0.2--2$\times10^{-8}$ for CH$_3$CN and 0.6--5$\times10^{-6}$ for CH$_3$OH.

\item[-] The two densest and most chemically-rich cores (core~A and B) show a coherent velocity field with a velocity gradient almost aligned with the dust elongated structure. Interestingly, the orientation of both velocity gradients is opposite, with red-shifted emission toward the north for core~B and toward the south for core~A. These velocity gradients are consistent with Keplerian disks rotating about central masses of 4~\mo\ for core~A and 18~\mo\ for core~B, \ie\ intermediate/high-mass stars.

\item[-] Core~B seems to power a precessing jet/outflow oriented in the NE-SW direction. The combination of the IRAC 4.5~$\mu$m, the UKIDSS H$_2$ 2.12~$\mu$m and the VLA 3.6~cm images together with our ALMA continuum and line maps suggests that the outflow/jet associated with core~B is undergoing precession, in agreement with the binary system hypothesized by \citet{sanchezmonge2013b}. Our ALMA high-velocity SiO\,(7--6) map together with the VLA 3.6~cm continuum map suggest that core~A might be powering an outflow in the east-west direction. Further observations of outflow/shock tracers are required to confirm this scenario.

\item[-] The emission of dense gas tracers such as H$^{13}$CO$^{+}$ and C$^{17}$O is also extended and coincident with the dust emission. This large molecular structure is consistent, in morphology and velocity, with the SMA maps reported by \citet{qiu2013}. Based on the velocity field, the orientation of the magnetic field (as reported by \citealt{qiu2013}), and the regularly spaced fragmentation ($\sim$0.023~pc), we interpret the elongated dust structure as the densest part of a 1D filament fragmenting and forming high-mass stars.

\end{itemize}

%----------------------------------------------------------------------
\begin{acknowledgements}
We are grateful to the Italian ARC node for the usage of their computer facilities during the cleaning and imaging process. We are also grateful to Ray Furuya for providing us the ASTE spectrum, and to Keping Qiu for providing the SMA spectra. \'A.\ S.-M.\ is grateful to Peter Schilke, Thomas M\"oller and Alexander Zernickel for helping with the analysis of the myXCLASS software. This publication makes use of data products from the Wide-field Infrared Survey Explorer, which is a joint project of the University of California, Los Angeles, and the Jet Propulsion Laboratory/California Institute of Technology, funded by the National Aeronautics and Space Administration.
\end{acknowledgements}

%-------------------------------------------------------------------

%-------------------------------------------------------------------
\appendix

%----------------------------------------------------------------------
\section{Hot core intensity and velocity maps\label{a:images}}

\begin{figure}[tp!]
\begin{center}
\begin{tabular}[b]{c c}
 \epsfig{file=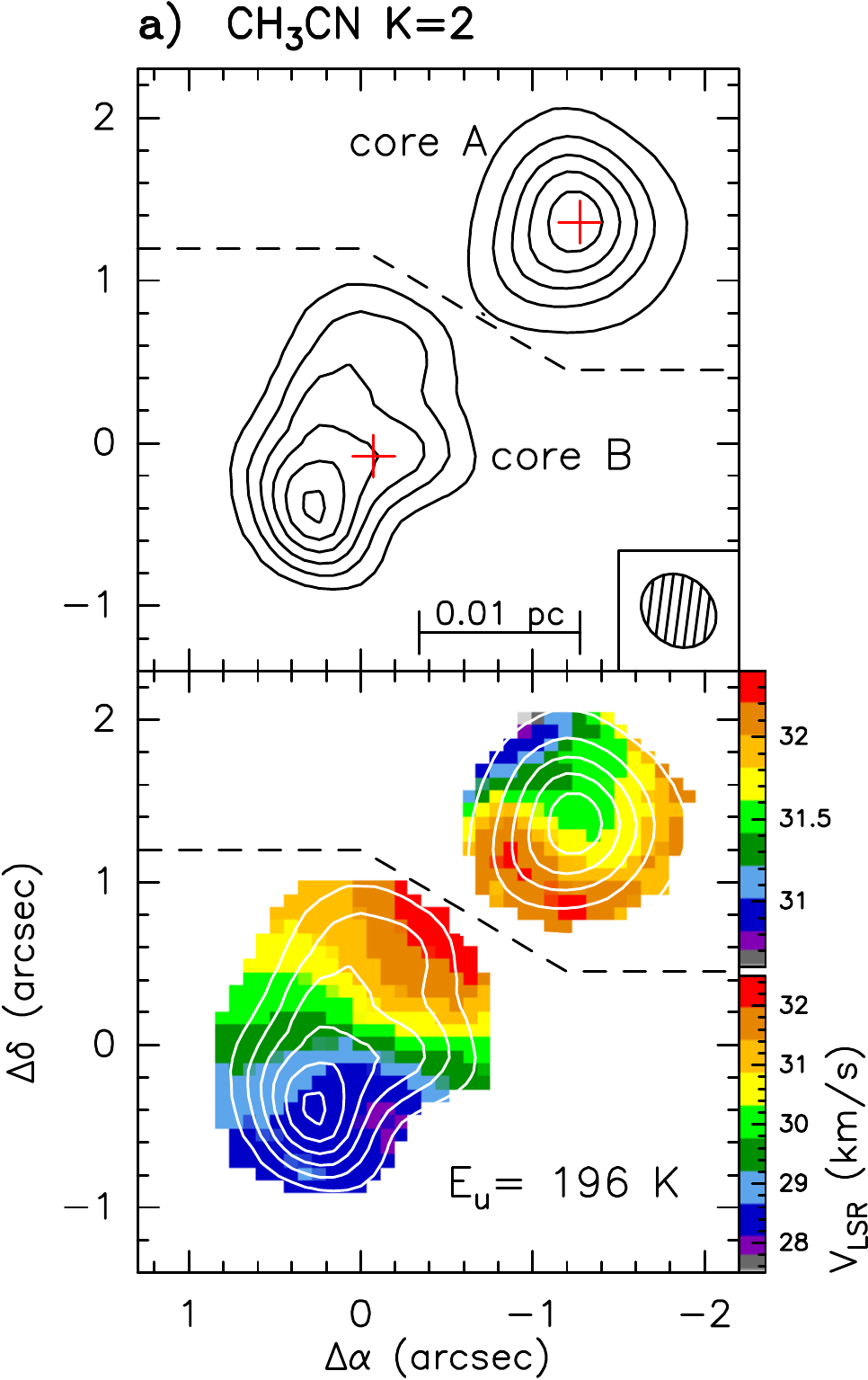, width=0.41\columnwidth, angle=0} &
 \epsfig{file=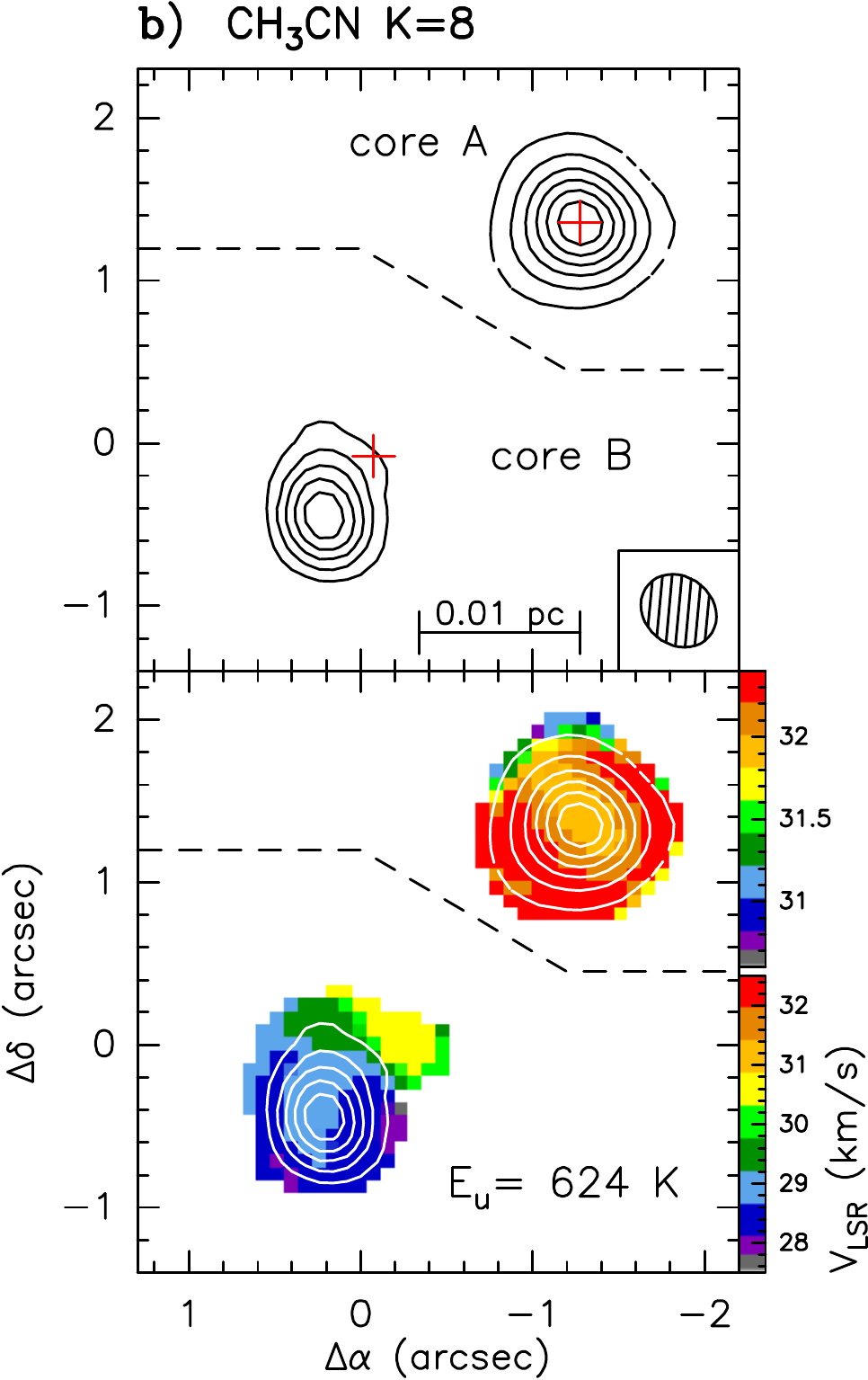, width=0.41\columnwidth, angle=0} \\
 \epsfig{file=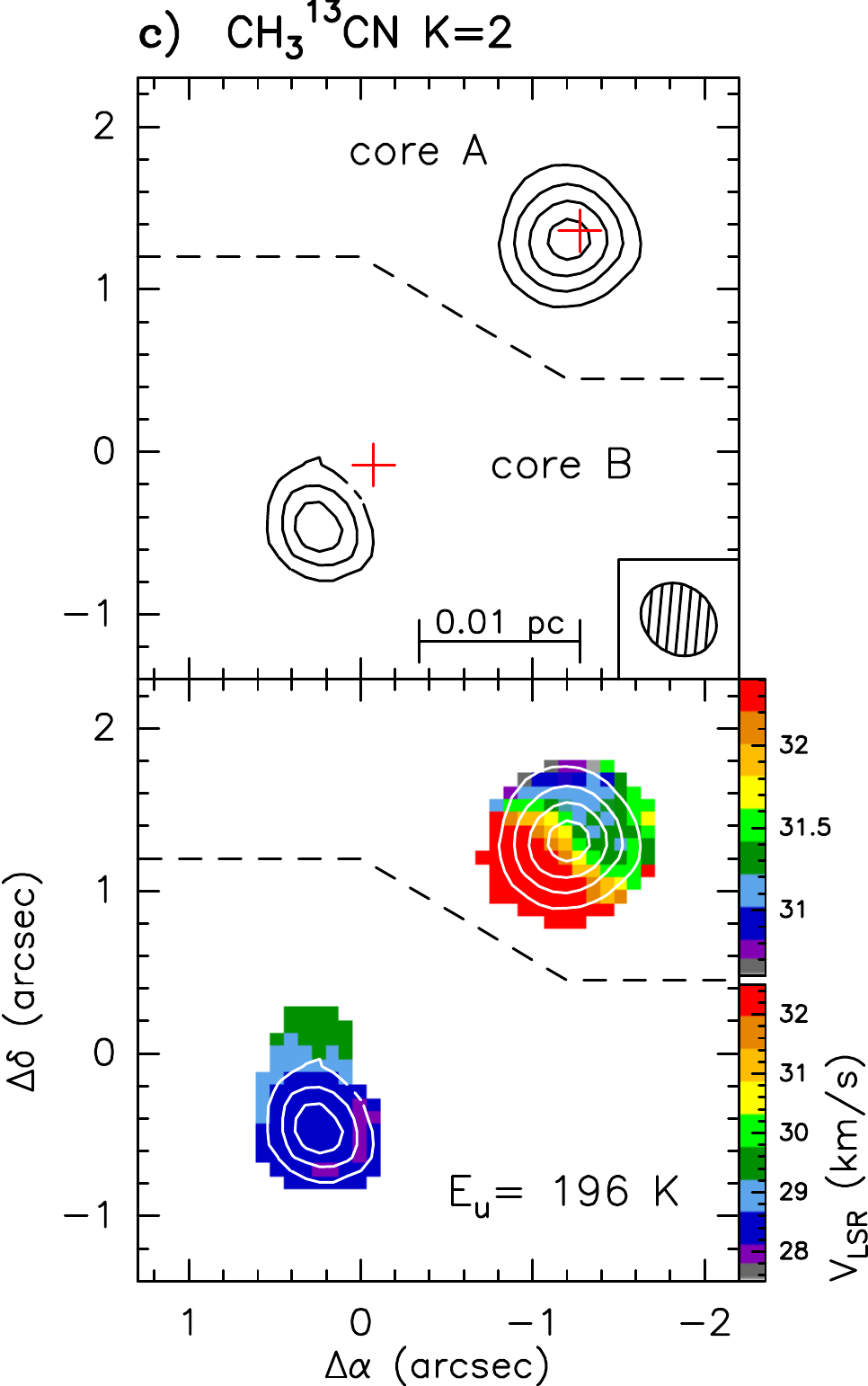, width=0.41\columnwidth, angle=0} &
 \epsfig{file=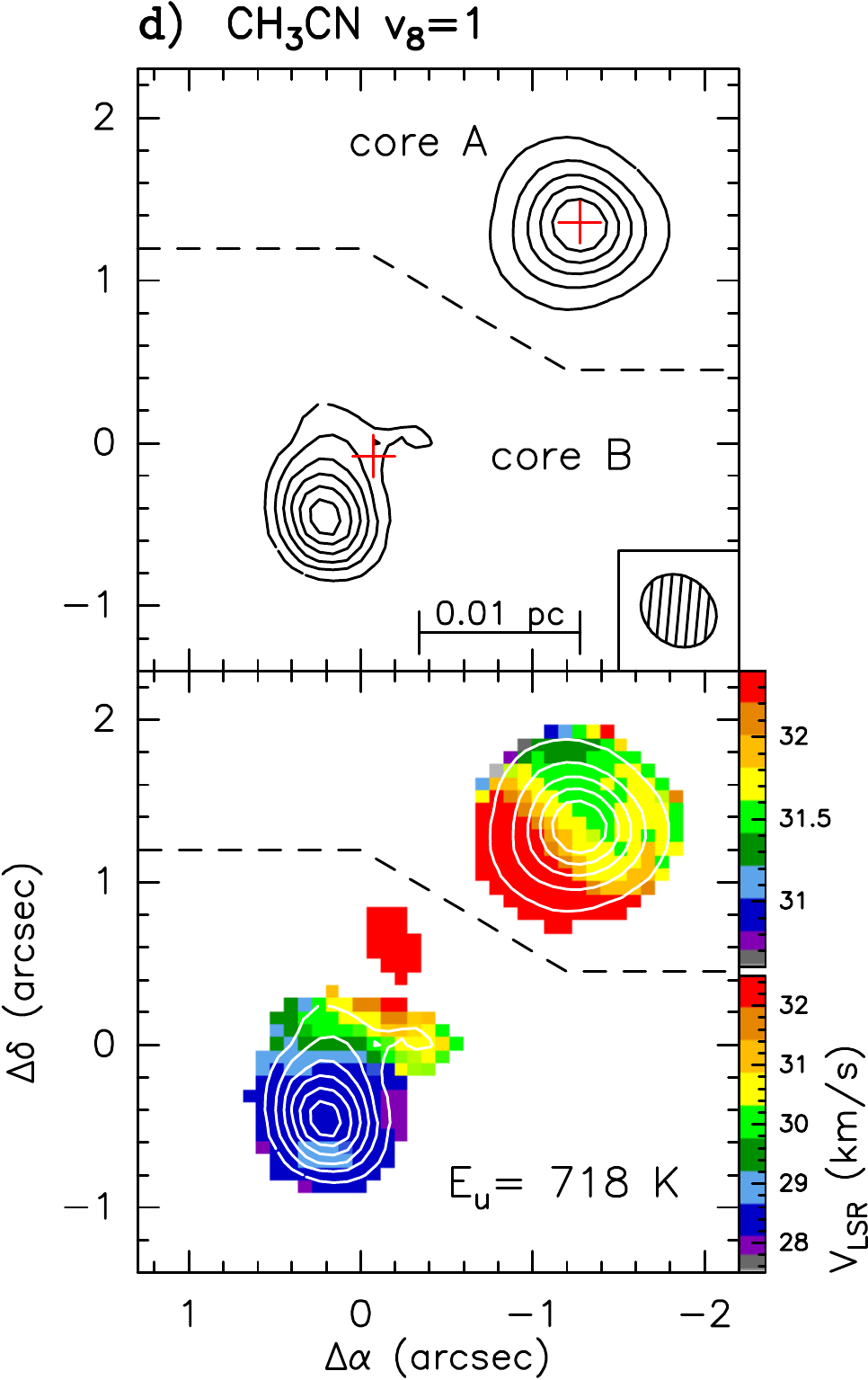, width=0.41\columnwidth, angle=0} \\
\end{tabular}
\caption{Maps of the integrated intensity (\emph{upper panels}) and velocity field (\emph{bottom panels}) of \textbf{a)} CH$_3$CN $K$=2 at 349.426~GHz, \textbf{b)} CH$_3$CN $K$=8 at 349.024~GHz, \textbf{c)} CH$_3$$^{13}$CN $K$=2 at 349.254~GHz, and \textbf{d)} CH$_3$CN $v_8$=1 $K,l$=3,1 at 350.552~GHz towards cores~A and B. For the different panels, the starting (and increasing contour levels in terms of $\sigma$, in \jpb~\kms) are 5$\sigma$ (10$\sigma$, with $\sigma$=0.15), 5$\sigma$ (10$\sigma$, with $\sigma$=0.15), 3$\sigma$ (3$\sigma$, with $\sigma$=0.10), and 5$\sigma$ (10$\sigma$, with $\sigma$=0.10) for core~A; and 5$\sigma$ (5$\sigma$, with $\sigma$=0.09), 4$\sigma$ (4$\sigma$, with $\sigma$=0.09), 2$\sigma$ (2$\sigma$, with $\sigma$=0.06), and 3$\sigma$ (3$\sigma$, with $\sigma$=0.06) for core~B. See details and description of symbols in Fig.~\ref{f:hotcore1}.}
\label{f:hotcore2}
\end{center}
\end{figure}

\begin{figure}[t!]
\begin{center}
\begin{tabular}[b]{c c}
 \epsfig{file=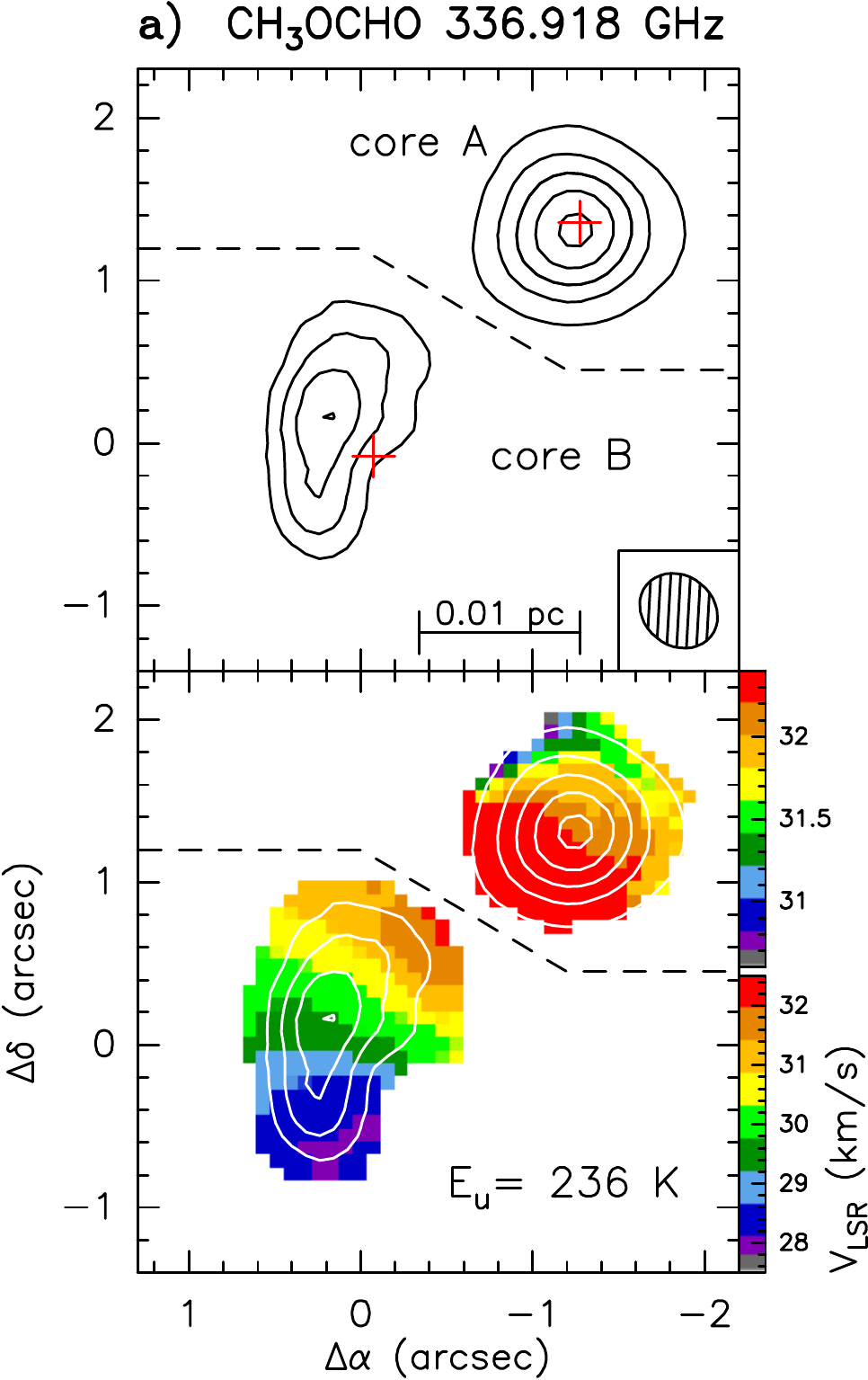, width=0.41\columnwidth, angle=0} &
 \epsfig{file=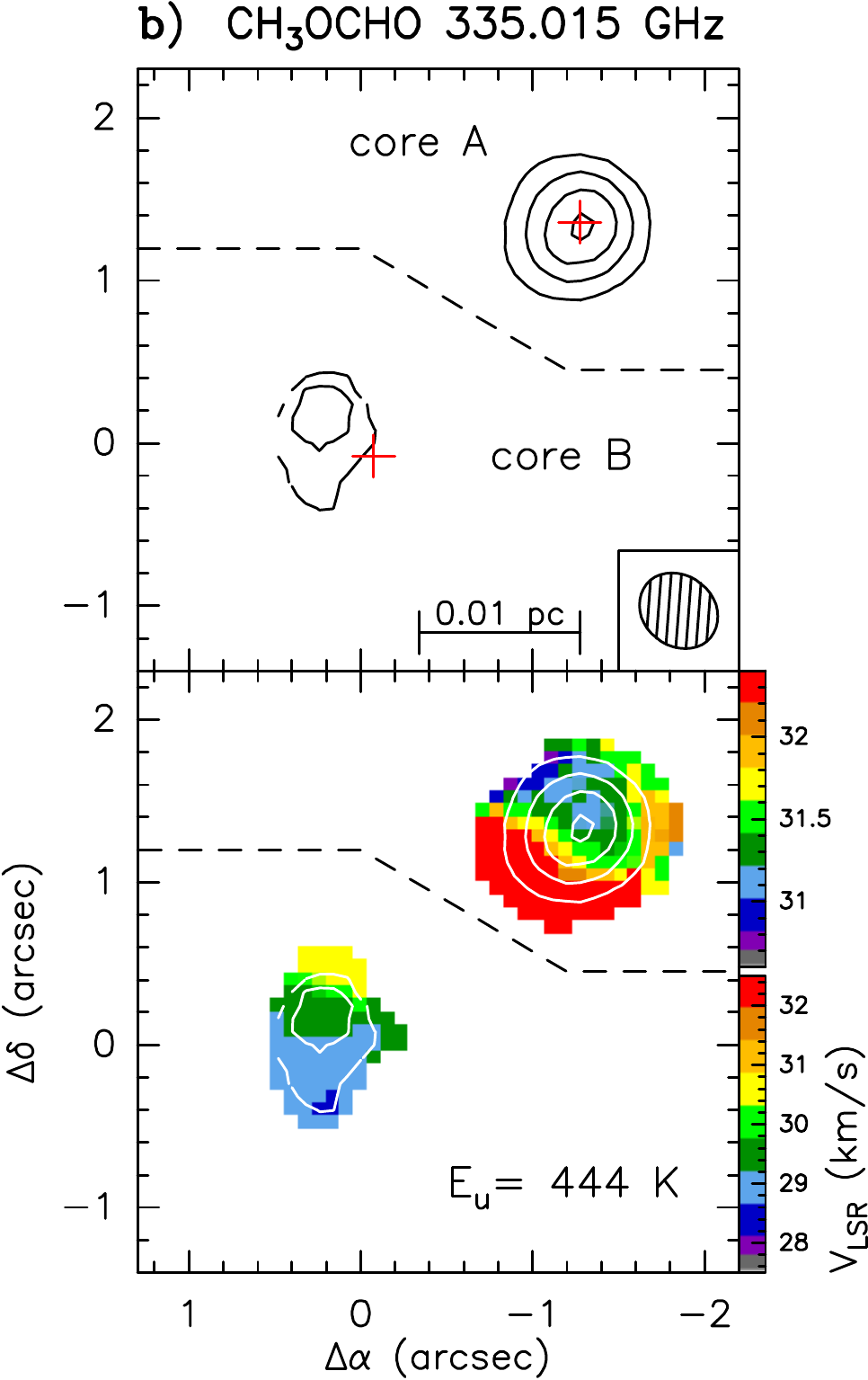, width=0.41\columnwidth, angle=0} \\
 \epsfig{file=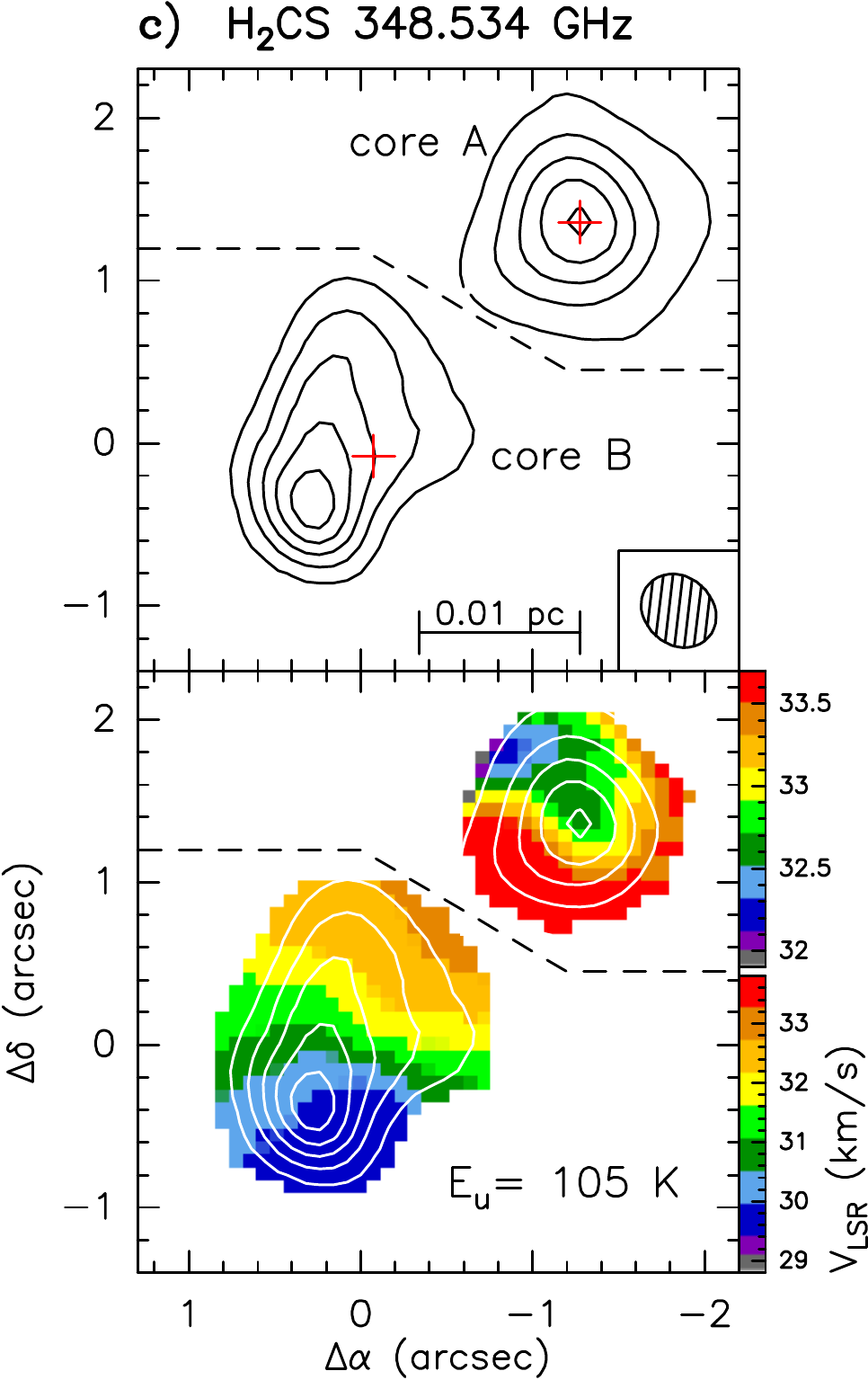, width=0.41\columnwidth, angle=0} &
 \epsfig{file=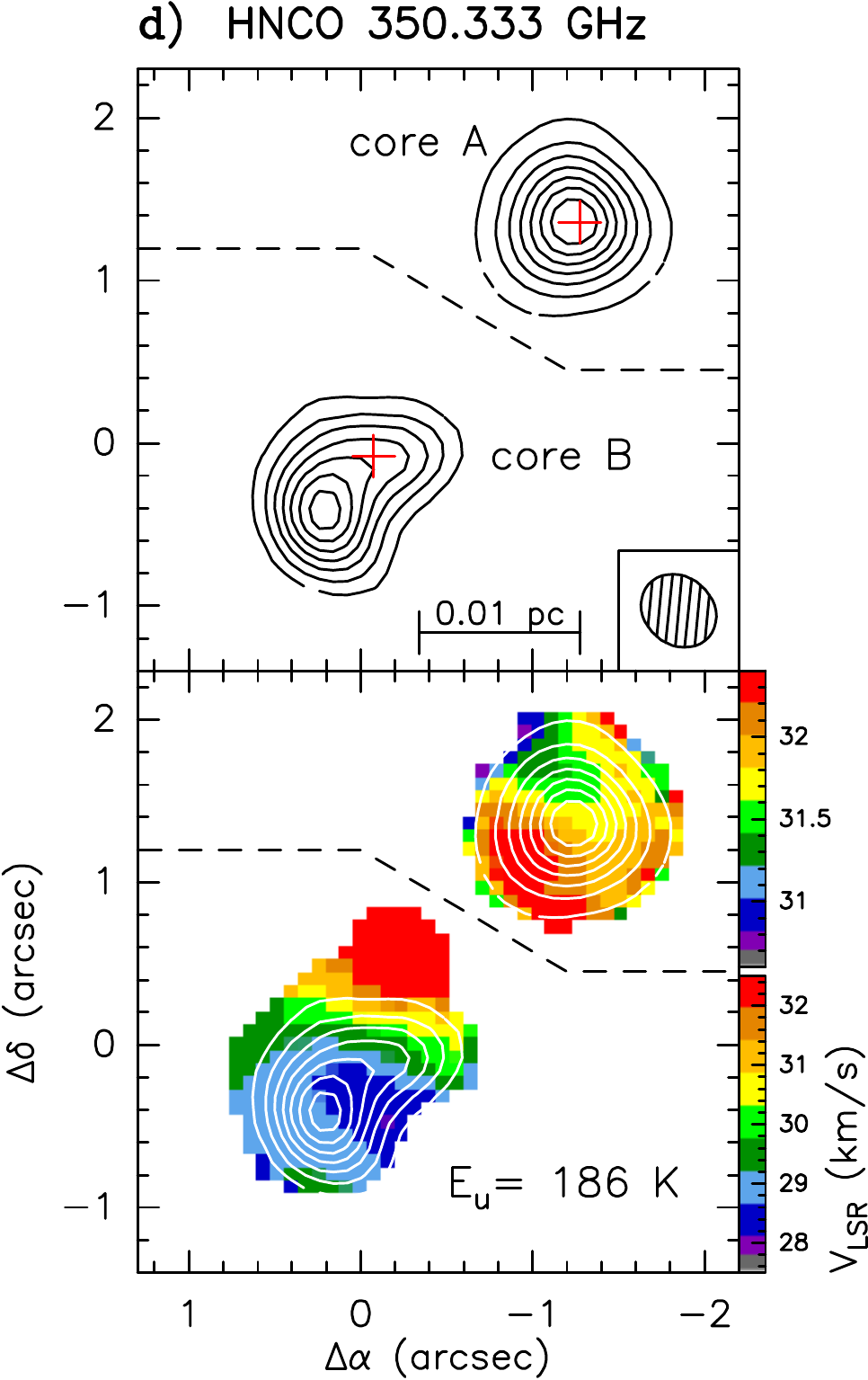, width=0.41\columnwidth, angle=0} \\
 \epsfig{file=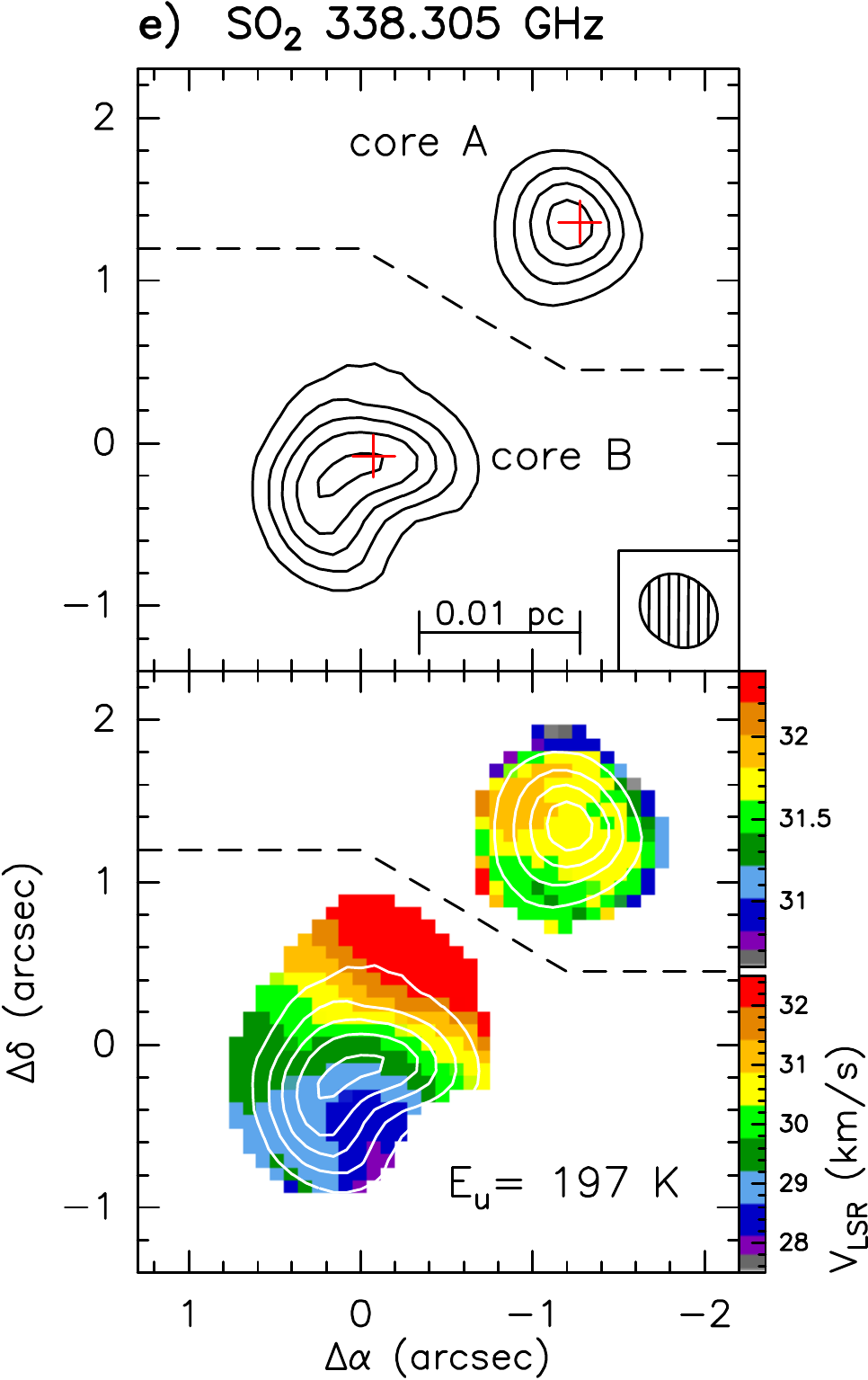, width=0.41\columnwidth, angle=0} &
 \epsfig{file=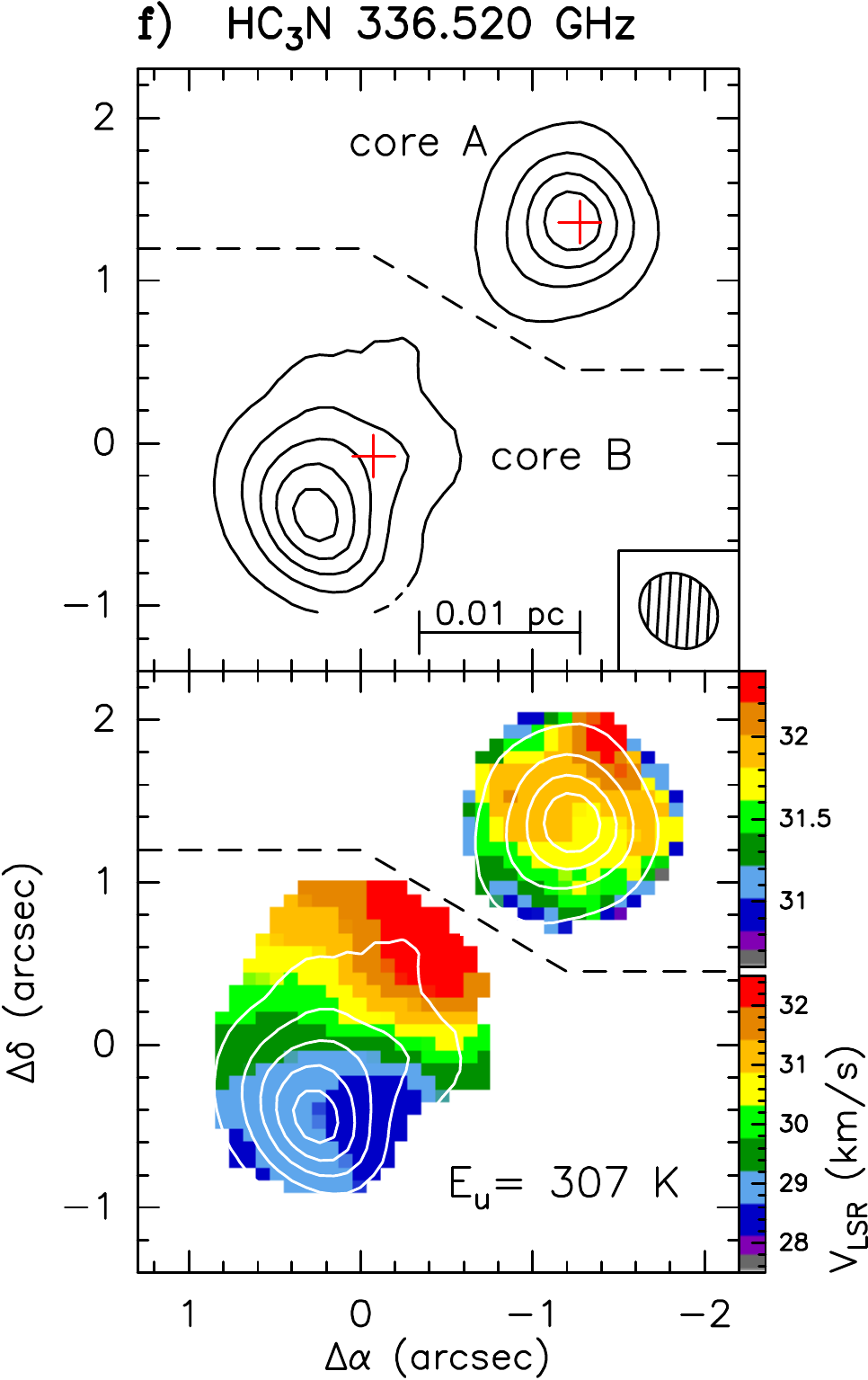, width=0.41\columnwidth, angle=0} \\
\end{tabular}
\caption{Maps of the integrated intensity (\emph{upper panels}) and velocity field (\emph{bottom panels}) of \textbf{a)} CH$_3$OCHO\,(26$_{6,20}$--25$_{6,19}$), \textbf{b)} CH$_3$OCHO\,(27$_{7,21}$--26$_{7,20}$), \textbf{c)} H$_2$CS\,(10$_{1,9}$--9$_{1,8}$), \textbf{d)} HNCO\,(16$_{1,16}$--15$_{1,15}$), \textbf{e)} SO$_2$\,(18$_{4,14}$--18$_{3,15}$), and \textbf{f)} HC$_3$N\,(37--36) towards cores~A and B. For the different panels, the starting (and increasing contour levels in terms of $\sigma$, in \jpb~\kms) are 5$\sigma$ (10$\sigma$, with $\sigma$=0.06), 3$\sigma$ (3$\sigma$, with $\sigma$=0.04), 5$\sigma$ (10$\sigma$, with $\sigma$=0.16), 5$\sigma$ (10$\sigma$, with $\sigma$=0.10), 5$\sigma$ (5$\sigma$, with $\sigma$=0.20), and 5$\sigma$ (10$\sigma$, with $\sigma$=0.11) for core~A; and 3$\sigma$ (3$\sigma$, with $\sigma$=0.03), 1$\sigma$ (1$\sigma$, with $\sigma$=0.02), 5$\sigma$ (5$\sigma$, with $\sigma$=0.09), 5$\sigma$ (5$\sigma$, with $\sigma$=0.06), 5$\sigma$ (5$\sigma$, with $\sigma$=0.10), and 5$\sigma$ (10$\sigma$, with $\sigma$=0.07) for core~B. See details and description of symbols in Fig.~\ref{f:hotcore1}.}
\label{f:hotcore3}
\end{center}
\end{figure}

In Figs.~\ref{f:hotcore2} and \ref{f:hotcore3}, we present the zeroth (integrated intensity) and first (velocity field) moment maps of various hot-core tracers detected towards cores~A and B. In Section~\ref{s:hotcores}, we describe how the moments have been obtained, while the symbols shown in these two figures are described in the caption of Fig.~\ref{f:hotcore1}.

%----------------------------------------------------------------------
%----------------------------------------------------------------------
%----------------------------------------------------------------------
%----------------------------------------------------------------------

%----------------------------------------------------------------------
\section{CH$_3$OH and CH$_3$CN myXCLASS fits\label{a:fits}}

\begin{figure*}[t!]
\begin{center}
\begin{tabular}[b]{c}
 \epsfig{file=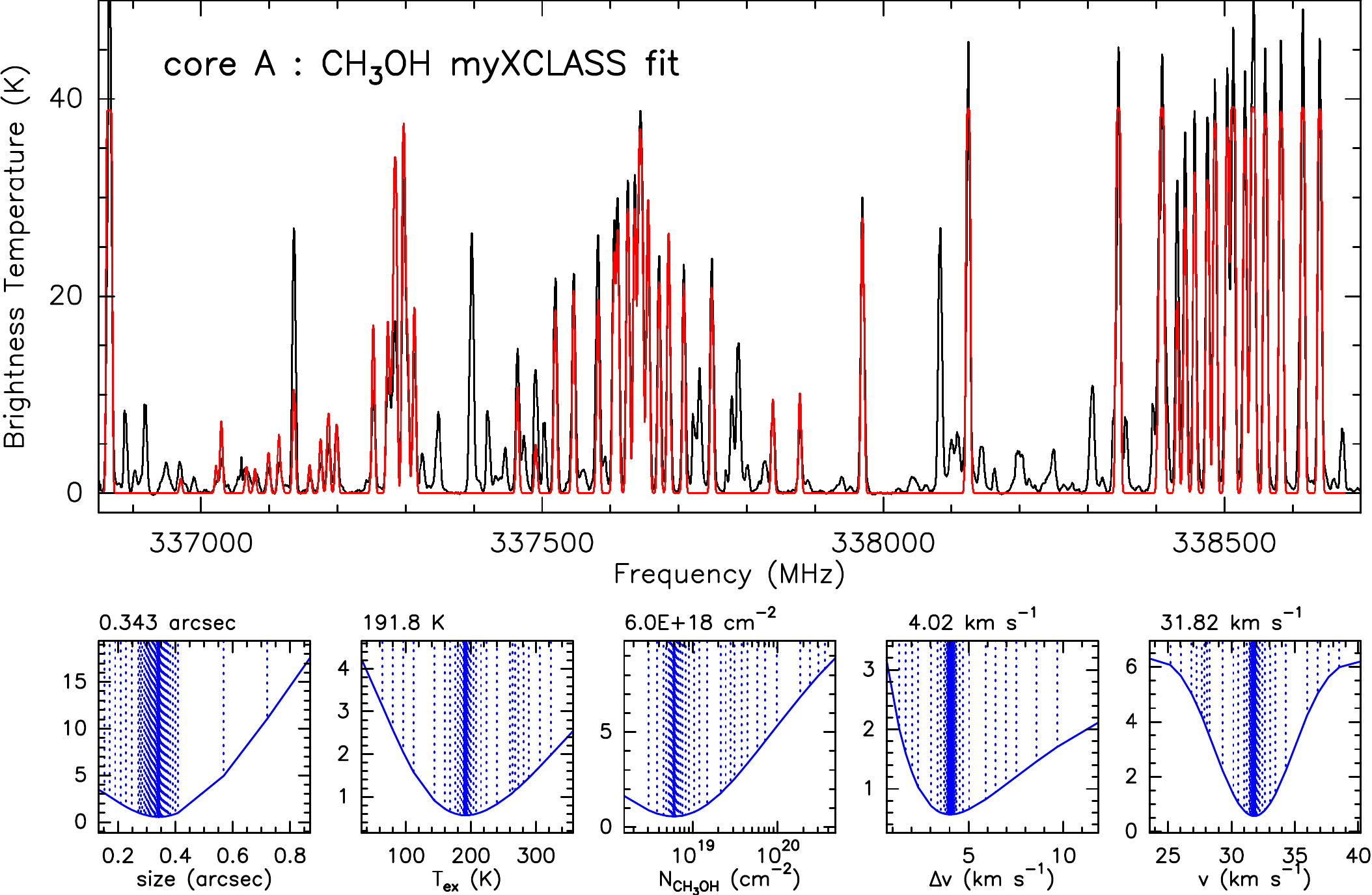, width=0.9\textwidth, angle=0} \\
 \\
 \\
 \epsfig{file=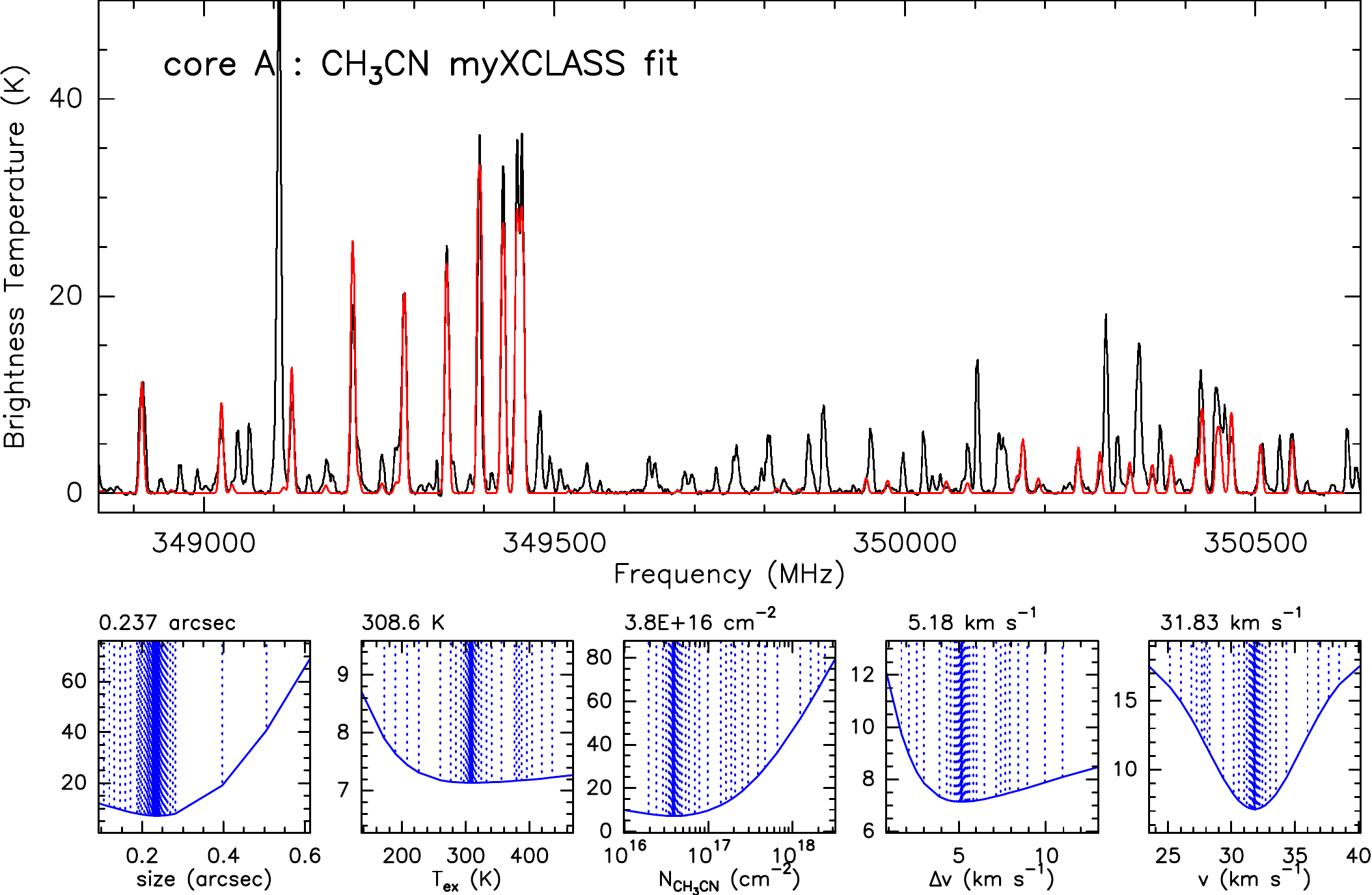, width=0.9\textwidth, angle=0} \\
\end{tabular}
\caption{The large panels show the observed CH$_3$OH (top panel) and CH$_3$CN (bottom panel) lines toward core~A in G35.20$-$0.74\,N. The red line shows the best fitted obtained by myXCLASS (see Section~\ref{s:temperature}. The fit in the top panel includes ground state CH$_3$OH and $^{13}$CH$_3$OH and torsionally excited CH$_3$OH lines, while the fit in the bottom panel includes ground state CH$_3$CN and CH$_3^{13}$CN and vibrationally excited CH$_3$CN lines in the frequency ranges shown in the panels. The small panels show the $\chi^2$ values obtained from the fits minimization. From left to right the panels show the size, the temperature, the column density, the linewidth and the LSR velocity.}
\label{f:coreAmyxclass}
\end{center}
\end{figure*}

\begin{figure*}[t!]
\begin{center}
\begin{tabular}[b]{c}
 \epsfig{file=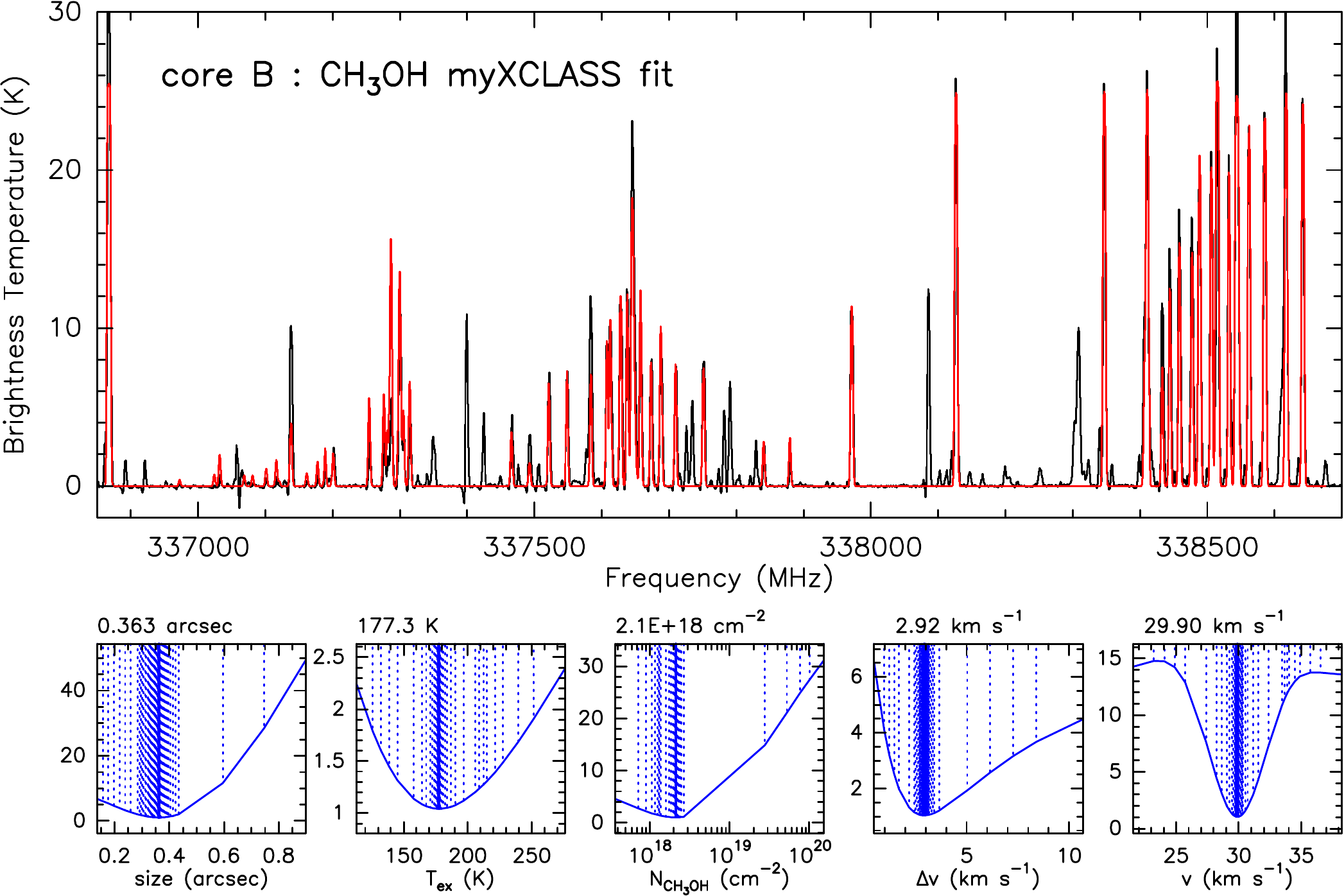, width=0.9\textwidth, angle=0} \\
 \\
 \\
 \epsfig{file=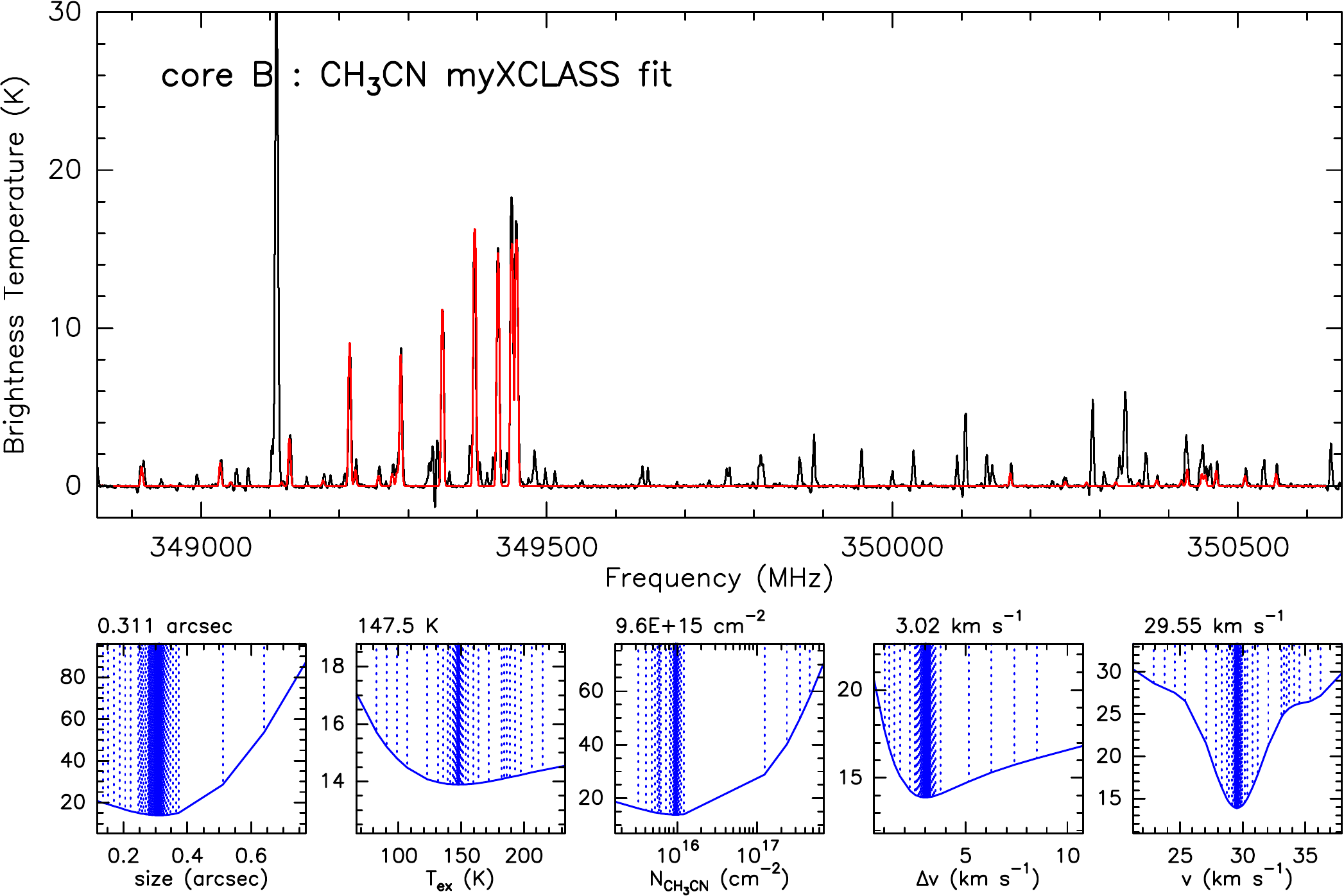, width=0.9\textwidth, angle=0} \\
\end{tabular}
\caption{Same as Fig.~\ref{f:coreAmyxclass} for core~B.}
\label{f:coreBmyxclass}
\end{center}
\end{figure*}

\begin{figure*}[t!]
\begin{center}
\begin{tabular}[b]{c}
 \epsfig{file=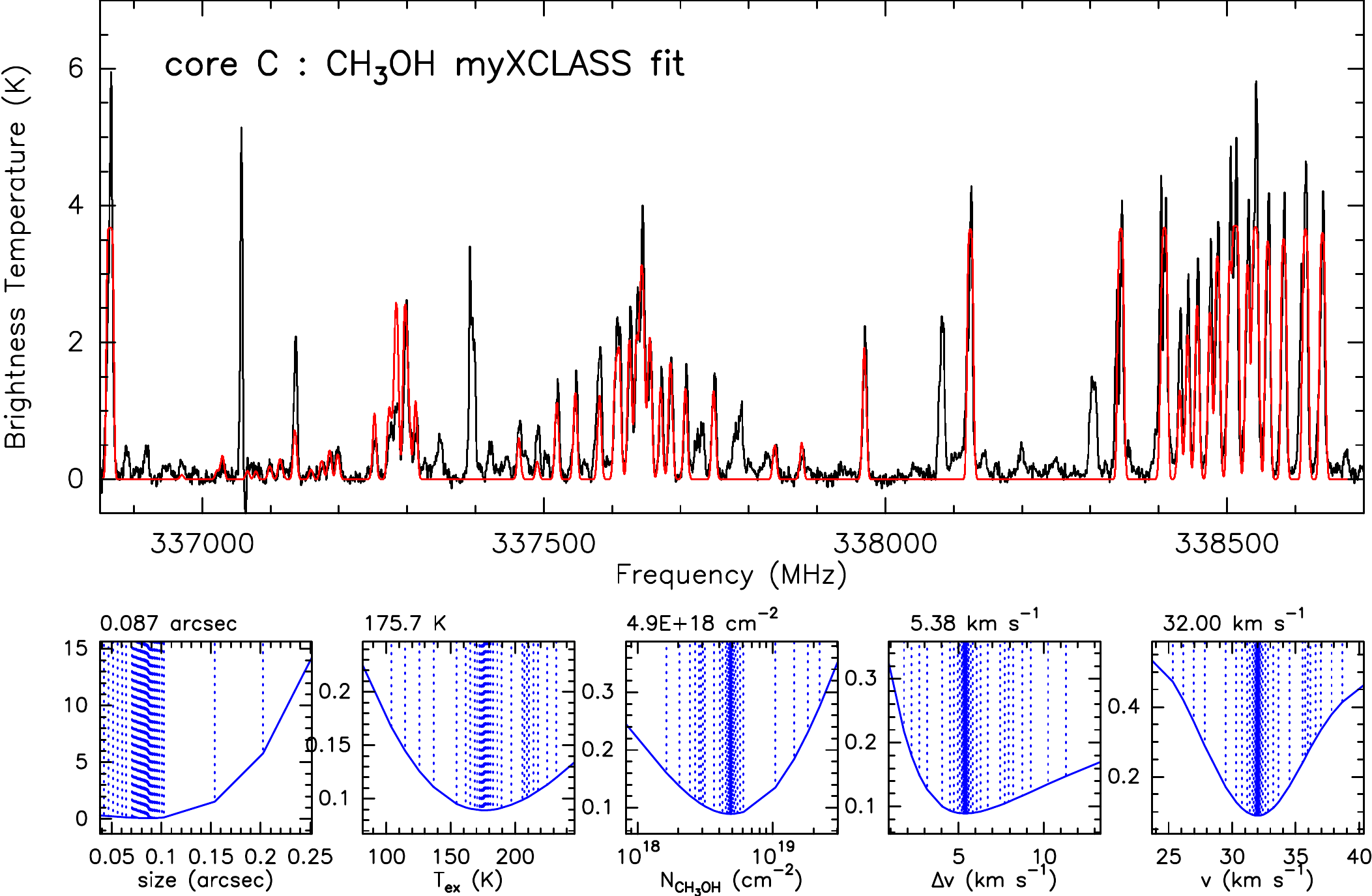, width=0.9\textwidth, angle=0} \\
 \\
 \\
 \epsfig{file=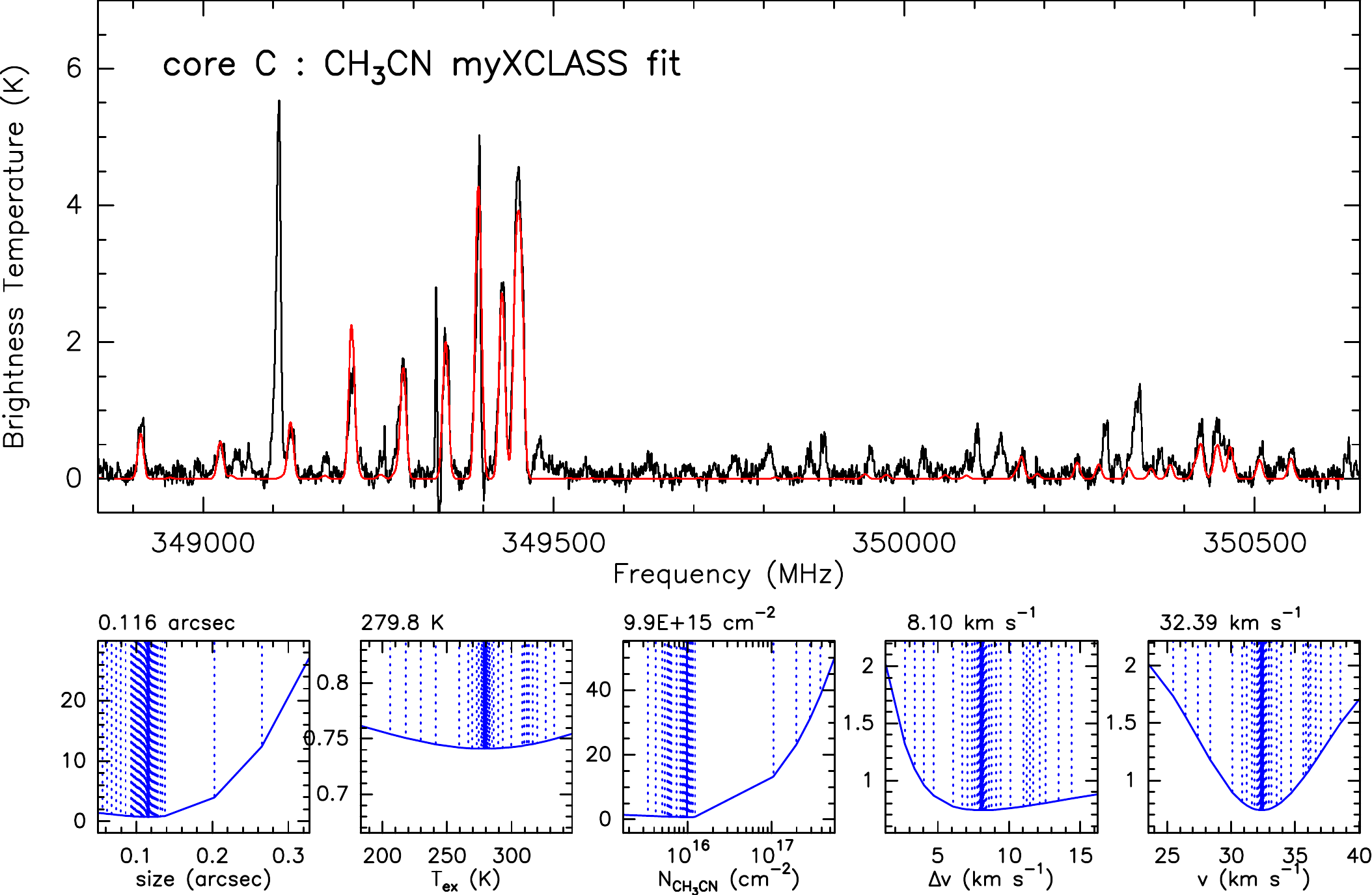, width=0.9\textwidth, angle=0} \\
\end{tabular}
\caption{Same as Fig.~\ref{f:coreAmyxclass} for core~C.}
\label{f:coreCmyxclass}
\end{center}
\end{figure*}

\begin{figure*}[t!]
\begin{center}
\begin{tabular}[b]{c}
 \epsfig{file=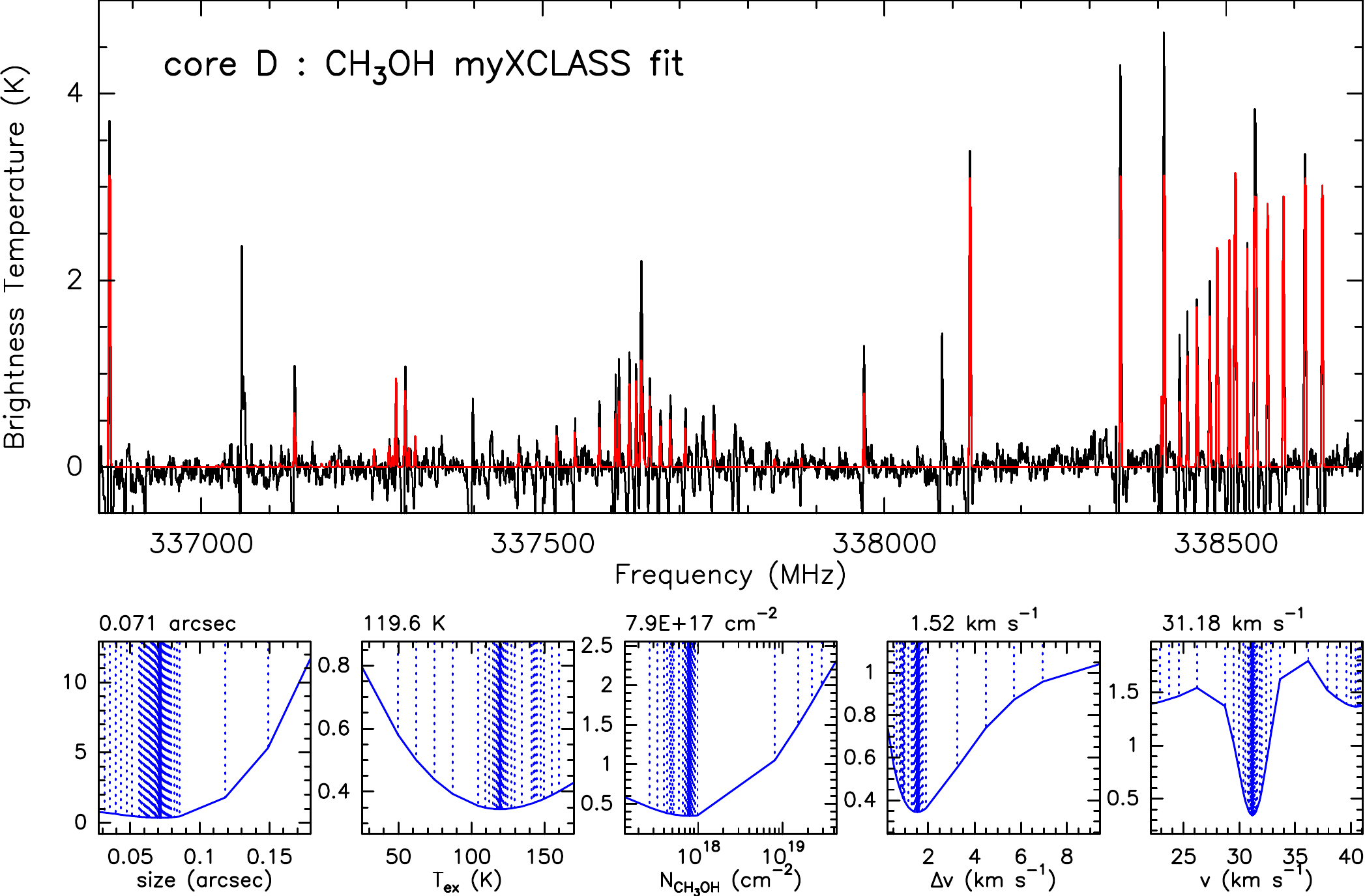, width=0.9\textwidth, angle=0} \\
\end{tabular}
\caption{Same as Fig.~\ref{f:coreAmyxclass} for core~D, and only for CH$_3$OH lines.}
\label{f:coreDmyxclass}
\end{center}
\end{figure*}

In Figs.~\ref{f:coreAmyxclass}--\ref{f:coreDmyxclass} we show the spectra of the bulge of methanol (CH$_3$OH, black line in the top panels) and methyl cyanide (CH$_3$CN, black line in the bottom panels) lines observed toward cores~A, B, C and D. For this last core, the methyl cyanide lines are too weak and cannot be fitted. The observed spectra have been fitted (red lines) using myXCLASS (see Section~\ref{s:temperature}). The procedure searches, by minimizing the $\chi^2$, for the best fit of five parameters: size (in arcsec), temperature (in K), column density (in cm$^{-2}$), linewidth (in \kms) and LSR velocity (in \kms). The blue lines in the small panels show the $\chi^2$ of the fits for different values of the parameters. In Table~\ref{t:myxclass}, we list the values of the parameters obtained in the best fits. We note that we have simultaneously fitted optically thick and thin transitions (\eg\ ground state, vibrationally excited and isotopologues of CH$_3$CN and CH$_3$OH), thus avoiding possible degeneracies \citep[see][]{zernickel2012} between the parameters of the fit (\eg\ column density and size).

%----------------------------------------------------------------------
%----------------------------------------------------------------------
%----------------------------------------------------------------------
%----------------------------------------------------------------------
%----------------------------------------------------------------------
%----------------------------------------------------------------------
%----------------------------------------------------------------------
%----------------------------------------------------------------------

\end{document}